\documentclass[%
nofootinbib,
 amsmath,amssymb,
 aps,
superscriptaddress
]{revtex4}

\usepackage[utf8]{inputenc}
\usepackage[T1]{fontenc}

\usepackage{slashed}
\usepackage{epsfig}
\usepackage{dsfont}
\usepackage{graphicx,color}
\graphicspath{{./} {./fig/}}

\usepackage{dcolumn}
\usepackage{bbold,bm}

\newcommand{\beq}{\begin{equation}}
\newcommand{\eeq}{\end{equation}}
\newcommand{\bea}{\begin{align}}
\newcommand{\eea}{\end{align}}

\newcommand{\cb}{\bar{c}}
\newcommand{\tr}{{\rm tr}}
\newcommand{\YM}{{\rm YM}}

\usepackage{slashed}
\usepackage[normalem]{ulem}
\usepackage{soul}

\definecolor{orange}{rgb}{.9,0.5,0.}
\definecolor{violet}{rgb}{.7,0.4,0.9}
\definecolor{vert}{rgb}{.3,0.7,0.}
\definecolor{bleu}{rgb}{.4,.7,1}
\definecolor{rouge}{rgb}{.8,0.2,0.2}
\definecolor{grey}{rgb}{.5,.8,.8}

\begin{document}

\title{A window on infrared QCD with
  small expansion parameters}

\author{Marcela Pel\'aez}%
\affiliation{%
 Instituto de F\'{\i}sica, Facultad de Ingenier\'{\i}a, Universidad de la Rep\'ublica, J.H.y Reissig 565, 11000 Montevideo, Uruguay.
}%

\author{Urko Reinosa}%
\affiliation{%
Centre de Physique Th\'eorique, CNRS, Ecole polytechnique, IP Paris, F-91128 Palaiseau, France.
}%
\author{Julien Serreau}%
\affiliation{%
 Universit\'e de Paris, CNRS, Astroparticule et Cosmologie, F-75013 Paris, France.
}%

\author{Matthieu Tissier}
\affiliation{LPTMC, Laboratoire de Physique Th\'eorique de la
  Mati\`ere Condens\'ee, CNRS UMR 7600, Sorbonne Universit\'e, \\ boite 121, 4 pl. Jussieu, 75252 Paris Cedex 05, France.
}
\author{Nicol\'as Wschebor}%
\affiliation{%
 Instituto de F\'{\i}sica, Facultad de Ingenier\'{\i}a, Universidad de la Rep\'ublica, J.H.y Reissig 565, 11000 Montevideo, Uruguay.
}%

\date{\today}

\begin{abstract}
  Lattice simulations of the QCD correlation functions in the Landau gauge have established two remarkable facts. First, the coupling constant in the gauge sector remains finite and moderate at all scales, suggesting that some kind of perturbative description should be valid down to infrared momenta. Second, the gluon propagator reaches a finite nonzero value at vanishing momentum, corresponding to a gluon screening mass. We review recent studies which aim at describing the long-distance properties of Landau gauge QCD by means of the perturbative Curci-Ferrari model. The latter is the simplest deformation of the Faddeev-Popov Lagrangian in the Landau gauge that includes a gluon screening mass at tree-level. 
  There are, by now, strong evidences that this approach successfully describes many aspects of the infrared QCD dynamics.
  In particular, several correlation functions were computed at one- and two-loop orders and
  compared with {\it ab-initio} lattice simulations. The typical error is of
  the order of ten percent for a one-loop calculation and drops to few
  percents at two loops. We review such calculations in the quenched
  approximation as well as in the presence of dynamical quarks. In the latter case, the spontaneous
  breaking of the chiral symmetry requires to go beyond a coupling expansion but can still be described in a controlled approximation scheme in terms of small parameters. We also review
  applications of the approach to nonzero temperature and
  chemical potential.
 \end{abstract}

\maketitle

\tableofcontents

\section{The ultraviolet Dr. Jekyll and the infrared Mr. Hyde}

Since the discovery of asymptotic freedom in the early 70's \cite{Politzer:1973fx,Gross:1973id}, a plethora of experimental and
theoretical works have firmly established quantum
chromodynamics (QCD) as the fundamental theory of strong
interactions. QCD seems paradoxical at first sight, however, being
a description of physically observable objects (the hadrons) in terms of unobservable
ones (the quarks and the gluons). The former are the
relevant excitations at length scales larger than about a Fermi
whereas the latter appear to be the relevant degrees of freedom at
smaller distances \cite{Weinberg:1996kr}. As is well-known, the coexistence of these
two, infrared (IR) and ultraviolet (UV) faces of the theory poses a
major difficulty for the detailed understanding of the properties and
interactions of hadrons. On the theory side, the dichotomy takes the
form of an essentially perturbative regime at short distances and/or
UV momenta versus a nonperturbative IR regime. This
picture actually extends to a wide class of QCD-like theories with
various quark contents, including the pure Yang-Mills (YM) case, with
no dynamical quarks, known as the quenched limit. It is based, on the one hand, on the phenomenon of
asymptotic freedom at UV scales and, on the other hand, on
the fact that standard perturbation theory predicts its own failure at
IR momenta in the form of a Landau pole, a finite energy scale
$\Lambda_{\text{QCD}}\sim 300$~MeV at which the coupling constant diverges.

Some essential aspects of the IR regime can be tackled from first principles using numerical simulations based on lattice gauge theory. In the QCD case, for instance, these simulations are able to reproduce the hadron spectrum with great accuracy using only the coupling constant and the quark masses as input parameters \cite{Fodor:2012gf}. More generally, lattice simulations give strong theoretical support in favor of two fundamental phenomena at IR scales, namely confinement---the fact that the physical excitations of the theory are massive colourless objects---and the dynamical breaking of the chiral symmetry for theories with not too many light quark flavours, including QCD \cite{Durr:2013goa}. Lattice simulations also study the rich phase structure of such theories under extreme conditions of temperature \cite{Borsanyi:2013bia,Bazavov:2014pvz} and density \cite{Sexty:2014dxa,Scorzato:2015qts,Gattringer:2016kco,Aarts:2016qrv}---relevant for quark-gluon plasma physics in high-energy nuclear collisions, astrophysics of ultra compact stars, or early-Universe cosmology---with, in particular, the possibility of a confinement-deconfinement transition and of a restoration of the chiral symmetry, depending on the details of the quark content. 

Although lattice simulations are by far the most powerful tool to
explore the IR sector of such theories, they remain limited in
at least two ways. First, they are based on the Monte Carlo sampling technique, which requires a positive definite measure of the (discretized) functional integral. They are thus essentially limited, so far, to the calculation of static quantities, that can be accessed from the Euclidean action when the latter is real. Typical dynamical quantities, which involve Minkowskian momenta (cross sections, transport coefficients, etc.), or cases where the Euclidean action is complex (for instance for some theories---including QCD---at nonzero chemical potential\footnote{This problem is usually known in the literature as the ``sign problem''.}) are still largely out of reach to present-day lattice technology. The second important limitation is that Monte Carlo simulations are, to a large extent, a black box from which it is not easy to pinpoint which are the fundamental phenomena at play. For instance, the basic dynamics of confinement remains, to date, largely unknown.

Alternatives to lattice methods typically involve functional quantum
field theory techniques, such as Dyson-Schwinger equations (DSE) \cite{Smekal1997a,Atkinson:1998zc,Alkofer:2000wg,Lerche:2002ep,Fischer:2002hna,Fischer:2006vf,Fischer:2008uz,Cyrol:2016tym}, the
functional renormlalisation group (FRG) \cite{Ellwanger:1996wy,Pawlowski:2003hq,Fischer:2004uk,Fischer:2008uz,Cyrol:2016tym}, and the  variational Hamiltonian
approach (VHA) \cite{Schleifenbaum:2006bq}. Although those, in principle, do not suffer from the
limitations mentioned above for lattice simulations, they come with
their own caveats. First, they necessarily rely on some approximations which
are often difficult to control in a systematic way in the
nonperturbative regime. Second, contrarily to lattice techniques, which
can directly access the physical observables of the theory, the
continuum approaches\footnote{In this article, for simplicity, we
  shall use the denomination ``continuum'' to encompass all theoretical approaches,
    except for lattice simulations.} are formulated in terms of the quark and gluon
degrees of freedom and it is often quite involved to extract information for physically observable quantities such as, {\it e.g.}, hadron masses.

\begin{figure}[t]
\centering
\includegraphics[width=.45\linewidth]{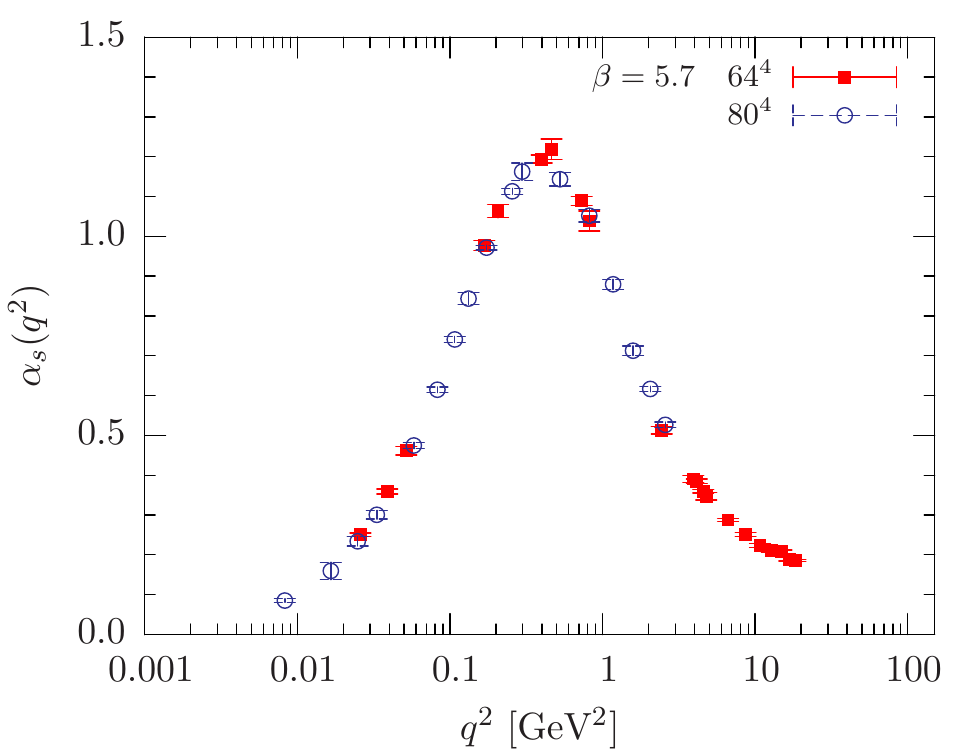}
 \caption{\label{Fig:lattice-Landau1} Lattice results for the Landau gauge Taylor coupling $\alpha_S(q^2)$ in the SU($3$) YM theory. Figure from Ref.~\cite{Bogolubsky:2009dc}.}
\end{figure}

Having brushed this broad panorama, let us now draw the main contours
of this review article. The above paradigmatic picture of a weakly
coupled high-energy regime versus a strongly coupled one in the IR
essentially relies on the Faddeev-Popov (FP) approach to perturbation
theory
\cite{Faddeev:1967fc} ,whose predictions are, however, in contradiction with some {\it ab-initio}
  results. In particular, the aforementioned
  Landau pole is most probably spurious. Indeed, numerical
  calculations of pure YM theories in the Landau gauge
show that, in the UV, the coupling\footnote{To be precise,
  the present statements concern the Taylor coupling, which
  characterizes the ghost-antighost-gluon vertex at vanishing ghost
  momentum \cite{Taylor:1971ff}; see Section \ref{sec_previous_semianalytical}. For a review of the strong coupling constant, see Ref. \cite{Deur:2016tte}.}  increases with
decreasing momentum scale as predicted by the FP perturbation theory, but
it remains finite at all scales and even decreases in the deep
IR \cite{Bogolubsky:2009dc}, see Fig.~\ref{Fig:lattice-Landau1}. Even more, one observes that the actual loop-expansion parameter---that is, in $d=4$ dimensions, $N_c\alpha_S/(4\pi)$, with $N_c=3$ the number of colours---never exceeds moderate values, of the order of $0.3$, thus potentially opening a novel perturbative way in the IR regime.
 This does not imply that the IR sector is fully perturbative--- spontaneous chiral symmetry breaking is one among many examples of phenomena that is not captured by a coupling expansion at any finite order---but this indicates that it is neither genuinely nonperturbative and that, at least in the Landau gauge, some aspects of the IR dynamics may admit a perturbative description. This stunning observation has, surprisingly, never reached to the broad audience it deserves and remains largely unknown even to the QCD community. It is one goal of the present article to advertise it to its full merit and to discuss in detail its far-reaching consequences.
 
 \begin{figure}[t]
\centering
 \includegraphics[width=.44\linewidth]{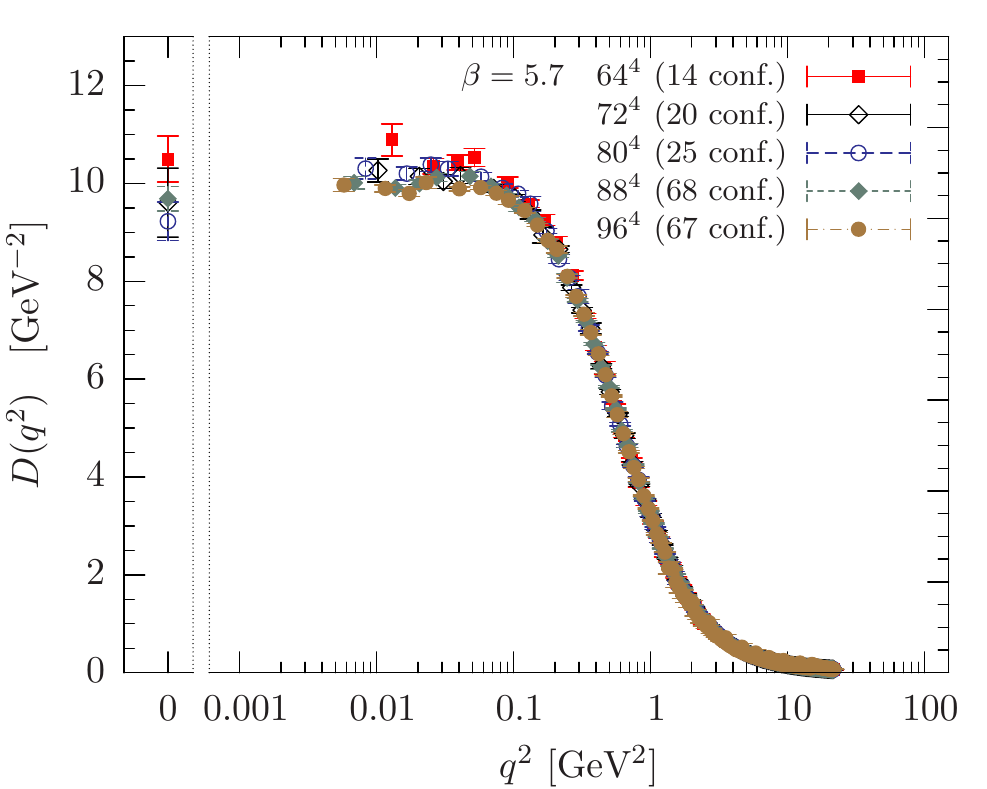} 
 \caption{\label{Fig:lattice-Landau2} Lattice results for the
   gluon propagator in the SU(3) YM theory in the
     Landau gauge. 
Figure from Ref.~\cite{Bogolubsky:2009dc}. In this reference, $D(q)$ stands for the gluon propagator, denoted $G(q)$ in the present article.}
\end{figure}

Because the FP theory predicts a Landau pole, an IR
 perturbative approach necessarily involves a modified starting point.
 One famous possibility is the Gribov-Zwanziger (GZ)
 quantization scheme \cite{Gribov:1977wm,Zwanziger:1989mf}, which aims at tackling the Gribov-Singer
 ambiguity \cite{Gribov:1977wm,Singer:1978dk} that plagues the FP theory in the IR. The
 approach reviewed in this article follows a different, more
 phenomenological route which exploits another important result of
 lattice simulations in the Landau gauge, namely, the fact that the
 gluon propagator reaches a finite nonzero value at vanishing
 momentum, corresponding to a nonzero screening mass, as shown in Fig.~\ref{Fig:lattice-Landau2}. This massive-like behaviour is of the utmost importance but, surprisingly again, has remained largely unknown even
 in the QCD community.
 
This suggests a simple massive deformation of the Landau gauge FP
 Lagrangian, with a tree-level gluon mass term \cite{Tissier:2010ts}, which actually
 corresponds to a particular case of the Curci-Ferrari (CF)
 Lagrangian \cite{Curci:1976bt}.\footnote{A possible relation between the gluon mass
   term and the Gribov-Singer ambiguity problem has been investigated
   in \cite{Serreau:2012cg,Nous:2020vdq}.} Note that the gluon mass is introduced in a gauge-fixed setting and thus does not, {\it per se}, break the gauge invariance of the theory.\footnote{The more subtle question of the Becchi-Rouet-Stora-Tyutin symmetry of the gauge-fixed theory is discussed in Sect.~\ref{sec_renorm}.} Among the interesting properties of this model
 let us mention i) its perturbative renormlalisability in four
 dimensions;  ii) the fact
   that one recovers the standard FP theory together with the
   associated phenomenology in the UV regime; and iii) the key observation
 \cite{Tissier:2011ey} that the mass term suffices to screen the IR
 divergences responsible for the Landau pole of the FP theory, thereby
 allowing for a well-defined perturbative expansion all the way from
 the UV to the deep IR. 

In this context, the working hypothesis is that the CF model provides
a good starting point for an efficient perturbative description of
various IR aspects of QCD-like theories. This is to be viewed as an effective description, where essential aspects of
the IR dynamics (in the Landau gauge) are efficiently captured by a
simple gluon mass term---which has to be fitted against some ({\it
  e.g.}, lattice or experimental) data---and where the residual interactions can then
be treated perturbatively. This idea has been first put forward in
Refs.~\cite{Tissier:2010ts,Tissier:2011ey}, where the Euclidean ghost
and gluon propagators of YM theories in the Landau gauge were computed
at one-loop order in the CF model and successfully compared to lattice
data for different gauge groups and different spacetime
dimensions. This exciting observation has triggered systematic studies
of modified (IR) perturbative descriptions based on the CF model. This
includes the genuine CF model, viewed as a phenomenological proxy of
a---yet to be found---IR completion of the FP Lagrangian
\cite{Pelaez:2013cpa,Reinosa:2013twa,Pelaez:2014mxa,Reinosa:2014ooa,Reinosa:2014zta,Pelaez:2015tba,Reinosa:2015gxn,Reinosa:2015oua,Reinosa:2016iml,Pelaez:2017bhh,Reinosa:2017qtf,Gracey:2019xom,Pelaez:2020ups,Reinosa:2020mnx,Barrios:2020ubx,Serreau:2020clz,Barrios:2021cks,Hayashi:2018giz,Hayashi:2020few,Kondo:2015noa,Kondo:2019rpa,Weber:2011nw,DallOlio:2020xpu,Kojo:2021knn,Song:2019qoh,Suenaga:2019jjv},
as well as the screened perturbation theory approach \cite{Siringo:2015jea,Siringo:2015aka,Siringo:2015wtx,Siringo:2016jrc,Siringo:2017svp,Comitini:2017zfp,Comitini:2020ozt},
where the gluon mass term is added and subtracted to the FP Lagrangian
and where the subtracted mass is treated, together with the coupling,
in a perturbative expansion around the CF Lagrangian. A large number
of quantities of physical interest have been computed, either in the
vacuum or at nonzero temperature and density, including propagators, three-point vertex functions, phase diagrams, etc. at one-loop order and, by now, also at two-loop order in many cases \cite{Gracey:2019xom,Barrios:2020ubx,Barrios:2021cks}. The results compare very well with the available lattice data, thereby confirming the above perturbative picture. The approach is also used to investigate quantities which are not directly accessible with lattice techniques. The present article aims at summarising these results and advertising them to a broad audience.

The article is organised as follows. In Section
  \ref{sec:gauge_fixing}, we introduce our notations and discuss some
  aspects of the gauge fixing in nonAbelian gauge theories that
  motivate the model studied in this review. Section \ref{sec:previous} presents an overview of the
status of the YM propagators in the Landau gauge. In Section \ref{sec:CF}, we briefly review the CF model and in Section
 \ref{sec:vac}, we present the results of perturbative calculations of
 propagators and three-point vertex functions of YM theories in this
 model at one- and two-loop orders and their comparison to lattice
 results. Similar calculations in the case of dynamical
 quarks are reviewed in Section \ref{sec:quarks}, including the case
 of light quarks, where the dynamical breaking of chiral symmetry
 plays a key role. We discuss the results of the perturbative CF model
 for the propagators and the phase diagram of the theory at nonzero
 temperature and chemical potential in Section \ref{sec:temp}. Section
 \ref{sec:Mink} reviews some results in the Minkowskian domain
 and Section \ref{sec:open} briefly discusses some of the open questions. Finally, we summarise and conclude in Section \ref{sec:concl}.\\

\section{QCD and gauge fixing}\label{sec:gauge_fixing}

\subsection{The QCD lagrangian}
The aim of this section is mainly to fix our notations.
QCD is  a field theory which involves fermionic (Dirac bispinor) fields $\psi$ for the quarks and a gluon field $A_\mu$ which takes values in the Lie algebra of the SU(3) gauge group. In this review, we mainly focus on Euclidean properties and it is therefore convenient to work with the Wick-rotated Lagrangian density $\mathcal L$, which can be written as the sum of a pure gauge term $\mathcal L_\YM$  and a matter term $\mathcal L_\psi$. The first one is the YM Lagrangian
\begin{equation}
  \label{eq_F2}
  \mathcal L_\YM=\frac 12 \tr\, F_{\mu\nu}^2.
\end{equation}
It involves the field strength $F_{\mu\nu}=\partial_\mu A_\nu-\partial_\nu A_\mu-i g_b[A_\mu,A_\nu]$, where $g_b$ is  the bare coupling constant. The matter term involves a sum over the quark flavours:
\begin{equation}
  \label{eq_Lpsi}
  \begin{split}
     { \cal L_\psi}= \sum_{i=1}^{N_f}\bar\psi_i\left (\slashed D + M_{b,i}\right)\psi_i ,
  \end{split}
\end{equation}
where the covariant derivative acting on a fermion is
$D_\mu\psi=(\partial_\mu-ig_b A_\mu)\psi$, the Feynman notation is
used ($\slashed D=\gamma_\mu D_\mu$ with $\gamma_\mu$ the Euclidean
Dirac matrices whose anticommutator is diagonal, $\{\gamma_\mu,\gamma_\nu\}=2\delta_{\mu\nu}\mathbb{1}$), and $M_{b,i}$ are the bare masses of the different
quark flavours. For completeness, we recall that the covariant
derivative acting on a field $X$ which takes values in the Lie algebra
reads $D_\mu X=\partial_\mu X-i g_b[A_\mu,X]$.

QCD involves, in principle, six quark normlaliss. For the present applications, the three heavier ones, $c$, $b$, and $t$, can be safely neglected and one considers only the three lighter ones, $u$, $d$, and $s$.
 It is however, interesting, for methodological
purposes, to consider other gauge groups, different number of quark
flavours  and even change the space(time) dimension. The QCD-like theories considered here are based on the gauge groups SU($N_c$) and involve $N_f$ quark normlaliss with various masses. The case
$N_f=0$ is interesting by its simplicity and because it is expected to exhibit several key properties of QCD. This pure YM or quenched limit can be viewed as QCD with all quark masses so large that there are no contribution from
fermionic fluctuations.

Under a gauge transformation, the fields transform as
\begin{align}
  &A_\mu\to A_\mu^U=UA_\mu U^\dagger+\frac i{g_b}U\partial_\mu U^\dagger,\qquad \psi\to\psi^U=U\psi,
\end{align}
where $U(x)$ is a field which takes value in the gauge group. This
causes the field strength and the covariant derivatives to transform
as $F_{\mu\nu} \to U F_{\mu\nu}U^\dagger$ and
$D_\mu \psi\to U D_\mu \psi$, which in turn implies that the QCD lagrangian is gauge invariant.

In actual calculations, it is often convenient to decompose a field
$X$ taking values in the Lie algebra (such as the gluon field) on a
basis $\{t^a\}$ of the Lie algebra: $X=X^a t^a$. We
choose the normlalisation of the generators $t^a$ such that $\tr\, t^at^b=\frac 12 \delta^{ab}$  so that the YM lagrangian reads (sum over indices implicit)
\begin{equation}
\mathcal L_\YM=\frac 14 (F_{\mu\nu}^a)^2.
\end{equation}
In this basis, the components of the covariant derivative, $(D_\mu X)^a=\partial_\mu X^a+g_bf^{abc}A_\mu^b X^c$, and of the field strength, $F_{\mu\nu}^a=\partial_\mu A_\nu^a-\partial_\nu A_\mu^a+g_bf^{abc}A_\mu^bA_\nu^c$, involve the structure constants $f^{abc}$ of the gauge group.

\subsection{Gauge fixing}
\label{sec_gauge_fixing}


Gauge invariance is a very powerful concept that highly constrains
the possible physical theories but it also comes with important
drawbacks. In particular, it implies that the propagator for the gluon
field, one of the building blocks of continuum quantum field theory
approaches, is not well defined. Technically, gauge invariance imposes
that the second derivative of the YM action is transverse, $q_\mu\delta^2 S_\YM/\delta A_\mu^a(q)\delta A^b_\nu(-q)=0$, and thus noninvertible, which makes apparent that there is no tree-level gluon propagator associated to the YM action.

To overcome this difficulty, the strategy used in virtually all
continuum approaches consists in fixing the
gauge. The underlying idea consists in dividing the space of all gauge
configurations in equivalence classes, called gauge orbits: two gauge configurations belong to the same equivalence
class if they are related by a gauge transformation. Gauge invariance
stipulates that all the field configurations within a gauge orbit bear
the same physical content. In the path integral version of quantum
field theory, summing over all gauge field configurations is redundant
and it is enough to retain one representative per gauge orbit. The
procedure which consists in restricting the path integral to one field
configuration per gauge orbit is called gauge fixing and the
representative is chosen according to a given gauge condition. The
Landau gauge, defined by the condition
\begin{equation}
    \label{eq_landau}
    \partial_\mu A_\mu ^a=0,
\end{equation}
is a very convenient and widely used choice, in particular, for what concerns nonperturbative approaches. This whole review concerns the Landau gauge.\footnote{We mention that the screened perturbation theory approach has also been applied to the case of linear covariant gauges \cite{Siringo:2018uho,Siringo:2019lmg}, which have recently also been investigated with lattice techniques \cite{Cucchieri:2009kk,Bicudo:2015rma,Cucchieri:2018doy}.} 

In continuum  approaches, the gauge-fixing procedure is, most often,
implemented through the famous FP procedure
\cite{Faddeev:1967fc}. It boils down to adding to the Lagrangian
density a gauge-fixing part expressed in terms of ghost fields $c$ and
$\bar c$ (which are Grassmann variables) and a Lagrange multiplier
$h$ (known as the
  Nakanishi-Lautrup field). For the Landau gauge, it reads
\begin{equation}
\mathcal L_{\text{FP}}=\partial_\mu \bar c^a(D_\mu c)^a+i h^a\partial_\mu A_\mu^a.
    \label{eq_L_gauge_fixing}
\end{equation}

The YM and FP actions are invariant under the
Becchi-Rouet-Stora-Tyutin (BRST) symmetry
\cite{Becchi:1974md,Becchi:1975nq,Tyutin:1975qk}, whose generator $s$
is characterized by
  \begin{align}\label{eq_BRST}
    s A_\mu^a=(D_\mu c)^a\,,\qquad s c^a=-\frac {g_b}2 f^{abc} c^b c^c\,,\qquad s\cb^a=ih^a\,,\qquad s(ih)^a=0.
  \end{align}
  The BRST symmetry is a crucial property of the FP gauge fixing procedure which is heavily used to prove renormalisability and discuss the unitarity of the theory (we will come back on these issues below). It has several interesting properties. First, the symmetry is nonlinearly realized (the variations of the fields are not linear in the fields). It is actually a supersymmetry, which transforms bosonic fields to Grassmann ones, and reciprocally. This implies that $s$ {\em anticommutes} with the Grassmann quantities. Finally, it is nilpotent, $s^2=0$.

Once the gauge is fixed, it becomes meaningful to compute
averages of quantities which are {\em not} gauge invariant. In fact,
the determination of physical observables in this context relies
inevitably on the previous evaluation of such quantities: the
correlation functions of the fundamental fields appearing in the Lagrangian. In the last two decades, an important activity has been devoted to characterize the basic QCD correlation functions by various methods, including lattice simulations. We stress though that it is by no means necessary to fix the gauge on the lattice in order to extract physical observables. Still, the calculation of correlation functions by means of gauge-fixed lattice simulations provides a very important insight. 

The results obtained by lattice simulations are described  in Sect. \ref{sec_Latt}, but before embarking on this discussion, let us recall why the Landau gauge is particularly useful in lattice simulations. In principle, fixing the gauge on the lattice requires to find the gauge transformation $U$ such that the gauge constraint $\partial_\mu A_\mu^U(x)=0$ is fulfilled. This represents a large set of nonlinear equations (there are $(N_c^2-1)$  such equations per lattice site) and this problem is highly nontrivial numerically. In the case of the Landau gauge, an alternative approach consists in extremising the functional \cite{Wilson1980,Mandula:1987rh,Mandula:1990vs}
\begin{equation}
f[A,U]=\int d^dx\,\text{tr}\left[A_\mu^U(x) A_\mu^U(x) \right]
    \label{eq_f}
  \end{equation}
with respect to the gauge transformation 
$U$ at fixed $A$. It can easily been shown that such an extremum fulfills
the Landau gauge condition. This extremisation
problem is still quite intricate because $f$ involves many variables but very powerful numerical methods are
available if one restricts to local minima. In any case, it is way easier to
address than the original root-finding problem.

\subsection{The Gribov ambiguity} 

The general philosophy behind the idea of gauge fixing presented in section \ref{sec_gauge_fixing} suffers from a major issue. As first pointed out by Gribov \cite{Gribov:1977wm}, the procedure of retaining one representative per gauge orbit is in practice more intricate than it may seem. Indeed, there exist distinct gauge-field configurations that fulfil the gauge condition \eqref{eq_landau} but that are the gauge transforms of one another. In other words, the functional \eqref{eq_f} admits many extrema. These are called Gribov copies. Later on, Singer \cite{Singer:1978dk} proved that this ambiguity is not restricted to the Landau gauge only but exists for a large class of gauge conditions. Moreover, Neuberger \cite{Neuberger:1986xz,Neuberger:1986vv} studied the influence on the Gribov copies on the
gauge-fixing property {\em \`a la} FP. He showed that, on a
lattice of finite size, the FP procedure is ill defined because physical observables are given by an undetermined $0/0$ ratio. The status of this gauge-fixing procedure is therefore
questionable at a nonperturbative level.

In a first attempt to overcome this ambiguity, Gribov
  \cite{Gribov:1977wm} proposed to limit the domain of functional
  integration over the gluon field to what is now called the first Gribov region,
  characterised by minima of the functional \eqref{eq_f}. Later, Zwanziger \cite{Zwanziger:1989mf}
  proposed a local, renormalisable field theory (the GZ theory)
  which implements this restriction of the domain of
  integration at the expense of adding a collection of auxiliary fields. It was clear from the seminal work of Gribov that the restriction to the first region successfully eliminates the infinitesimal Gribov copies, {\it i.e.}, copies that are infinitesimally close to one another.  Unfortunately, it was also pointed out that this may not be sufficient to fully resolve the Gribov ambiguity, that is, to completely fix the gauge. In fact, Van Baal \cite{vanBaal:1991zw} proved
  that there exist Gribov copies within the first Gribov region. An unambiguous gauge fixing, called the absolute Landau gauge, would consist in restricting to gauge configurations belonging the fundamental modular region, which corresponds to absolute minima of the functional \eqref{eq_f}. This is however a very difficult numerical task and no continuum or lattice technique exists to implement this constraint in an efficient way. Nonetheless, some geometric characterisations of the fundamental modular region lead to the conclusion that constraining the path integral to such field configurations would modify correlation functions in the UV by exponentially small contributions $\propto \exp(-a/g_b^2)$ with some
constant $a$. Taking into account the running of the coupling constant, it is therefore expected that Gribov copies have no role in the UV but may influence the IR properties of QCD.\footnote{This is indeed what is observed in lattice simulations \cite{Sternbeck:2005tk,Mehta:2014jla}. It was also argued that the constant $a$ in the exponential above tends to $0$ in the IR, which explains why the copies cannot be neglected in this range \cite{DellAntonio:1989wae}.}

  Another strategy for tackling the Gribov ambiguity consists in
  summing over {\em all} Gribov copies---be they minima, maxima or
  saddle points of the functional \eqref{eq_f}---with a non-flat weight function that lifts the degeneracy between the (equivalent) Gribov copies and compensates their multiple counting in the path integral \cite{Serreau:2012cg}; see also
  \cite{Serreau:2013ila,Serreau:2015yna,Tissier:2017fqf,Nous:2020vdq}. By properly choosing
  the weight function, the gauge-fixed theory can be
  written in terms of a local, renormalisable action using standard auxiliary fields techniques.

Several other proposals have been put forward to address the Gribov issue in continuum approaches, none of which is completely satisfactory, and we refer the reader to the literature for a more detailed description \cite{Parrinello:1990pm,Zwanziger:1990tn,Fachin:1991pu,Schaden:1998hz,vonSmekal:2008en, Maas:2009se}. Finally, we stress that the Gribov-Singer ambiguity is easily resolved in lattice calculations, for instance by arbitrarily choosing only one of the numerous extrema (in practice a minimum) of the discrete version of the functional \eqref{eq_f} \cite{Mandula:1987rh,Mandula:1990vs}. Other choices are possible as well \cite{Maas:2009se}.

The existence of the Gribov-Singer ambiguity has far-reaching
consequences. In particular, the textbook FP construction described
above is justified only as long as there exists only one
representative per gauge orbit but is {\em a priori} invalid when
there are more. Consequently, predictions based the FP action must be
taken with a grain of salt. For instance, the aforementioned Landau
pole could be the consequence of an ill-defined gauge-fixing
procedure.  A second example concerns the BRST symmetry
  described by Eq.~(\ref{eq_BRST}) which is indeed a symmetry of the
  FP action but whose status is questionable at a nonperturbative
  level \cite{Neuberger:1986vv,Neuberger:1986xz}.

\section{Yang-Mills correlation functions: previous results}\label{sec:previous}

\label{sec_previous}

Computing gauge-invariant quantities
in continuum approaches heavily relies on the knowledge of
correlation functions which do depend on the gauge condition. Here, 
we focus on the Landau gauge, which has been the most studied in this context. We first describe
the results from semi-analytical methods and, in the following
section, the results from lattice Monte-Carlo simulations.

\subsection{Correlation functions from continuum approaches }
\label{sec_previous_semianalytical}
The Landau gauge condition (\ref{eq_landau})
  imposes a transversality condition: any correlation function vanishes if the Lorentz index of an
  external gluon leg is contracted with the corresponding momentum. This drastically reduces the number
  of tensorial structures that may appear in a given correlation
  function. For instance, the gluon propagator in the Landau gauge is transverse with respect to the gluon momentum. In the Euclidean domain, it reads
  \begin{equation}\label{glpropstruc}
    G_{\mu\nu}^{ab}(q)=\delta^{ab} P^\perp_{\mu\nu}(q)G(q)\quad\text{with}\quad P^\perp_{\mu\nu}(q)=\delta_{\mu\nu}-\frac{q_\mu q_\nu}{q^2}\,,
  \end{equation}
where $G(q)$ is a scalar function of $q^2$. In general ({\it e.g.}, in linear gauges), the gluon propagator also involves a longitudinal part, proportional to $P^\parallel_{\mu\nu}=\delta_{\mu\nu}-P^\perp_{\mu\nu}$.

The pioneering semi-analytical studies
\cite{Mandelstam:1979xd,BarGadda:1979cz,Brown:1987ca,Brown:1988bn} for
the IR behaviour of QCD were guided by the idea of ``IR
slavery'' and were based on the conjecture of a gluon propagator with
a singular IR behaviour
\begin{equation}
 G^{\text{IR slavery}}(q)\sim \frac{1}{q^4}\,,
\end{equation}
that, in a one-gluon exchange approximation, could justify the existence of a confining linear potential
between static quarks. This behaviour
was indeed obtained as a solution of the DSE for the gluon propagator in a very simple approximation. 

However, the analysis of more elaborate truncations within 
the DSE, the FRG, and the VHA,
including not only the gluon but, also, the ghost propagator, led to the discovery of solutions with a completely different IR 
behaviour \cite{Alkofer:2000wg} where the gluon propagator tends to zero at small momentum
\begin{equation}
 G^{\text{scaling}}(q)\sim q^\alpha, \qquad\text{with }\qquad \alpha>0\,,
\end{equation}
while the ghost propagator is more singular than the bare one in the
IR
regime \cite{Smekal1997a,Atkinson:1998zc,Alkofer:2000wg,Lerche:2002ep,Fischer:2002hna,Pawlowski:2003hq,Fischer:2004uk,Fischer:2006vf,Schleifenbaum:2006bq,Huber:2007kc,Fischer:2008uz,Fischer:2009tn,Huber:2012zj,Huber:2012kd,Quandt:2013wna,Quandt:2015aaa,Huber:2016tvc,Cyrol:2016tym,Huber:2020keu},
\begin{equation}
 G_{c\bar c}^{\text{scaling}}(q)\sim \frac{1}{q^{2+\beta}}, \qquad \text{with (in $d=4$)}\qquad \beta=1+\frac{\alpha}{2}.
\end{equation}
These two correlation functions behave as power laws in the
long-distance regime, hence the name ``scaling'' solution.\footnote{A nonsingular solution for the gluon propagator was
  observed even before in Ref.~\cite{Ellwanger:1996wy} but the authors considered this behaviour as an artefact of their
  approximations.}
    
Another class of solutions, referred to as ``decoupling'', was identified some years later,
where the gluon propagator saturates in the IR (it tends to a
strictly positive constant at small momentum) and the ghost
correlation function behaves just as its tree-level
expression, up to nonsingular corrections \cite{Aguilar:2004sw,Aguilar:2006gr,Boucaud:2006if,Aguilar:2007ie,Aguilar:2008xm,Boucaud:2008ji,Boucaud:2008ky,Huber:2012kd}:
\begin{align}
& G^{\text{decoupling}}(q)\sim cst.,\\
& G_{c\bar c}^{\text{decoupling}}(q)\sim \frac{1}{q^{2}}.
\end{align}
  
In the Landau gauge, the coupling constant  can be defined as the ghost-antighost-gluon vertex at vanishing ghost momentum. This choice, called the Taylor scheme, is particularly interesting because it receives no loop corrections \cite{Taylor:1971ff}. As a consequence it is essentially determined from the ghost and gluon propagators:
\beq
\label{defalpha}
 \alpha_S(q^2)=\frac{g_0^2}{4\pi}D(q^2)F^2(q^2)\,,
\eeq
where we have introduced the gluon and the ghost dressing functions 
\begin{equation}\label{dressing}
 D(q)=q^2G(q)\quad{\rm and}\quad F(q)=q^2 G_{c\bar c}(q)\,,
\end{equation} 
and where $g_0$ is the renormalised coupling defined at the same renormalisation point as $D(q)$ and $F(q)$. 
Both the scaling and the decoupling solutions show  a regular coupling
constant in the IR, which tends to a constant in the former case and decreases to zero in the latter.

The IR behaviour of correlation functions has also been studied at leading order in the GZ approach. The early implementations of the original GZ model gave a scaling solution \cite{Gribov:1977wm,Zwanziger:1989mf,Zwanziger:1993dh}. However, it was soon realized that nontrivial condensates can naturally appear in this model, which, when included in a refined version of the model \cite{Dudal:2008sp,Vandersickel:2012tz}, led to a decoupling solution.

An important property of the scaling and the
decoupling solutions is that they are both incompatible with the existence of a K\"{a}ll\'en-Lehmann
representation with a positive spectral density for transverse
gluons \cite{Alkofer:2000wg,Cucchieri:2004mf,Cyrol:2016tym}. It was argued that such positivity
  violations are somehow related to confinement and to the presence of negative norm states which
  cannot appear in the physical spectrum of a unitary theory \cite{Alkofer:2000wg}. This will be discussed in more detail in Sec.~\ref{Sec_propagators}.
  
  These positivity violations were then observed in first-principle Monte-Carlo simulations
\cite{Cucchieri:2003di,Cucchieri:2004mf,Cucchieri:2008fc,Bowman:2007du,Maas:2007uv,Bogolubsky:2009dc} and
there is no doubt of their existence. However, their interpretation is
far from settled. In particular, it is clear that a {\it
  nonpathological} model with negative norm states must necessarily include
some form of confinement. That is, there must exist, within the set of
possible states, a subspace of physical states with positive definite
norm with an $S$-matrix that involves only such states. This effective decoupling of negative (and null)
norm states must occur either because there is some set of symmetries
that allows to characterise a physically acceptable subspace
and/or due to some selection rule of purely dynamical origin. Proving
the existence of a physical space with these characteristics and the
unitarity of the $S$-matrix in this space is as hard as 
proving confinement. What is clear, in the present state of
affairs,  is that, just as positivity violations cannot be invoked as an
indication of confinement, neither can a model be ruled out on the sole basis of their  existence. It could happen that we simply do not know the
true physical subspace with positive norm and unitary $S$-matrix.

Both the scaling and decoupling solutions unveiled a remarkable surprise: far from being very strong (not to mention with Landau
pole-type singularities), the featured correlations stay modest in the
IR. Surprisingly, these results were initially received with some
indifference by the QCD community. This may be due to various reasons. First, initial studies focused on correlations functions
which are not directly physical observables. Second, these solutions were found on the basis of approximation schemes that did not
rely on a small parameter that would ensure their robustness. In fact,
these were considered {\em nonperturbative} but the criterion used to retain or neglect a vertex was similar to what would be done in perturbation theory. Typically, two-point correlation functions were initially
 computed with either bare three- or four-point vertices or with dressed expressions based on educated guesses. Higher-order vertices were systematically neglected. More
recently, richer approximations have been considered and the results have
shown to be reasonably robust (see, for example, \cite{Huber:2020keu}) but, again, the vertices of order
greater than four have always been neglected, which still bears similarities in spirit to a higher-order perturbative analysis.

At this point, we mention that a plausible simple explanation of the success of
approximations which, although considered as nonperturbative, closely resemble perturbation theory is the existence of a relatively moderate coupling constant. This idea actually underlies the work reviewed in this article.

\subsection{Lattice results: propagators and three-point vertices}\label{sec_Latt}
The development of the semi-analytical methods presented in the previous section stimulated an important activity in computing correlation functions by means of lattice Monte-Carlo techniques. Using the gauge-fixing
  procedure described in Sect.~\ref{sec_gauge_fixing}, various two-
  and three-point correlation functions have been simulated in
  YM theory
  \cite{Mandula:1987rh,Bonnet:2000kw,Bonnet:2001uh,Cucchieri:2007rg,Bogolubsky:2009dc,Bornyakov:2009ug,Iritani:2009mp,Boucaud:2011ug,Maas:2011se,Oliveira:2012eh}
  and in QCD \cite{Bowman:2004jm,Bowman:2005vx,Silva:2010vx} (see
  Sect. \ref{sec_quarks}).
Once the gauge has been properly fixed, the computation of gluon correlators is
straightforward. It is also possible to compute correlation
functions involving ghosts, assuming that the terms in the gauge-fixed action that depend on these fields have the form of the
FP Lagrangian. The (Gaussian) functional integration over
the ghost field can be carried out exactly, which results in a
nonlocal functional measure in the gauge fields. An explicit
expression given by Wick's theorem for Grassmann variables is obtained
that involves products of the inverse of the FP operator times the
determinant of that same operator. The same holds for correlators involving quark fields. 

 \begin{figure}[t]
\centering
\includegraphics[width=.45\linewidth]{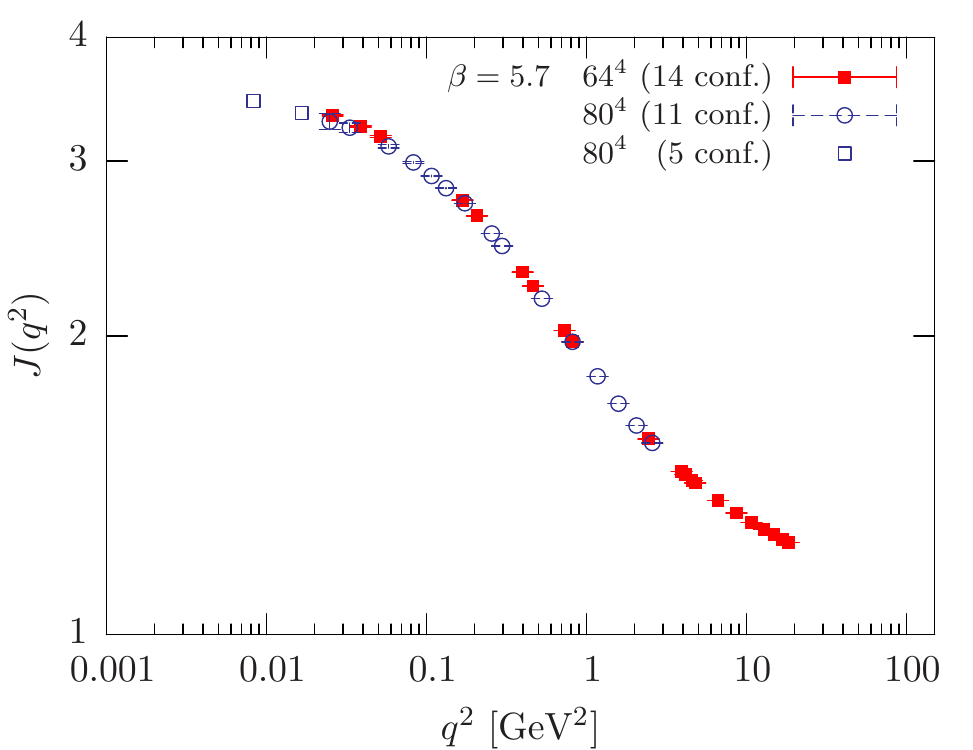}
 \caption{\label{Fig:lattice-Landau2b} The ghost dressing function---the propagator multiplied by $q^2$---in the SU(3) YM theory in the Landau gauge.
Figure from Ref.~\cite{Bogolubsky:2009dc}. In that reference, $J(q)$ stands for the ghost dressing function, denoted $F(q)$ in the present article; see Eq.~(\ref{dressing}).}
\end{figure}

The results of lattice simulations have confirmed the essential features
obtained from the continuum approaches, in particular, the
absence of strong correlations in the IR. Moreover, they allowed
to resolve, on a first-principle basis, the controversy about the
scaling versus decoupling solutions in YM or
in QCD. 

Simulations of the gluon and ghost two-point correlators have become very precise and clearly show a decoupling type solution in $d=3$ and $d=4$ dimensions both for pure YM and for QCD \cite{Bonnet:2000kw,Bonnet:2001uh,Cucchieri:2007rg,Bogolubsky:2009dc,Bornyakov:2009ug,Iritani:2009mp,Maas:2011se,Oliveira:2012eh}. This is illustrated in Figs.~\ref{Fig:lattice-Landau1}, \ref{Fig:lattice-Landau2}, and \ref{Fig:lattice-Landau2b}, which show the lattice results of Ref.~\cite{Bogolubsky:2009dc} for the Taylor coupling \eqref{defalpha},  the gluon propagator, and the ghost dressing function in $d=4$ for the SU($3$) YM theory. Concomitantly to the saturation of the gluon propagator at small momenta, the ghost propagator behaves as that of a massless excitation in this limit, that is, the dressing function remains finite in the deep IR. Simulations have also been made for other values of the number of colours $N_c$ and of the number of quark flavours $N_f$ with similar results.  In $d=2$ dimensions, instead, the lattice results show a scaling solution \cite{Cucchieri:2007rg,Maas:2011se}. Furthermore, as already mentioned, the detailed analysis of the gluon
propagator reveals, with no ambiguity, that, if it admits a K\"{a}ll\'en-Lehmann
representation, then the spectral function cannot be positive definite \cite{Cucchieri:2004mf,Bowman:2007du} (see
  Sect.~\ref{Sec_propagators} for a detailed discussion of this
  point).

As already mentioned, the relevant perturbative expansion parameter is not $\alpha_S(q)$ but,  in $d=4$, \cite{Weinberg:1995mt} 
\begin{equation}
\label{deflambda}
\lambda(q)=\frac{N_c g^2(q)}{16\pi^2}=\frac{N_c \alpha_S(q)}{4\pi}.
\end{equation}
Since the Taylor coupling $\alpha_S(q)$ never exceeds $1.3$, see
Fig.~\ref{Fig:lattice-Landau1}, the  expansion parameter $\lambda(q)$
is bounded by $0.3$.  This is a remarkable observation.\footnote{Similarly, the coupling is bounded in $d=3$ and $d=2$, with a moderate although slightly larger value in $d=3$ than in $d=4$ and significantly larger value in $d=2$.} In the absence of any prejudice, it indicates that some sort of perturbation theory should apply in the IR, at least
in the SU(3) YM theory. In the context of YM theories and QCD,
this natural conclusion, however, came at odds with the common wisdom of a genuinely nonperturbative IR regime and has remained largely unknown. This is the seed of a new paradigm.

Three-point correlators have also been simulated on the lattice and bring important information despite the fact that the statistical and systematic errors are much larger than for the propagators. One of the most important results is that the various coupling constants that can be extracted from the
ghost-gluon vertex (in various configurations of  momenta) take, again, moderate values for all momenta, including the IR limit, where they typically show a slow decrease towards zero \cite{Ilgenfritz:2006he,Sternbeck:2006rd,Cucchieri:2008qm,Bogolubsky:2009dc,Boucaud:2011ug,Maas:2019ggf,Aguilar:2021lke}. Remarkably, the smallness of the couplings is
compatible with a perturbative expansion. These
observations  include, as a particular case, the configuration of momenta corresponding to the Taylor scheme shown in Fig.~\ref{Fig:lattice-Landau1}. 

As for the three-gluon vertex, lattice results clearly show a ``zero-crossing'' in $d=3$ \cite{Cucchieri:2008qm,Maas:2020zjp} and, although not as clearly, also in $d=4$ dimensions \cite{Cucchieri:2008qm,Boucaud:2013jwa,Boucaud:2017obn,Sternbeck:2017ntv,Maas:2020zjp}.
That is, for the various
configurations of momenta studied, there always seems to be a sufficiently small momentum at which the correlation function changes sign and becomes negative. In the regime in which all
momenta tend to zero the correlation function seems to diverge towards minus infinity for $d=3$. However, the coupling constant that can be extracted from this vertex also remains rather small for all momenta.\footnote{The couplings 
  constant must be extracted, in general, from combinations of three-
  or four-point functions and propagators.} 
  
Lattice simulations of correlators involving quarks have been performed both in the quenched limit $N_f=0$ and in the presence of dynamical quarks $N_f>0$ \cite{Bowman:2004jm,Bowman:2005vx,Sternbeck:2012qs,Oliveira:2018lln}. The most important result observed in these simulations is that the quark propagator shows a significant dynamical mass generation, which is a signature of the spontaneous breaking of the chiral symmetry. That is, even in cases where the running mass is very small at the microscopic level (of the order of a few MeV), its zero momentum limit---the constituent mass---is of the order of several hundred MeV.

Finally, the quark-gluon vertex has been simulated in Refs.~\cite{Skullerud:2002ge,Skullerud:2003qu,Skullerud:2004pt,Kizilersu:2006et}. Among the more striking results, the associated coupling constant can get up to two-to-three times larger than the one in the pure gauge sector. This has far-reaching consequences as it indicates that, unlike the YM sector, the dynamics of light quarks is strongly coupled. This is consistent with the early observation \cite{Atkinson:1988mv} that spontaneous chiral symmetry breaking requires a sufficiently large quark-gluon coupling in the IR.


\section{The Curci-Ferrari model}\label{sec:CF}

One of the striking features of the studies reported in the previous
section is the saturation of the gluon propagator in the IR, which
shows the dynamical generation (in the Landau gauge) of what is called
a screening mass---not to be confused with a pole mass, as discussed in Sec.~\ref{sec:unitarity} below. This phenomenon is, by now, well established. 
As already mentioned, the FP perturbation theory is unable to describe the generation of this screening mass  and the simplest deformation of the FP Lagrangian that includes it is a particular case of a class of Lagrangians known as the CF Lagrangians. We review the latter
and their main properties, including symmetries and renormalisability, in the present section.

 \label{themodel}
\subsection {The Curci-Ferrari Lagrangian}

In the 70's, Curci and Ferrari proposed an alternative to the Higgs mechanism  that would provide a consistent theory of massive  vector bosons in the
presence of nonAbelian symmetries \cite{Curci:1976bt}. Although this
original motivation has been abandoned (mainly for reasons related to the issue of unitarity, see Sec.~\ref{sec:unitarity} below),  the model has received a renewed interest in the context of IR QCD \cite{Tissier:2010ts,Tissier:2011ey}. In this section, we consider the original model, which goes beyond the case of the Landau gauge, for completeness. The Lagrangian density\footnote{In their
   original article \cite{Curci:1976bt}, Curci and Ferrari considered a
   more general model. We will limit ourselves here to the subset of
   parameters that are compatible with a massive renormalisable
   model.} reads
 \begin{equation}
   \label{eq_fullcf}
\mathcal{L}=\mathcal{L}_{\rm{YM}}+\mathcal{L}_{\rm{CFDJ}}+\mathcal{L}_{\rm{m}},
\end{equation}
where $\mathcal{L}_{\rm{YM}}$ is the YM Lagrangian density (\ref{eq_F2}) and where the gauge-fixing  and mass contributions read, respectively, 
\begin{equation}
\label{cfaction}
\mathcal{L}_{\rm{CFDJ}}=
\frac{1}{2}\partial_\mu \bar c^a (D_\mu c)^a
+\frac{1}{2}(D_\mu \bar c)^a \partial_\mu c^a+\frac{\xi_b}{2}h^ah^a 
+ih^a\partial_\mu A_\mu^a 
-\xi_b\frac{g_b^2}{8}(f^{abc}\bar c^bc^c)^2,
\end{equation}
and 
\begin {equation}
\label{lagmass}
\mathcal L_{m}=m_b^2\left[\frac 12 (A_\mu^a)^2+\xi_b \bar c^a c^a\right].
\end {equation}
Here, $g_b$, $\xi_b$, and $m_b$ are the bare coupling, gauge-fixing parameter, and gauge boson mass, respectively. The Lagrangian (\ref{cfaction}) is the first
example in the literature of a nonlinear gauge fixing {\it \`a la} FP, sometimes referred to as the Curci-Ferrari-Delbourgo-Jarvis gauge \cite{Delbourgo:1981cm}.
An important technical aspect is that the mass term (\ref{lagmass}) is introduced at tree level in a gauge-fixed
version of the Lagrangian density, which, notably, guarantees its
perturbative renormalisability (see Sect. \ref{sec_renorm}). The principal reason is that the tree-level gluon propagator $G_0(p)$ decreases as $1/p^2$ at large momentum. This is at odds with what happens when a gauge boson mass is directly added to the YM Lagrangian, where $G_0(p)\sim{\rm const.}$ 

We present here the Euclidean version of the model with the notations that are most commonly used nowadays
and which differ slightly from those originally employed in \cite{Curci:1976bt}.
In particular, we consider the model in the presence of a Nakanishi-Lautrup field $h^a$
that simplifies the writing of the (modified) BRST symmetry (see Sect. \ref{sec_renorm}).
We do not consider matter fields for now, but their inclusion is
straightforward, see Sect. \ref{sec_quarks}.
The main interest of the CF lagrangian in the form
(\ref{cfaction}) is that the ghost-antighost exchange symmetry is
simple and that it preserves the linear realization of some
continuous symmetries \cite{Curci:1976ar,Delbourgo:1981cm,Tissier:2008nw}. This is not the case in the nonsymmetric version of the model:
\begin{equation}
\label{jaugenonsym}
\mathcal{L}_{\rm{GF}}^{\rm {ns}}=
\partial_\mu \bar c^a (D_\mu c)^a
+\frac{\xi_b}{2}h^ah^a +ih^a \partial_\mu A_\mu^a
-i\frac {\xi_b} 2 g_b f^{abc} h^a \bar c^bc^c
-\xi_b\frac{g_b^2}{4}(f^{abc}\bar c^bc^c)^2,
\end{equation}
which is obtained from  Eq.~(\ref{cfaction}) by the field redefinition $ih^a\to ih^a+\frac{g_b}{2}f^{abc}\bar c^b c^c$ and which proves more convenient for actual calculations. The case $\xi_b= 0$, where the ghost mass and the four-ghost interaction are absent, corresponds to the the simple massive extension of the Landau gauge considered in this review.

\subsection{Symmetries and renormalisability}
\label{sec_renorm}
In their original work, Curci and Ferrari observed that the Langrangian~(\ref{jaugenonsym}) is invariant under the following generalisation of the---at the time recently discovered---BRST transformation
\begin{align}
\label{symmbrst}
s A_\mu^a=(D_\mu c)^a\,, \quad s c^a= -\frac{g_b}{2} f^{abc} c^bc^c\,, \quad s \bar c^a =ih^a\,, \quad s(i h^a)= m_b^2 c^a\,.
\end{align}
In the massless case, $m_b=0$, the
symmetry (\ref{symmbrst}) is the standard nilpotent BRST symmetry, see
Eq.~(\ref{eq_BRST}). In the massive case, the Lagrangian (\ref{cfaction}) can be seen
as a deformation of the standard FP Lagrangian and
the transformation (\ref{symmbrst}) as the corresponding deformation of the BRST symmetry.
These deformations modify the behaviour of the
model for momenta comparable to or smaller than the renormalised gluon mass. 

In Ref. \cite{Curci:1976bt}, Curci and Ferrari proved the renormalisability of the theory by making
  use of the symmetries of the model, in particular, the modified BRST
  symmetry (\ref{symmbrst}). Later, de Boer {\it et al.} \cite{deBoer:1995dh} computed the renormalisation factors at one-loop order, considering the five renormalisation factors
\begin{equation}
\label{renormfact}
A^{\mu,a}=\sqrt{Z_A} A_R^{\mu,a},\hspace{0.5cm}
c^{a}=\sqrt{Z_c} c_R^{a},\hspace{0.5cm}
\bar{c}^{a}=\sqrt{Z_c} \bar{c}_R^{a},\hspace{0.5cm}
g_b=Z_g g,\hspace{0.5cm}
m_b^2=Z_{m^2} m^2,\hspace{0.5cm}
\xi_b=\frac{Z_{A}}{Z_{\xi}} \xi
\end{equation}
 as independent.
The symmetries of the model imply that the divergent part of the renormalisation factors are constrained by the relations
\begin{equation}\label{eq:nonren}
\sqrt{Z_A} Z_c Z_{g}=Z_{\xi}^2,\hspace{0.5cm}Z_A Z_c Z_{m^2}=Z_{\xi}^2,
\end{equation}
which reduces the number of independent renormalisation factors to
three, which can all be extracted only from the two-point correlation
functions.  The first relation generalizes Taylor's nonrenormalisation
theorem \cite{Taylor:1971ff}, previously known in the particular case
of the standard Landau gauge ($\xi_b=0$, $m_b=0$). The constraints (\ref{eq:nonren}),
first conjectured in Refs.~\cite{Browne:2002wd,Gracey:2002yt}, have been
proven to all orders of perturbation theory \cite{Wschebor:2007vh} and
were, in fact, shown to be direct consequences of gauged
supersymmetries of the Lagrangian \eqref{lagmass}
\cite{Tissier:2008nw}.  The Landau gauge condition $\xi=0$ is stable
under renormalisation, {\it i.e.}, $Z_\xi(\xi = 0) = 1$ to all orders
of perturbation theory,\footnote{For $\xi_b=0$, the
    FP lagrangian Eq.~(\ref{cfaction}) is invariant under
  $c\to c+ \text{Cst.}$ The four-ghost interaction proportional to $\xi_b$
  breaks this symmetry. This is at the heart of the nonrenormalisation theorem $Z_\xi(\xi = 0) = 1$.}  so the number of independent
renormalisation factors is further reduced to two.

\subsection{Unitarity}\label{sec:unitarity}

Despite the interesting properties described above, the CF model was
soon discarded as a pertinent description of a massive Higgs boson
because of issues related with unitarity. The aim of this
section is to give a critical overview of this topic and discuss the possibility of  a unitary CF model for describing strong interactions. We first recall
the standard proof of unitarity, in the framework of the (massless) FP
gauge fixing \cite{Becchi:1975nq,Becchi:1974md,ZinnJustin:1974mc,Kugo:1977zq,Kugo:1979gm}. 

One of the main goals of a field theory is to describe the scattering
amplitudes between in- and out-states. In the simple cases
({\em e.g.} for the $\phi^4$ theory), the space of in-states is obtained by
applying the creation operator $a_{\text{in}}^\dagger(\vec q)$ to the
vacuum. It can be checked that these states have a positive norm, a
necessary property for the state space to be a {\em bona fide} Hilbert
space of a quantum theory. The situation is more involved when
considering a (Lorentz-covariant) gauge-fixed field theory
because some of the states obtained in this procedure have negative
norm (this is the case of gluons with a polarization in the time
direction and of the ghosts). Such a field theory makes sense at a quantum level only if one can define a subspace (called the
physical subspace), which a) contains only states with a positive norm
and b) is stable under the time evolution. Within this physical
subspace, the theory has all the necessary ingredients to represent a
quantum theory.

In order to impose condition b), a natural idea is to characterize
the physical subspace by using symmetry arguments. The standard
strategy consists in considering the kernel of the BRST symmetry. One
however finds by inspection that there exist states with null norm, all
of them belonging
to the image of BRST \cite{Kugo:1977zq,Kugo:1977yx,Kugo:1977mk}.\footnote{Note that, since the standard BRST symmetry is
  nilpotent, $s^2=0$, the image of BRST belongs to the kernel.} To
account for these states, one defines the physical space as
the cohomology of BRST, that is, the kernel of the BRST transformation modulo
any element in the image. One must then explicitly check that the cohomology only involves states of positive norm, hence
ensuring property a). It has been shown to be the case at all orders of perturbation theory for the FP Lagrangian.\footnote{The proof  consists in considering the axial gauge which is not Lorentz covariant but for which both the positivity of the state space and unitarity are explicit; see {\it e.g.} \cite{Weinberg:1996kr}.} This textbook construction is, however,
not completely satisfactory for QCD because, as such, the physical
subspace would contain coloured states, while confinement implies that
only colour-neutral states should appear as acceptable states. An
extra restriction, yet to be uncovered, should be used to define a
physical subspace with only hadronic states.

How could this discussion be adapted to the CF model? A first
difficulty is that the BRST symmetry is not nilpotent anymore
($s^2\neq 0$). It is, nevertheless, possible to work in the kernel of $s^2$,
which is a symmetry of the CF action. In this subspace, the BRST
symmetry is again nilpotent and we can adapt the procedure described
above. There is, however, a more thorny issue: It has been shown that this subspace contains negative-norm states
\cite{Ojima:1981fs,deBoer:1995dh}. On this basis, the CF model was
discarded as being non-unitary. This conclusion is valid if the
gauge field is associated with an observable particle ({\it e.g.}, in the
context of weak interactions). However, it should be mitigated if one
considers the CF model as a theory for strong interactions. Indeed, it could very well be that the CF model is, in fact, confining.
More precisely, there could exist a subspace, yet to be found, in which properties a)
and b) listed above hold. We recall that, even in the standard FP
case, the construction of a satisfactory physical subspace, composed
of singlet states, is yet to be built and it is conceivable that
a fix to this issue could also resolve the unitarity problem in the CF
model.\footnote{An attempt in this direction has been undertaken in the context of the GZ model \cite{Schaden:2014bea}.} This intuition is based on the observation that the states of
negative norm unveiled in \cite{Ojima:1981fs,deBoer:1995dh} are
coloured. If all the negative norm states would be coloured, building
a subspace where these states are removed would kill two birds with
one stone: we would define a physical subspace consistent with
confinement and where unitarity would be ensured. To date, the issue is still
open and the CF model cannot be discarded on the basis of unitarity
arguments.

The issue of unitarity discussed in this section has strong
connections with the property of positivity violation discussed
above. Indeed, a theory with only positive norm states would admit a
K\"{a}ll\'en-Lehmann representation with a positive spectral
density. Positivity violation for the transverse gluons indicates that
(at least some of) these modes are unphysical and should be removed
from the physical subspace.

\subsection{A minimal deformation of the Faddeev-Popov Lagrangian}

To conclude this section, we stress that the CF model in the Landau gauge is the simplest
extension of the FP Lagrangian that:
\begin{itemize}
 \item preserves standard perturbation theory in the UV, in particular, remains renormalisable;
 \item maintains the linearly
realised symmetries of the FP Lagrangian;
\item has the same field content as the FP Lagrangian;
\item renounces to the standard, nilpotent BRST symmetry (that, anyway, seems to be broken beyond perturbation theory
\cite{Gribov:1977wm,Singer:1978dk,Neuberger:1986xz}).
\end{itemize}
Indeed, in order to keep standard perturbation theory and linearly realised symmetries in the UV, the strictly renormalisable couplings must be identical to those of the FP Lagrangian and the only admissible modifications involve couplings with positive mass dimensions. Such deformations, which do not modify the UV, are called ``soft''. In the Landau gauge, the only possibility is the gluon
mass term.\footnote{Of course, the class of possible soft deformations is much larger if the field content of the model is enlarged as, {\it e.g.}, in the GZ approach.} In this sense, the Landau gauge CF model is the minimal extension of the FP approach with a soft breaking of the BRST symmetry in the IR.

Here, we want to warn the reader against a common misinterpretation of the CF model in the QCD context.The gluon mass term is not meant as an explicit modification of the theory---as the resemblance of the FP and CF Lagrangians might wrongly suggest---but, rather, as an effective way to capture actual features of Landau gauge QCD that are missed by the FP perturbative approach. Although such an effective deformation of the gauge-fixed Lagrangian may induce actual modifications of the original theory, the latter must remain under control if the model is to be a good description. In the remainder of this article, we review large pieces of evidence demonstrating that the CF model indeed captures many features of the YM and QCD-like theories at a relatively low computational cost.


\section{Yang-Mills Correlation functions in the vacuum}
\label{sec:vac}

 The working hypothesis underlying the use of the CF model in the Landau gauge is that it provides an efficient starting point for a reliable and controllable perturbative approach to the IR dynamics of the YM fields. This hypothesis has been put to test, by now, in a large number of cases, by comparing the results of actual perturbative calculations at one- and two-loop orders to lattice data in the Landau gauge, when available. The present section reviews these results in the case of the YM correlation functions in the vacuum.

Before we proceed, a word of caution is in order, which applies in fact to the remainder of the review. Our primary aim is to review those works in the literature which actually postulate perturbation theory in the IR. The main line of investigation concerns a genuine perturbative expansion within the CF model \cite{Tissier:2010ts,Tissier:2011ey,Kondo:2015noa,Hayashi:2018giz,Weber:2011nw,Song:2019qoh,Suenaga:2019jjv}.\footnote{We mention that in some publications, the
CF model is sometimes called a ``massive YM'' model. We find
this confusing because this terminology wrongly suggests that one adds
a mass to the YM Lagrangian before gauge fixing (which is well-known to be
nonrenormalisable). Note, first, that the gluon mass does not break the gauge symmetry of the theory because it is introduced in the gauge-fixed Lagrangian. Second, it  does not result in an actual massive vector asymptotic state, even at a perturbative level. Instead, as reviewed here, loop effects lead to spectral positivity violations, already at one-loop order, which guarantee that the (massive) gluon cannot be part of the physical spectrum of the model. For these reasons, we refrain from using the terminology ``massive YM'' model and we prefer ``massive Landau gauge'', or, simply, the CF model.} The latter is considered either as an actual candidate or as a proxy for an IR completion of the FP theory. In the former case, the gluon mass term is expected to be either related to the issue of gauge fixing and of the Gribov problem, or it is to be eventually self-consistently determined. In the latter case, the gluon mass is simply an additional, phenomenological input parameter which encodes some unknown aspects of the IR physics. 

Another approach \cite{Siringo:2015wtx} that revives the original screened perturbation theory \cite{Karsch:1997gj} of finite temperature field theory and extends it to the vacuum case, postulates the validity of the FP Lagrangian at all scales and uses the CF Lagrangian simply as a shifted expansion point for perturbation theory. In this case, one adds and subtracts a gluon mass term and formally treats the subtracted mass together with the coupling in a (double) perturbative expansion around the CF Lagrangian.\footnote{One interesting aspect of this approach is that the gluon mass parameter could, in principle, be self-consistently determined \cite{Comitini:2017zfp,Siringo:2019qwx}.}  So, although they rely on different hypothesis, the two approaches appear very similar in practice and they actually give similar results when it comes to comparing with lattice data. In order to avoid confusion in the present review, we focus on the
strict perturbative CF model and we mention the results of the screened perturbation theory when appropriate.

\subsection{Renormalisation and renormalisation group}

As  discussed in Sect.~\ref{sec_renorm}, the CF model is renormalisable in 
$d \leq 4$. The divergent parts of the renormalisation factors in $d=4$ dimensions are constrained by the two nonrenormalisation theorems (\ref{eq:nonren}), where $Z_\xi$ can be set to $1$ in the case of the Landau gauge. To fully determine the renormalisation factors and, in particular, their finite parts, it is necessary to choose a renormalisation scheme. In order to unify the presentation  as much as possible in this review, we
choose to focus on results obtained within a single scheme. We mention, though, that other schemes have been considered
as well and that the scheme dependence of the results has been---in some cases thoroughly---investigated
\cite{Tissier:2011ey,Weber:2011nw,DallOlio:2020xpu}. As noted previously, one important feature of the CF model is that one can
devise IR-safe schemes,\footnote{We call IR-safe those schemes where the flow remains regular at all scales without the need to introduce other parameters than those of the original Lagrangian.} for which the flow is regular at all scales
\cite{Tissier:2011ey,Weber:2011nw,DallOlio:2020xpu}. We choose the one
such scheme for which most of the existing CF model calculations have
been performed, namely, the one originally proposed in Ref. \cite{Tissier:2011ey}, which we shall refer to as the IRS scheme for simplicity. It is defined by the conditions 
\begin{equation}
G^{-1}(p=\mu)=m^2+\mu^2, \hspace{.4cm} F(p=\mu)=1, \hspace{.4cm}
Z_g\sqrt{Z_A} Z_c=1,  \hspace{.4cm} Z_{m^2} Z_A Z_c=1,
\label{rencond}
\end{equation}
where $G(p)$ and $F(p)$ are the (renormalised) scalar part of the gluon propagator and the ghost dressing function, defined in Eqs.~\eqref{glpropstruc} and \eqref{dressing}, respectively. 
Note that the two last conditions in Eq.~\eqref{rencond} apply to both the divergent parts and the finite parts of the renormalisation factors.

In order to correctly describe the logarithmic UV tails of the
correlation functions, it is important to take into account
renormalisation group (RG) effects. Indeed, although the coupling  stays limited, its running is significative. We thus define the standard RG  beta functions and anomalous dimensions as
\begin{align}
\beta_g =\frac{dg}{d\ln\mu}\Big|_{g_b, m^2_b},\qquad\beta_{m^2}=\frac{dm^2}{d\ln\mu}\Big|_{g_b, m^2_b},\qquad\gamma_A=\frac{d\ln Z_A}{d\ln\mu}\Big|_{g_b, m^2_b},\qquad\gamma_c=\frac{d\ln Z_c}{d\ln\mu}\Big|_{g_b, m^2_b},
\end{align}
where the $\mu$-derivatives are taken at 
fixed bare parameters.
The renormalisation conditions \eqref{rencond} imply that the beta functions
can be fixed in terms of the ghost and gluon anomalous dimensions as \cite{Tissier:2011ey}
\begin{align}
 \beta_g= g\left(\frac{\gamma_A}{2}+\gamma_c\right)\qquad{\rm and}\qquad
 \beta_{m^2}=m^2\left(\gamma_A+\gamma_c \right) .
\label{eq_flows}
\end{align}

The two-point correlation functions in the YM theory have been 
calculated in this model not only at one loop 
\cite{Tissier:2010ts,Tissier:2011ey} but also at two loops 
\cite{Gracey:2019xom}.  An important point to be mentioned is that due to the gluon mass the model gives a well-behaved perturbative expansion. Not only is the model renormalisable but it also features no IR divergences for non-exceptional configurations of Euclidean momenta at all orders of perturbation theory (this applies to any vertex function for any $d>2$, see Appendix B in Ref.~\cite{Tissier:2011ey}). This result is not entirely trivial due to the presence of massless ghosts modes. In the case of exceptional configurations of Euclidean momenta, some IR divergences are present \cite{Tissier:2011ey}. The case of Minkowksian momenta
would require a separate analysis. The one-loop calculations are easily performed and already exhibit the main nontrivial features of the model~\cite{Tissier:2011ey,Reinosa:2017qtf}. In the IRS scheme \eqref{rencond}, the one-loop anomalous dimensions read, for $d=4$,
\begin{equation}
\label{eq_gamma_c}
\begin{split}
\gamma_c=-\frac{g^2 N_c}{32 \pi ^2 t^2} \left[2(t+1)t -t^3 \ln t+(t+1)^2 (t-2) 
\ln (t+1)\right],
\end{split}
\end{equation}
\begin{equation}
\begin{split}
\gamma_A=&-\frac{g^2 N_c}{96 \pi ^2 t^3}  \Bigg[(17 t^2-74t+12)t-t^5 \ln t+(t-2)^2 (2 t-3)
  (t+1)^2 \ln (t+1)\\
&+t^{3/2} \sqrt{t+4} \left(t^3-9
   t^2+20 t-36\right) \ln
   \left(\frac{\sqrt{t+4}-\sqrt{t}}{\sqrt{t+4}+\sqrt{t}
   }\right)\Bigg],
\end{split}
\end{equation}
where $t=\mu^2/m^2$. The two-loop calculation is way more involved and requires the use of symbolic 
programming \cite{Gracey:2019xom}.
\begin{figure}[t]
\includegraphics[width=0.45\textwidth]{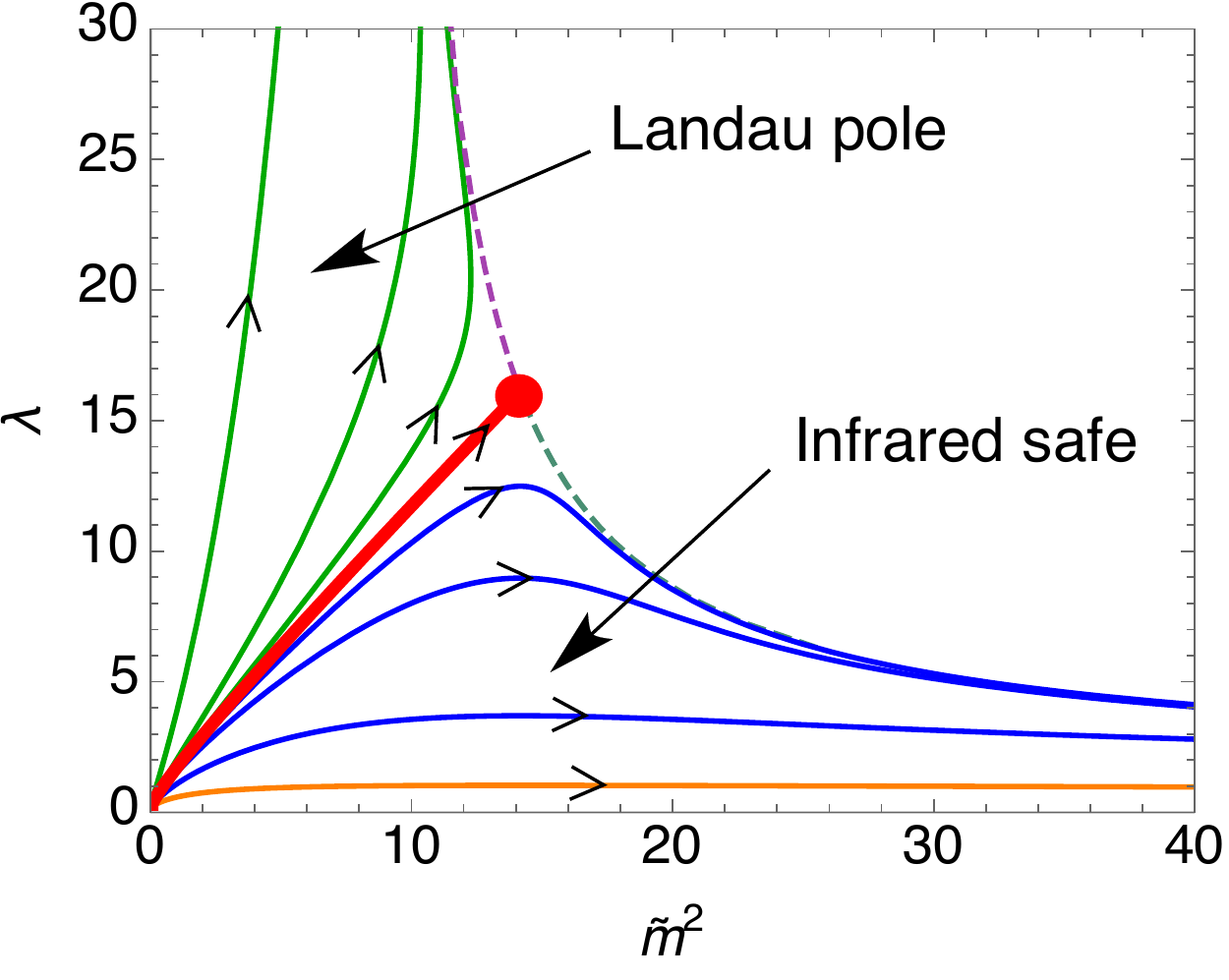}\hglue9mm
\caption{The RG flow of the CF model in the plane $(\lambda=N_cg^2/(16\pi^2), \tilde m^2=m^2/\mu^2)$ in the IRS
scheme for $N_c=3$ and $d=4$ at one-loop order. From Ref.~\cite{Reinosa:2017qtf}. The two-loop corrections have been computed in Ref.~\cite{Gracey:2019xom}. The green trajectories present a Landau pole, while the blue ones are IR safe. The separatrix (red curve) describes a scaling solution and ends at a nontrivial IR fixed point located at $(\lambda^*,\tilde m^*)\sim (16.11, 3.77)$ at one loop and $(\lambda^*,\tilde m^*)\sim (1.64, 0.6)$ at two loops.
The orange line is the trajectory that describe well the lattice results for the propagators of the SU($3$) YM theory. Figure from \cite{Reinosa:2017qtf}.}
\label{Fig:lambda_IS}
\end{figure}
The RG flow is obtained by
integrating the beta functions \eqref{eq_flows} with initial
conditions $m_0=m(\mu_0)$ and $g_0=g(\mu_0)$ at a given scale
$\mu_0$. The $d=4$ one-loop flow is shown in Fig.~\ref{Fig:lambda_IS}
in terms of the dimensionless mass $\tilde m^2= m^2/\mu^2=1/t$ and of
the coupling \eqref{deflambda}.  In the UV regime, $t\gg 1$,
  the beta functions behave as
  \beq
  \frac{\beta_g}g\sim -\frac{11}3\frac{N_cg^2}{16\pi^2}+\mathcal
  O(1/t)
 \qquad {\rm and }\qquad
 \frac{\beta_{m^2}}{m^2}=-\frac{35}{6}\frac{N_c g^2
  }{16\pi^2}+\mathcal O(1/t).
  \eeq 
  All relevant trajectories start from the Gaussian fixed point $g=t^{-1}=0$ which
  is attractive in the UV. As anticipated, asymptotic freedom is recovered and the one-loop
  beta function for the coupling constant takes its universal form. As
one flows towards the IR, three types of trajectories are
observed. For a given coupling constant in the UV, if the gluon mass
is large enough, the flow is driven towards a fully attractive
  IR fixed point, the vicinity of which is characterized by
  \beq
  \frac{\beta_g}g\sim \frac 16 \frac{N_c g^2}{16\pi^2}+\mathcal O(t)
  \qquad{\rm and}\qquad
  \frac{\beta_{m^2}}{m^2}\sim \frac 13 \frac{N_c
    g^2}{16\pi^2}+\mathcal O(t).
    \eeq 
    In this case, the flow is IR safe
  (there is no Landau pole) and the propagators are regular for
  arbitrary momentum. As will be seen later, the trajectories that
correctly describe the data from the numerical simulations are of this
type. These trajectories correspond to the decoupling solutions
described in Sect.~\ref{sec_previous_semianalytical}. On the other
hand, if the gluon mass is taken below a certain threshold the RG flow
is singular and presents a Landau pole.\footnote{The general structure
  obtained here, with a regime of regular {\it vs.}
  singular solutions as a function of the gluon mass parameter at fixed
  coupling is also observed in nonperturbative continuum approaches
  \cite{Pawlowski:2003hq,Fischer:2004uk,Fischer:2008uz} although the scaling exponents at the IR scaling
  fixed point are different than those obtained here in the IRS scheme
  \cite{Reinosa:2017qtf}.} There is a limiting trajectory which
separates these two behaviours. It connects the Gaussian fixed point
in the UV to an IR nonGaussian fixed point. This corresponds to a
scaling solution as described in
Sect.~\ref{sec_previous_semianalytical}, with, here,
$\alpha=\beta=d-2$ (at all orders of perturbation theory
\cite{Reinosa:2017qtf}). This is called the Gribov scaling. The
structure of the flow is robust against two-loop corrections, the main
difference being the location of the IR scaling fixed point
\cite{Gracey:2019xom}. We stress that the latter involves large couplings for which the present perturbative analysis is not reliable. This, for instance, is reflected in the important change of the location of this fixed point from one to two loops. In contrast, the RG trajectory that describes best the lattice results is under perturbative control.
Finally, the one-loop RG flow in the IRS scheme has also been studied for general dimensions \cite{Tissier:2011ey,Reinosa:2017qtf}. Explicit expressions for the anomalous dimensions $\gamma_A$ and $\gamma_c$, of similar complexity as above were obtained also for $d=3$ and $d=2$. For $d>2$, the structure of the flow is the same as the one described above in $d=4$. Instead, the case $d=2$ is qualitatively different as there are no IR-safe trajectories, at one-loop order at least.

Once the running of the mass and coupling constant are
  determined, the RG-improved expressions for a vertex functions involving $n_A$
  gluon fields and $n_c$ ghost fields are obtained, as usual, by
  solving the RG equation
  \begin{equation}
    \label{eq_RG_eq}
    \left(\mu\partial_\mu-\frac{n_A\gamma_A+n_c\gamma_c}{2}+\beta_g\partial_g+\beta_{m^2}\partial_{m^2}\right)\Gamma^{(n_A,n_c)}=0,
  \end{equation}
  whose solution relates the vertex function at different scales:
  \begin{equation}
    \label{eq_integration_RG}
    \Gamma^{(n_A,n_c)}(\{p_i\},\mu,g(\mu),m^2(\mu))=z_A^{n_A/2}(\mu)z_c^{n_c/2}(\mu)\Gamma^{(n_A,n_c)}(\{p_i\},\mu_0,g_0,m^2_0).
  \end{equation}
  This relations involves the $z$ factors:
  \begin{align}
     z_A(\mu)&=\exp\int_{\mu_0}^\mu \frac{d\mu'}{\mu'}\gamma_A(\mu')=\frac{g_0^2}{g^2(\mu)}\frac{m^4(\mu)}{m_0^4}\\
     z_c(\mu)&=\exp\int_{\mu_0}^\mu \frac{d\mu'}{\mu'}\gamma_c(\mu')=\frac{g^2(\mu)}{g_0^2}\frac{m_0^2}{m^2(\mu)}
  \end{align}
 where we used Eq.~(\ref{eq_flows}) in the last equalities to relate these to the running
  coupling constant $g(\mu)$ and mass $m(\mu)$, obtained by integrating
  the RG flow.

\subsection{Fitting procedure}
\label{sec_fit}
The calculation of the RG-improved correlation functions relies on
integrating the RG flow. In the standard FP theory, the initialisation of the coupling constant at some RG scale
$\mu_0$ is merely a scale-definition (this is the phenomenon of
dimensional transmutation). For a multidimensional flow, as in the CF model, the
initialisation process has far-reaching consequences because it
specifies one of the infinitely-many RG trajectories (see Fig.~\ref{Fig:lambda_IS}) and a change of initial condition is in
general not a simple scale redefinition.

Now, following the philosophy that the gluon mass is a
phenomenological parameter that we do not try to determine from first
principles, its value should be fixed by using external
information. The strategy is the following. One initialises the RG flow
at some scale $\mu_0$, integrates it, compute the RG-improved correlation function by using Eq.~(\ref{eq_integration_RG}) and compares it with available lattice data. One then changes the
initialisation parameters so as to minimise an error function and obtain the best agreement with
lattice simulations. In general, one uses the strategy of fitting
simultaneously all available data. A less stringent test would consist
in fitting independently different correlation functions. In this last
situation, the minimum of the error function typically lies at
different points in the parameter space for different correlation functions and one would obtain
better agreement with lattice simulations for each one separately.

Finally, we mention that when comparing CF results with lattice data, yet another parameter
must be fixed: the overall normalisation of the correlation function under study.\footnote{The normalisation of correlation functions depends on both the regularisation and the renormalisation schemes and is thus different from one calculation to another, even from one lattice simulation to another. The relation between different normalisation factors could be computed in principle but this is a very difficult task in practice.} Consider the example of the YM propagators described in the next section. To obtain a set of curves, one needs to fix four parameters:
the initial values $g_0$ and  
$m_0$  of the running coupling parameters as well as two multiplicative normalisation factors, for the gluon and for the ghost propagators.

\subsection{Yang-Mills propagators}
\label{Sec_propagators}

The ghost and gluon propagators in YM theories have been computed at
one-loop order in the perturbative CF approaches in
Refs.~\cite{Tissier:2010ts,Tissier:2011ey,Kondo:2019rpa,DallOlio:2020xpu} and in the screened perturbation approach in \cite{Siringo:2015jea,Siringo:2015aka,Siringo:2015wtx,Siringo:2019qwx}. They give a good agreement with existing lattice data for appropriate values of the gluon mass and gauge coupling parameters. Recently, two-loop corrections have been computed \cite{Gracey:2019xom}, which clearly improve the agreement with lattice data and greatly strengthen the confidence in the validity of the CF perturbative approach. We review those latest results here.

Evaluating
\eqref{eq_integration_RG} for the two-point vertices at $\mu=p$ and
taking into account the renormalisation conditions \eqref{rencond},
the gluon and ghost propagators can be expressed in terms of the running mass and
coupling constant as
\begin{equation}
\label{eq_int_za_IRsafe}
G(p)=\frac{m^4(p)}{m^4_0}\frac{g^2_0}{g^2(p)}\frac{1}{p^2+m^2(p)}\,, 
\quad F(p)=\frac{g^2(p)}{g^2_0}\frac{m^2_0}{m^2(p)}.
\end{equation}

\begin{figure}[t]
\includegraphics[width=0.5\textwidth]{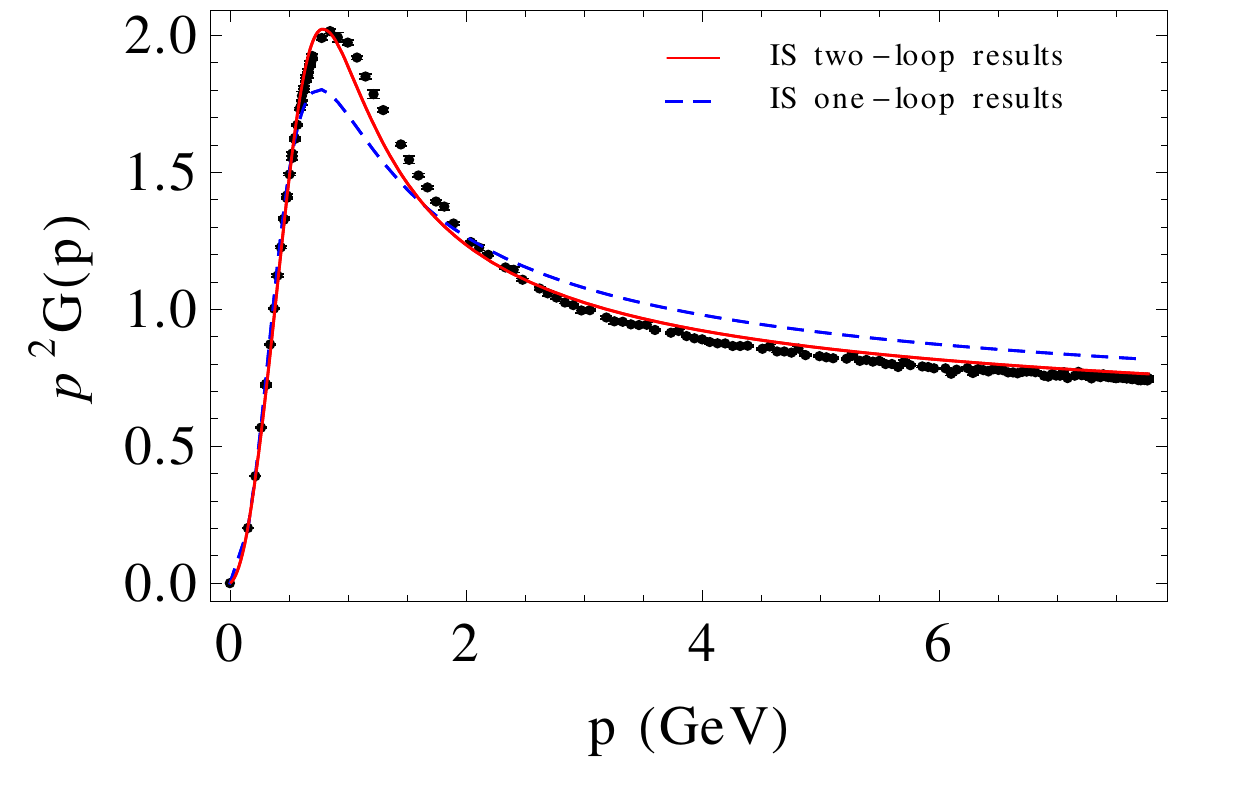}\hglue4mm
\includegraphics[width=0.5\textwidth]{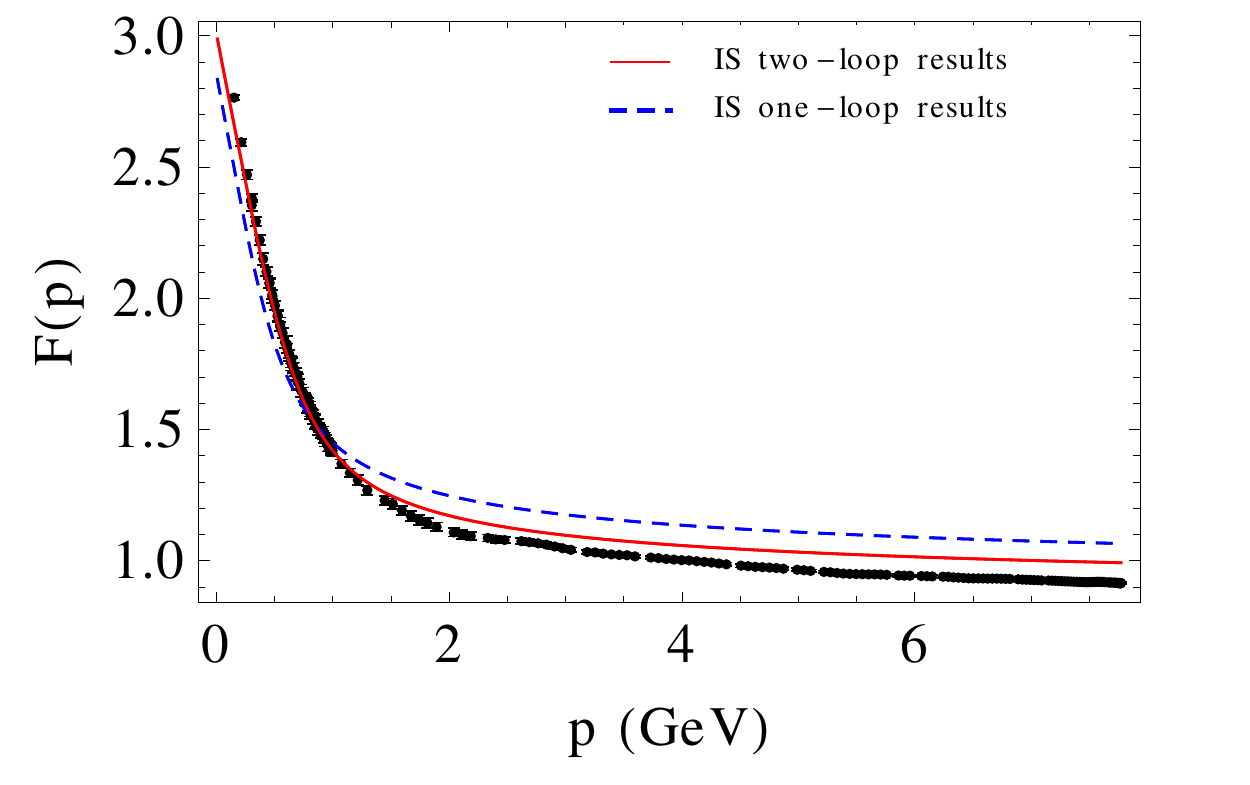}\\
\includegraphics[width=0.5\textwidth]{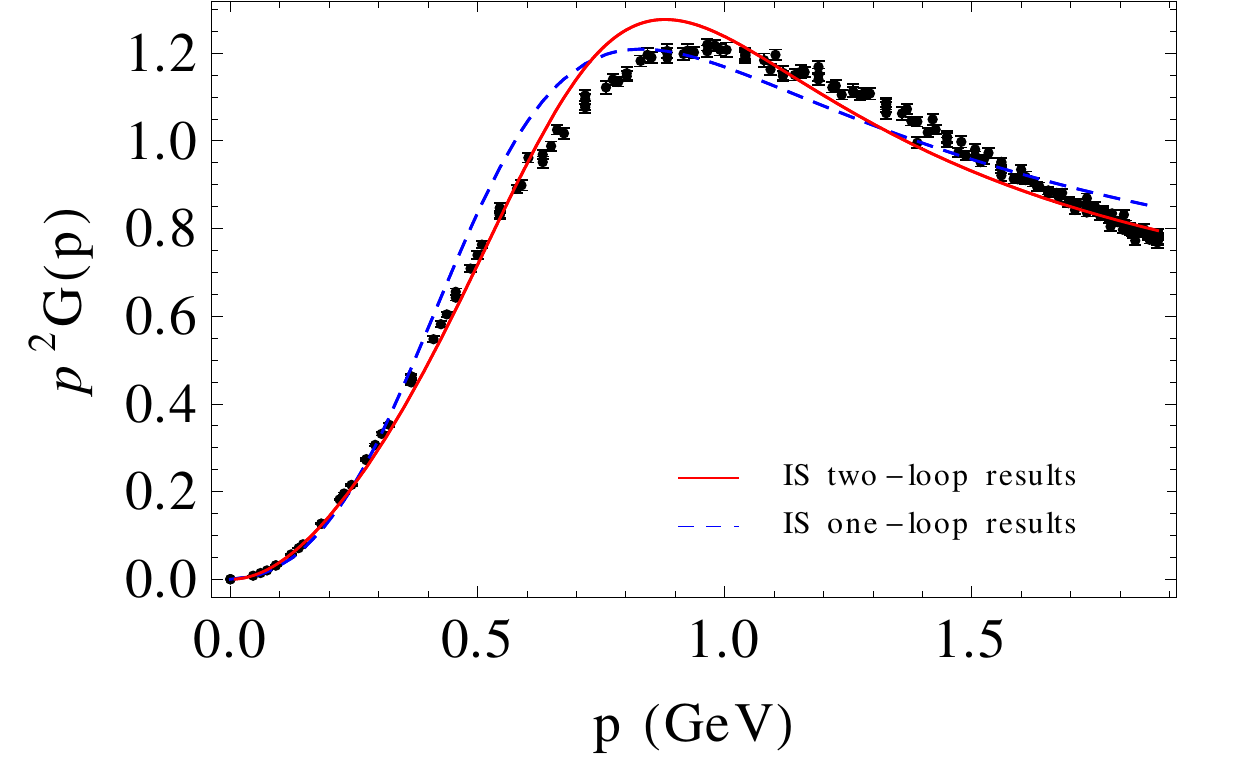}\hglue4mm
\includegraphics[width=0.5\textwidth]{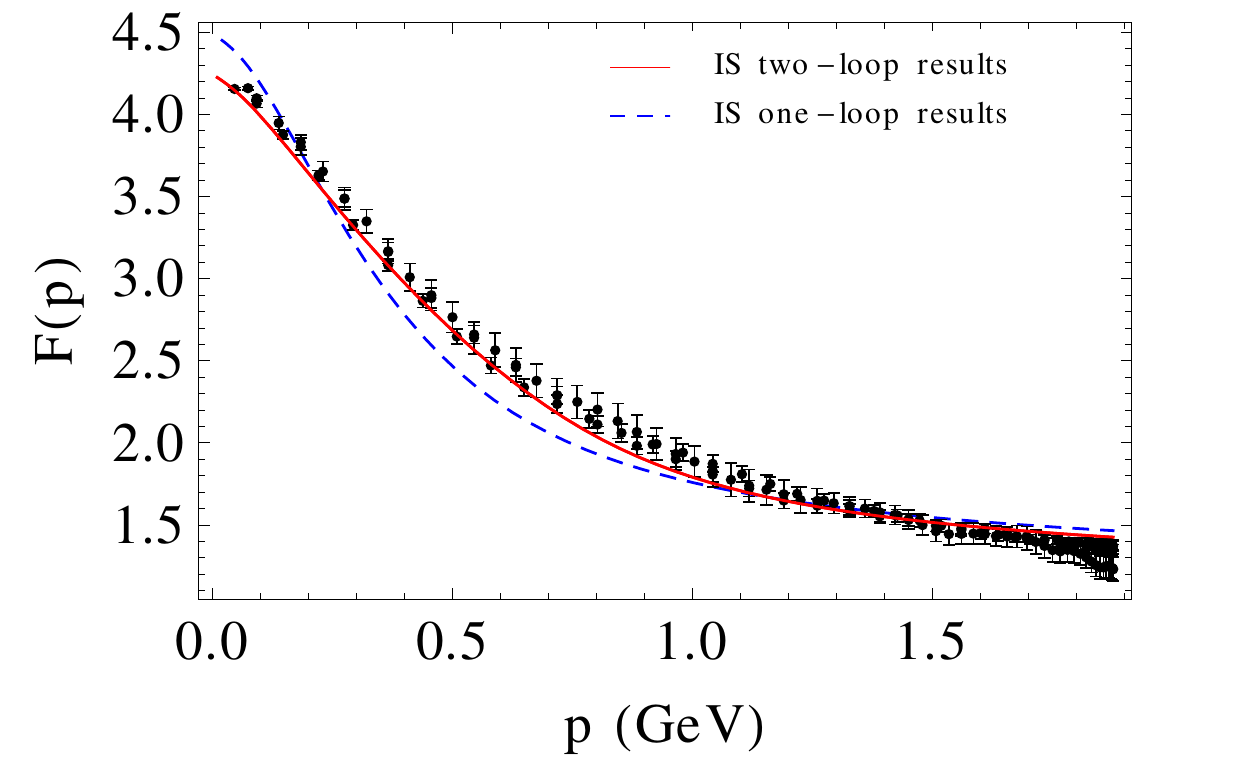}
\caption{The gluon (left) and ghost (right) dressing functions in $d=4$ for the SU($3$) (top) and SU($2$) (bottom) YM theories. The lines are the one- and two-loop results in the CF model. The squares are the lattice data from Ref.~\cite{Duarte:2017wte} for SU($3$) and from Ref.~\cite{Cucchieri:2008qm} for SU($2$). Figures from Ref.~\cite{Gracey:2019xom}.}
\label{Fig:SU3_IS_2y1L}
\end{figure}

The one- and two-loop gluon and ghost dressing functions are compared to lattice data in
Fig.~\ref{Fig:SU3_IS_2y1L} in $d=4$ dimensions for $N_c=3$ and $N_c=2$. The first remarkable point is that the
comparison is very good already at one-loop order, with a global error of about $7\%$ for $N_c=3$ and $10 \%$ for $N_c=2$ (and a maximal error of about $15 \%$). For $N_c=3$, the inclusion of the two-loop contributions 
clearly improves the agreement (with a global error of $4 \%$), whereas the improvement is less significant ($6 \%$) for $N_c=2$ \cite{Barrios:2020ubx}. This can be understood from the fact that the 
coupling constant that controls the perturbative expansion is slightly
larger for $N_c=2$ than for $N_c=3$. 
 We mention that these comparisons have also been done 
within other schemes and generically lead to similar results 
\cite{Tissier:2011ey,Gracey:2019xom,DallOlio:2020xpu}. Interestingly, it is also possible to devise optimized IR safe schemes which give excellent agreement with the data already at one-loop order \cite{DallOlio:2020xpu}. Finally, one observes that the scheme dependence is reduced when going from 
one to two loops and that the improvement is more pronounced in the SU($3$) case, 
possibly for the reason mentioned above.

\begin{figure}[t]
\includegraphics[width=0.5\textwidth]{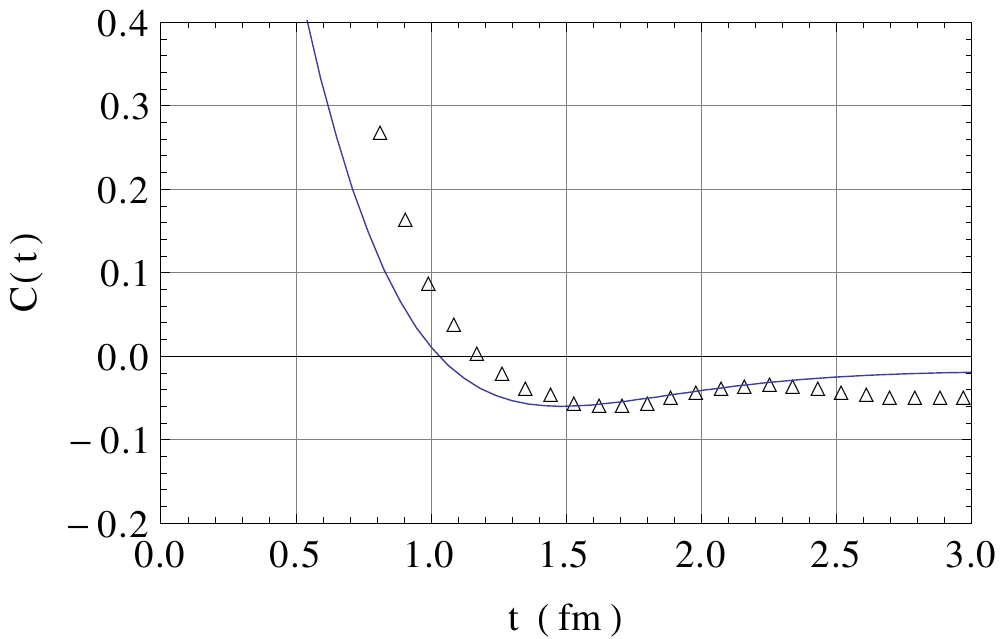}
\caption{The function $C(t)$ defined in Eq.~(\ref{functCt}) as obtained from lattice simulations \cite{Bowman:2007du} (triangles)
compared to CF result \cite{Tissier:2010ts} (dots) in the $N_c=3$ case. The parameters are fixed by fitting the Euclidean propagators. A normalisation factor has been introduced to account for the multiplicative renormalisation
of the lattice propagator. }
\label{Fig:positivityviolation}
\end{figure}

An important point, mentioned in Sect.~\ref{sec_Latt}, is that, as clearly seen from lattice simulations, the gluon propagator shows violations of positivity. To be precise, in a unitary,
Lorenz-invariant theory with only positive norm states, one can prove the K\"all\'en-Lehmann
representation, that is, the two following properties:
\begin{itemize}
\item the Euclidean propagator admits the integral representation
\begin{equation}
\label{KLrep}
 G(p)=\int_0^\infty \frac{d\mu}{2\pi} \frac{\rho(\mu)}{p^2+\mu^2}
\end{equation}
\item the spectral density $\rho(\mu)$ is positive or zero:
\begin{equation}
\label{rhopositivity}
 \rho(\mu)\ge 0
\end{equation}

\end{itemize}
It is difficult to test separately both properties by only having access to the Euclidean propagator. However, lattice simulations show that they cannot be satisfied simultaneously
for the (transverse) gluon propagator. To test this, lattice simulations study the Fourier transform \cite{Cucchieri:2004mf,Bowman:2007du}
\begin{equation}
\label{functCt}
C(t)=\int_{-\infty}^{+\infty} \frac{dp}{2\pi} \mathrm{e}^{ipt}G(p)=\int_0^\infty \frac{d\mu}{2\pi} \frac{\rho(\mu)}{2\mu}\mathrm{e}^{-\mu t}
\end{equation}
where, in the last expression, the representation (\ref{KLrep}) was assumed. If, on
top of (\ref{KLrep}), the inequality (\ref{rhopositivity}) is satisfied, then the function $C(t)$ is {\it positive}. In Fig.~\ref{Fig:positivityviolation} this function calculated in the lattice simulation \cite{Bowman:2007du} is compared to its one-loop expression in the CF model \cite{Tissier:2010ts}
for the $N_c=3$ case . One observes a good agreement and, clearly, the function is not positive. 
Similar lattice results were obtained for the $N_c=2$ case \cite{Cucchieri:2004mf}.
The violation of this
  positivity condition on $C(t)$ can have two origins. Either the
  propagator cannot be expressed in the form (\ref{KLrep}), or the
  representation (\ref{KLrep}) is valid, but with a density
  $\rho(\mu)$ which takes negative values.

\begin{figure}[t]
\includegraphics[width=0.6\textwidth,angle=0]{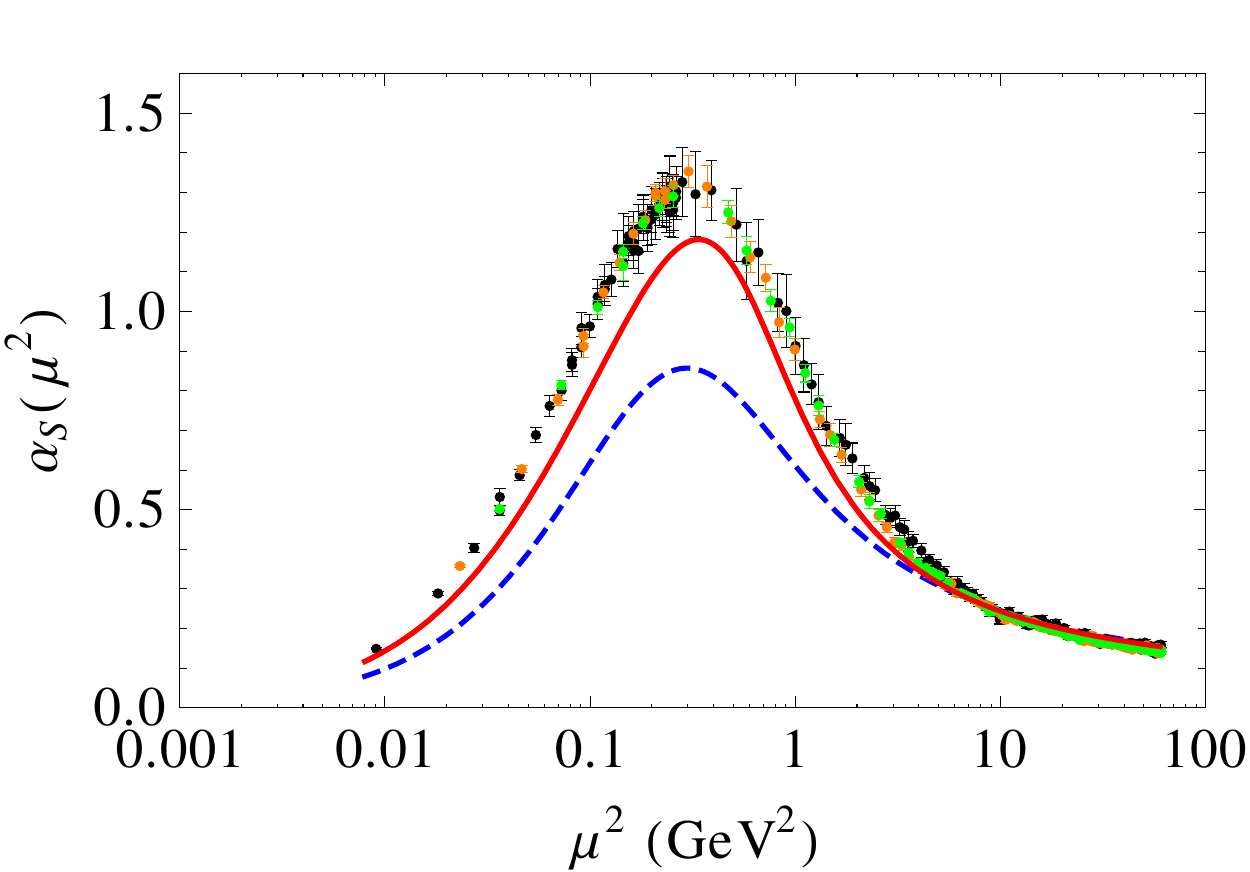}
\caption{The strong coupling constant in the Taylor scheme. The data points are those of Ref.~\cite{Duarte:2016iko}. The dashed and plain lines correspond to the one-loop and two-loop CF results in the IRS scheme, respectively. In both cases, a normalisation factor is applied so that the value $\alpha_S(\mu_0)$ agrees with the lattice result at $\mu_0=10$ GeV; see Ref.~\cite{Gracey:2019xom} for details.}
\label{fig:alphaCF}
\end{figure}

We also show, in Fig.~\ref{fig:alphaCF}, the Taylor coupling \eqref{defalpha} as a function of momentum at one and two loops, compared to the lattice data for $d=4$ and $N_c=3$. We find a good agreement with the lattice results and an apparent convergence  of the
successive perturbative orders. As mentioned before, a similar analysis can be made for $N_c=2$, where, however, the maximum value of the coupling is slightly
larger. Although the perturbative analysis is still valid, this seems to slower the apparent convergence \cite{Gracey:2019xom,Barrios:2020ubx}.

The propagators have also been evaluated at one loop in $d = 3$ dimensions 
\cite{Tissier:2011ey,Pelaez:2013cpa} and compared to existing numerical simulations for $N_c=2$. 
Again, a fairly good agreement is obtained although the results are
not as good as for $d = 4$. Finally, simulations have also been
carried out for $d=2$ but, as mentioned previously, the perturbative
approach considered here is not applicable in that case. Also
noteworthy is the fact that the positivity violations mentioned
earlier are already produced at one-loop order in $d>2$
\cite{Tissier:2010ts,Tissier:2011ey,Reinosa:2017qtf}, thus implying
that the tree-level mass term in the CF Lagrangian does not correspond
to an actual massive excitation in the spectrum.

We thus see that a perturbative description is able to correctly
describe nontrivial features of the YM propagators even at IR
momenta.  This validates---at the level of propagators for the
moment---our working hypothesis concerning the perturbative CF
model. This challenges the standard paradigm of the nonperturbative IR
regime.

\subsection{Yang-Mills three-point functions}\label{sec:YM3}

In order to further test the ability of the CF model to reproduce
the YM correlators, and now that the parameters have been fixed from the fits to two-point functions, the next natural step is to study  the predictions of the model regarding higher correlation functions. In  this section, we consider  the three-gluon vertex and the ghost-gluon vertex, that have both
been computed in lattice simulations for the YM theory in the Landau gauge for particular configurations of momenta, for the SU($2$) gauge group in $d 
= 4$ and $d = 3$ dimensions \cite{Cucchieri:2008qm,Maas:2019ggf} and, more 
recently, also for the SU($3$) gauge group in $d = 4$ \cite{Ilgenfritz:2006he,Sternbeck:2006rd,Sternbeck:2017ntv,Boucaud:2017obn,
Zafeiropoulos:2019flq,Cui:2019dwv,Aguilar:2019uob}.\footnote{We mention that more precise
  data for the three-point correlators have been produced since then
  \cite{Aguilar:2021lke}.}

\subsubsection{The three-gluon vertex}
\label{sec_3gluons}

In linear covariant gauges (including the Landau gauge), the colour 
structure of this vertex is proportional to the structure constants $f^{abc}$ 
at all orders of perturbation theory \cite{Smolyakov:1980wq}. It can be decomposed in terms of six tensor components associated to the Lorentz 
group. We use the decomposition proposed by 
Ball and Chiu \cite{Ball:1980ax}:
\begin{equation} 
\Gamma_{A^a_\mu A^b_\nu 
A^c_\rho}^{(3)}(p,k,r)=-ig_0f^{abc}\Gamma_{\mu\nu\rho}(p,k,r).
\end{equation}
with
\begin{widetext}
\begin{equation}
\label{eq_gammaAAA}
  \begin{split}
\Gamma_{\mu\nu\rho}(p,k,r)&=A(p^2,k^2,r^2)\delta_{\mu\nu}(p-k)_\rho+ 
B(p^2,k^2,r^2)\delta_{\mu\nu}(p+k)_\rho 
-C(p^2,k^2,r^2)(\delta_{\mu\nu}p.k-p_{\nu}k_{\mu})(p-k)_\rho\\
&+\frac{1}{3}S(p^2,k^2,r^2)(p_{\rho}k_{\mu}r_{\nu}+p_{\nu}k_{\rho}r_{\mu})+  
F(p^2,k^2,r^2)(\delta_{\mu\nu}p.k-p_{\nu}k_{\mu})(p_{\rho}k.r-k_{\rho}p.r)\\
&+H(p^2,k^2,r^2)\left[-\delta_{\mu\nu}(p_{\rho}k.r-k_{\rho}p.r)+\frac{1}{3}(p_{
\rho}k_{\mu}r_{\nu}-p_{\nu}k_{\rho}r_{\mu})\right]+\text{perm.,}  
  \end{split}
\end{equation}
\end{widetext}
where ``perm.'' stands for the simultaneously cyclic permutations of the momenta $(p,k,r)$ and the indices $(\mu,\nu,\rho)$.   
The scalar functions $A$, $C$, and $F$ are symmetric under permutation of their first two arguments, whereas
$B$ is antisymmetric. The function $H$
is completely symmetric and $S$ is completely antisymmetric. All the tensorial components above have been computed at one-loop order in the CF model, for arbitrary momenta \cite{Pelaez:2013cpa}. It has been checked that the one-loop expressions reduce to the known ones in the FP model \cite{Davydychev:1996pb} in the limit $m\to0$. The expressions are rather cumbersome and we shall not reproduce them here. Instead, we focus on the comparison with lattice data.

It is worth emphasising that what is  measured in lattice
simulations are correlators, which involve a contraction of the external legs of 
the vertex (\ref{eq_gammaAAA}) with (transverse) gluon
propagators. It follows that the longitudinal structures, controlled by the 
functions $B$ and $S$, are not accessible with the existing simulations.
\begin{figure}[t]
\includegraphics[width=0.45\textwidth]{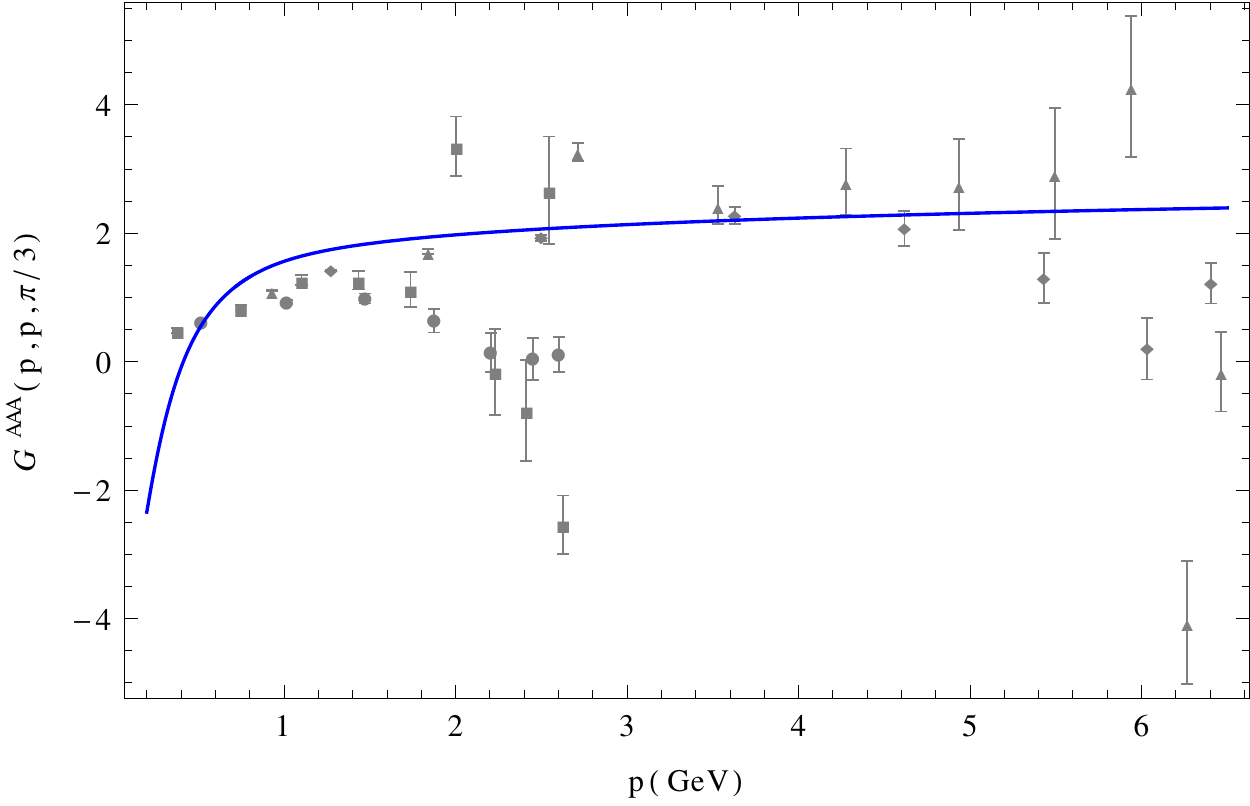}\qquad
 \includegraphics[width=0.45\textwidth]{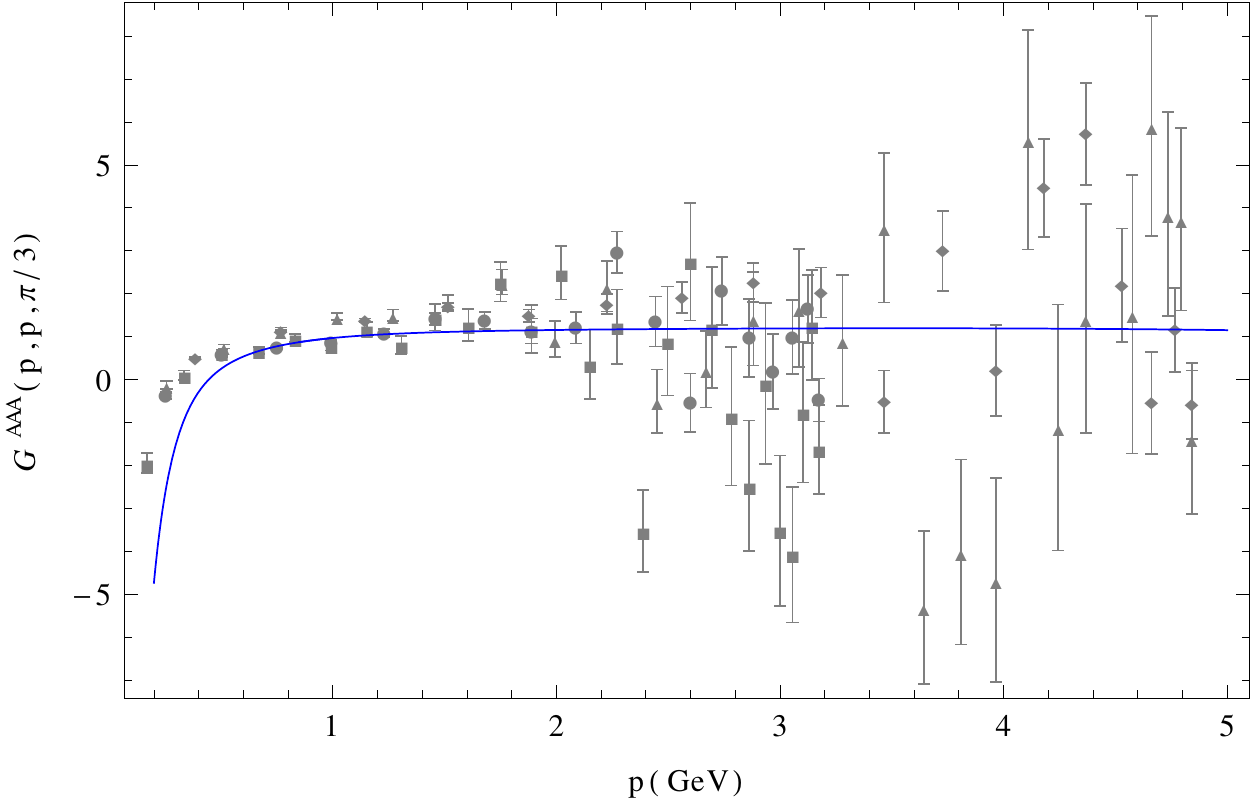}
\caption{Three gluon correlation function $G^{AAA}(p,k,r)$ for
  $p^2=k^2=r^2$ (employing the notations of
  Ref.~\cite{Pelaez:2013cpa}) in $d=4$ (left) and  $d=3$
  (right). The blue curves correspond to the IRS scheme whereas the red dashed curves correspond to a vanishing momentum scheme. Lattice data from Ref.~\cite{Cucchieri:2008qm,Maas:2019ggf}.
  Figure extraced from Ref.~\cite{Pelaez:2013cpa}.}
\label{fig_AAA_4d}
 \end{figure}
What has actually been calculated in the Monte-Carlo simulations for the SU($2$) theory is the following quantity  
\begin{equation}
\label{GAAA}
G^{AAA}(p,k,r)=\frac{\lambda_{\mu \nu \rho}^{\rm tree}(p,k,r)\Gamma_{\mu\nu\rho}(p,k,r)}
{\lambda_{\mu \nu \rho}^{\rm tree}(p,k,r) \lambda_{\mu \nu \rho}^{\rm tree}(p,k,r)}\,.
\end{equation}
where
\begin{equation}
\lambda_{\mu \nu \rho}^{\rm tree}(p,k,r)= P^\perp_{ \mu\alpha}(p) P^\perp_{\nu\beta }(k) P^\perp_{\rho\gamma}(r) \left[\delta_{\beta\gamma}(k-r)_\alpha+ \delta_{\alpha\gamma}(r-p)_\beta+\delta_{\alpha\beta}(p-k)_\gamma\right] .
\end{equation}
Eq.~\eqref{GAAA} corresponds to the transverse part of the three-gluon vertex projected along the tree-level
tensorial component $[\delta_{\beta\gamma}(k-r)_\alpha +\text{cyclic 
permutations}]$, normalised by the same combination at tree level.

In Fig.~\ref{fig_AAA_4d}, we show a comparison between the one-loop results in the CF model and the lattice data for the combination \eqref{GAAA} for a particular configuration of momenta in both  $d=4$ and $d=3$ dimensions. The overall agreement is satisfactory, although not as good as for the propagators. We stress that the lattice data show important statistical and (even larger) systematic uncertainties. This is clearly visible in Fig.~\ref{fig_AAA_4d} where the various lattice points are statistically incompatible among each other. For this reason, it is not obvious to assess the actual accuracy of the CF prediction. The one-loop expressions of Ref.~\cite{Pelaez:2013cpa} have been compared to lattice data for all measured configurations of momenta, which yields a similar level of accuracy. The SU($3$) case has also been treated in lattice simulations \cite{Boucaud:2017obn} and compared to the one-loop predictions of the CF model \cite{Figueroa2021}, with results of similar quality to those presented in Fig.~\ref{fig_AAA_4d}. We recall that these comparison involve no fitting parameters except for an overall normalisation.\footnote{The normalisation is not fixed by fitting the (renormalised) propagators because the data presented here \cite{Cucchieri:2008qm,Maas:2019ggf} concern the bare three-point correlator.}

In Fig.~\ref{fig_AAA_4d}, one clearly observes, for $d=3$, what has been referred to as a {\it zero crossing}: The vertex function becomes negative for small enough values of momenta. The same behaviour is observed also for other configurations of momenta. This feature, first observed in lattice simulations \cite{Cucchieri:2008qm}, is a very simple prediction of the perturbative CF model at one-loop order.\footnote{In fact, the zero crossing has been first explained in a continuum approach in the CF model calculation of \cite{Pelaez:2013cpa} and has later been observed in other continuum approaches \cite{Aguilar:2013vaa,Blum:2014gna,Eichmann:2014xya,Athenodorou:2016oyh,Boucaud:2017obn}.} It simply comes from the fact that, at very low momenta, the three-gluon vertex is dominated by the ghost-loop diagram which comes with the opposite sign as compared to the tree-level term. This implies, not only a zero crossing, but a divergence towards negative infinity when all momenta vanish. The same behaviour is also predicted in $d=4$ but for much smaller values of momenta, the divergence being only logarithmic \cite{Pelaez:2013cpa}. This explains why this is more difficult to see in lattice simulations \cite{Aguilar:2021lke}.

Finally, we note that the phenomenon of ghost dominance at low momenta, responsible for the above-mentioned zero crossing, is of 
a more general scope. For instance, it is also valid for other continuum approaches, such as the DSE or the FRG. It also plays a pivotal role for finite temperature physics, as discussed in Sec.~\ref{sec:temp}. Finally, it provides a simple description of the dominant low momentum behaviour of vertex functions \cite{Tissier:2011ey}. Let us describe this last point here.
Charaterizing correlation functions with more external legs becomes very
involved because more and more Feynman diagrams must be computed and
because the number of independent tensorial structures increases rapidly with the number of external legs. It is, however, possible to
extract some information concerning the deep IR regime (that is,
when all external momenta are small compared to the gluon mass) at a
low computational cost \cite{Tissier:2011ey}. In this regime, the
vertex is dominated by the Feynman diagrams with the smallest
number of (massive) gluon propagators. Moreover, at fixed number of
external legs, the Feynman diagrams with more and more loops are
suppressed for IR momenta. As a result, the leading IR behaviour is
given by a one-loop diagram (barring an accidental compensation of
diagrams). 
It is then possible to predict the leading IR behaviour of a
vertex. For instance, the gluon self-energy is dominated by a one-ghost-loop diagram, which behaves, in $d=4$, as ${\rm const.}+ p^2 \ln (p^2/m^2)$ in the deep
IR. Similarly, the three-gluon vertex behaves as $p \ln (p^2/m^2)$,
which explains the zero crossing described in above. In general, the $n$-gluon vertex with $n>2$ behaves as
$p^{4-n}$ modulo logarithms.

\subsubsection{The ghost-gluon vertex}

The ghost-antighost-gluon vertex---usually dubbed ghost-gluon vertex for short---has a much simpler Lorentz structure than the three-gluon vertex. It can be decomposed in terms of two vectorial components as
\begin{equation}
  \label{eq_vertex_cbcA_def}
  \Gamma^{(3)}_{c^a\cb^b A_\mu^c}(p,k,r)=-i g_0 f^{abc} 
\left[k_{\mu}V(p^2,k^2,r^2)+r_{\mu}W(p^2,k^2,r^2)\right]\,,
\end{equation}
where $p$, $k$, and $r$ are the (incoming) momenta of the ghost, antighost and 
gluon, respectively. Only the scalar function $V(p^2,k^2,r^2)$ is measurable in Landau gauge lattice simulations for it is the only term that contributes when the vertex is contracted with a (transverse) gluon propagator.

Both vectorial components in Eq.~\eqref{eq_vertex_cbcA_def} have been computed at one-loop order in the CF model for arbitrary momenta in $d=3$ and $d=4$ and for any $N_c$ \cite{Pelaez:2013cpa}. The various symmetries of the model \cite{Tissier:2008nw} give rise to (Ward and Slavnov-Tayor) identities that, 
first, constrain the three-gluon and the ghost-gluon vertices separately and, second, imply 
nontrivial relations between those. It has been checked that the one-loop CF expressions verify these identities and that they reduce to the known one-loop expressions in the FP model \cite{Davydychev:1996pb} in the limit $m\to0$. 

The one-loop CF results have been compared to the existing lattice data.
In a similar way as for the three-gluon vertex, what has been measured in the Monte-Carlo simulations is the following quantity: 
\begin{equation}
\label{GccbA}
 G^{c\cb A}(p,k,r)=\frac{k_\mu P^\perp_{\mu\nu}(r)\left[k_{\nu}V(p^2,k^2,r^2)+r_{\nu}W(p^2,k^2,r^2)\right]} {k_\mu P^\perp_{\mu\nu}(r)k_\nu }=V(p^2,k^2,r^2).
\end{equation}
Again, this corresponds to the transverse part of the ghost-gluon vertex projected on the tree-level
tensorial structure, normalised to the same combination at tree level. Thanks to Taylor's nonrenormalisation theorem, the expression \eqref{GccbA} 
associated with the ghost-gluon vertex is UV finite and is identical to its tree-level form when the ghost momentum vanishes.  This is to be contrasted with the quotient \eqref{GAAA}, which has a multiplicative UV divergence and needs to be renormalised.

On top of the full one-loop expressions of Ref.~\cite{Pelaez:2013cpa}, two-loop corrections to the ghost-gluon vertex have been calculated in Ref.~\cite{Barrios:2020ubx}. The calculation is quite involved for general momentum configurations and the two-loop contribution has only been computed for the case of vanishing gluon momentum, namely,
\begin{equation}
 v(k^2)=V(k^2,k^2,0),
\end{equation}
in $d=4$ and for both $N_c=2$ and $N_c=3$. Even for this particular momentum configuration, the calculation requires the use of symbolic programming, similar to that used in the case of the two-loop propagators (see Sect.~\ref{Sec_propagators}). It has been explicitly verified that the two-loops expressions fulfil several nontrivial consistency checks \cite{Barrios:2020ubx}:
\begin{itemize}
 \item The divergent parts are compatible with the Taylor 
nonrenormalisation theorem.
 \item In the limit $m\to0$, the expressions reduce to those of the FP theory, obtained for the same 
configuration of momenta in Ref.~\cite{Davydychev:1997vh}.
\item The one- and two-loops contributions vanish in the limit of zero ghost momentum, that is, $v(k^2=0)=1$.
\end{itemize} 
We stress that this last property which is clearly observed in lattice simulations is not retrieved from the usual (massless) calculations because of IR divergences which induce a nonanalytic behaviour at small momenta. Specifically, fixing first the gluon momentum to zero and later taking the limit of vanishing ghost momentum leads to a divergence of the vertex, at odds with the lattice data. The CF mass removes this IR divergence and yields the correct nonrenormalisation behaviour.

\begin{figure}[t]
\includegraphics[width=0.45\textwidth]{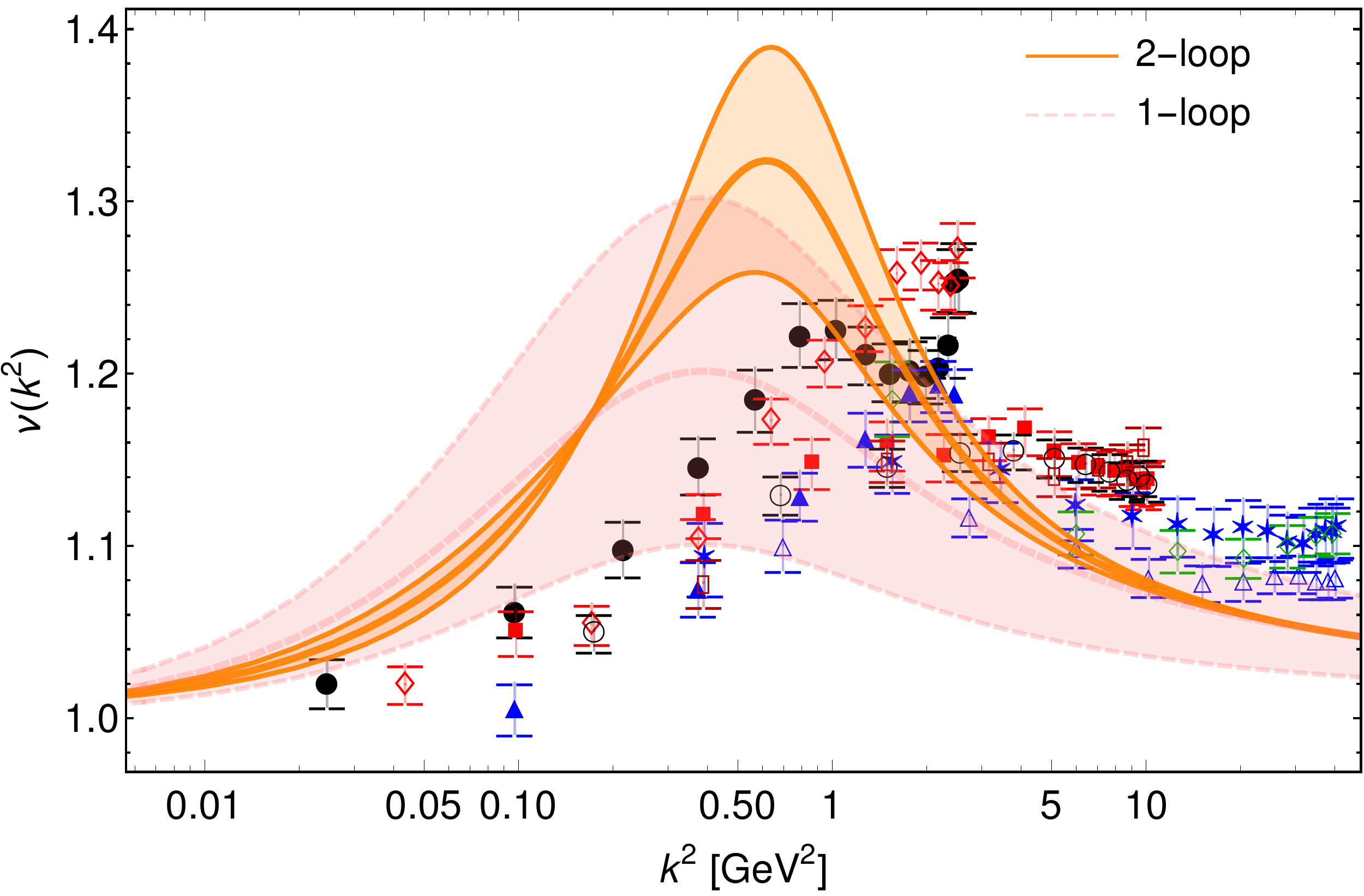}\hglue4mm
\includegraphics[width=0.45\textwidth]{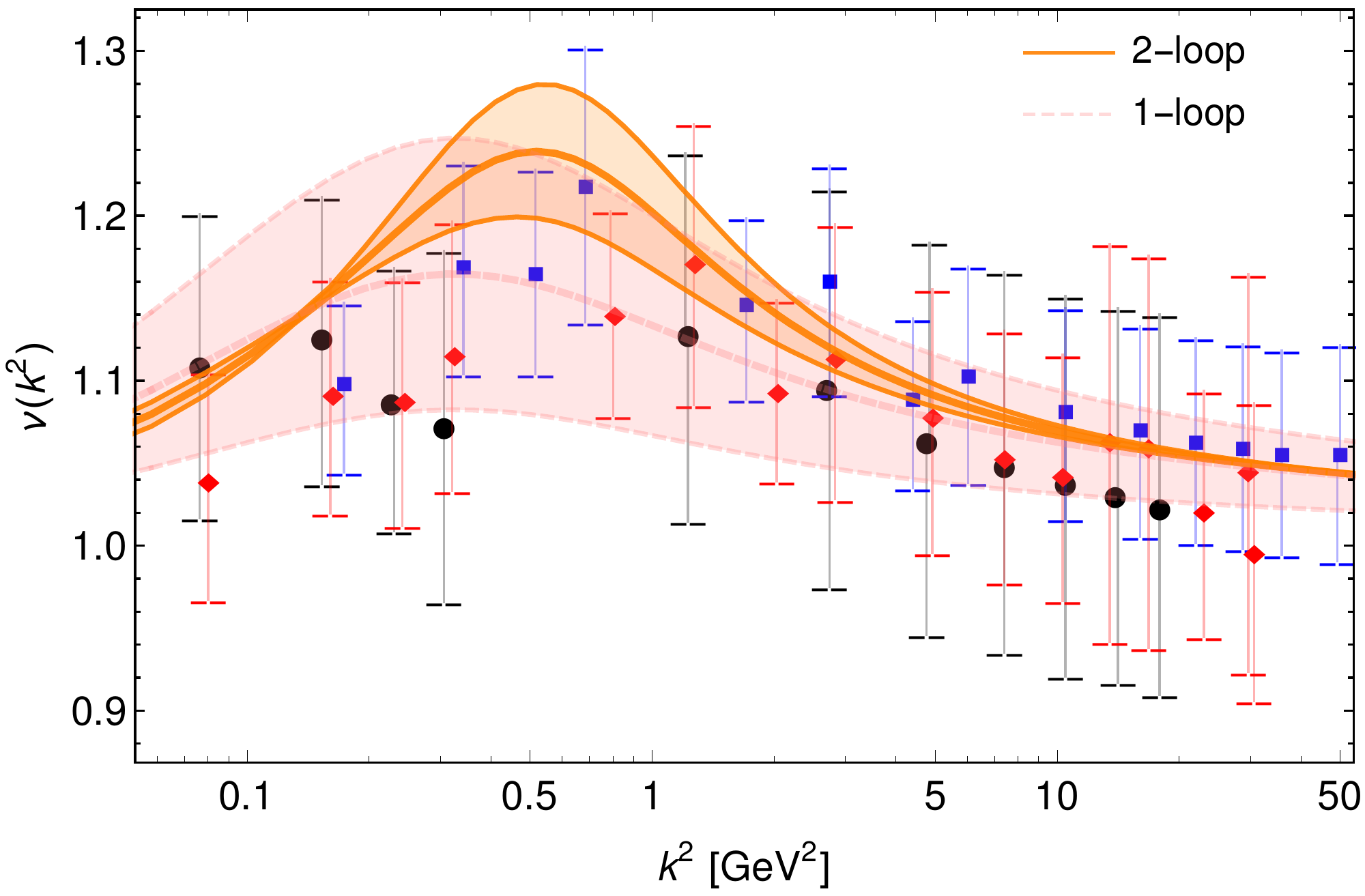}
\caption{The CF model prediction for the function $v(k^2)$ 
in the SU($2$) (left) and SU($3$) case (right) in the IRS scheme, compared to 
the lattice data in the
Taylor scheme from Ref.~\cite{Maas:2019ggf} for SU($2$) and from Ref.~\cite{Ilgenfritz:2006he,Sternbeck:2006rd}  for SU($3$). The parameters $m$ and $g$ 
at the initial scale $\mu_0$ where previously
determined from the fits of the gluon and ghost propagators. Figures from Ref. \cite{Barrios:2021cks}.}
\label{Fig:vertice_twoLoops}
\end{figure}
\begin{figure}
\hglue-6mm\includegraphics[width=0.51\textwidth]{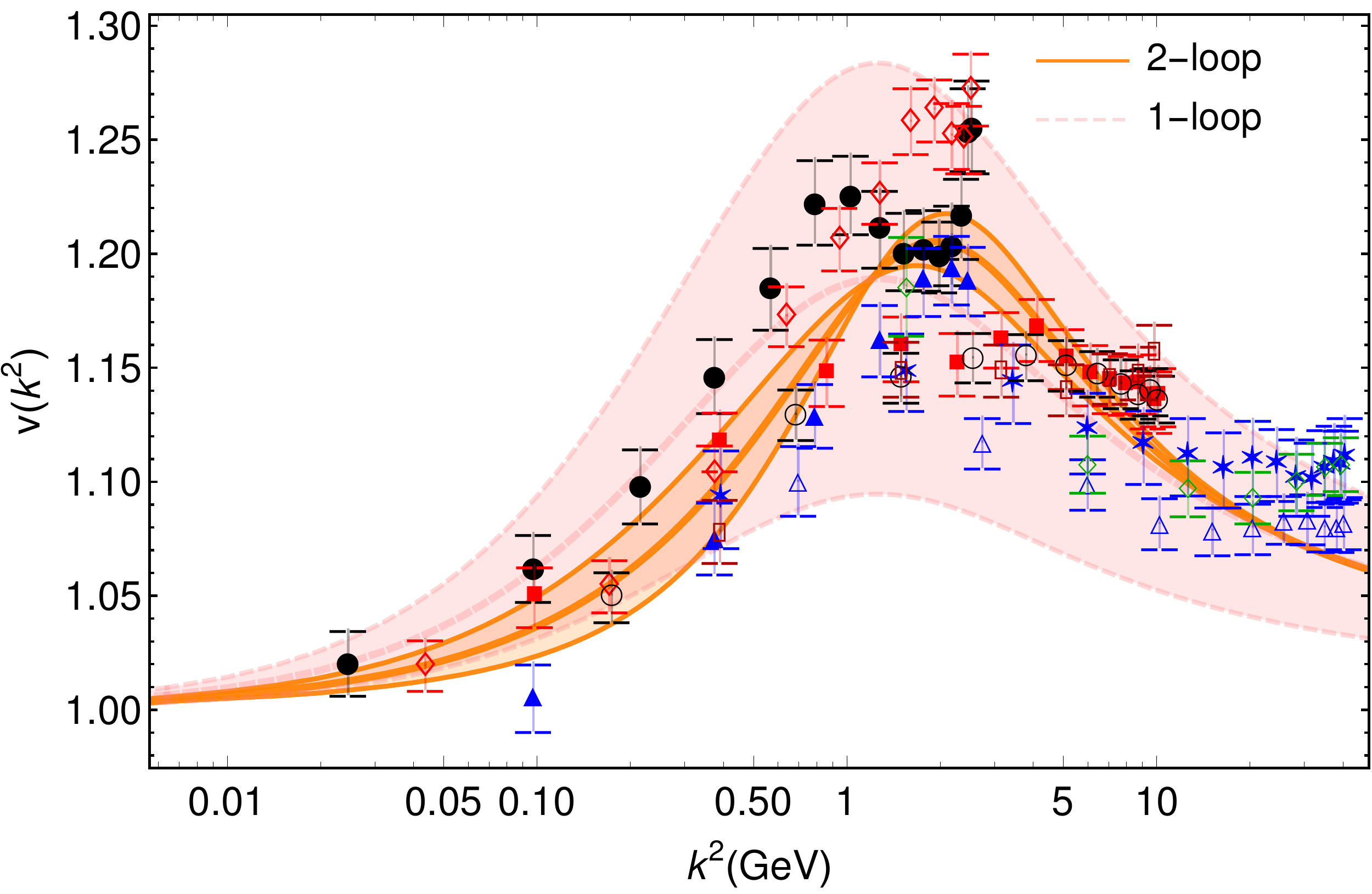}
\caption{The best fit for the vertex function $v(k^2)$ in the SU($2$) case at one- and two-loop
orders \cite{Pelaez:2013cpa,Barrios:2020ubx} in the IRS scheme, when compared to the lattice data in the Taylor
scheme \cite{Maas:2019ggf}. Figure from Ref. \cite{Barrios:2021cks}.}
\label{Fig:bestfit_SU2}
\end{figure}

The one- and two-loop expressions of Refs.~\cite{Pelaez:2013cpa,Barrios:2020ubx} are compared with the lattice data from 
Refs.~\cite{Maas:2019ggf} for $N_c=2$ and from Refs.~\cite{Ilgenfritz:2006he,Sternbeck:2006rd} for $N_c=3$ in Fig.~\ref{Fig:vertice_twoLoops}. It should be noted that this comparison is done without any further parameter adjustment than the one used to fit the propagators (not even an overall normalisation factor).  Because  the fit of the latter for the SU($2$) group is of lower quality than that for the SU($3$) group, due, in part, to possible systematic errors in the lattice simulations, this can introduce a significant error in the parameter estimation. For this reason a second type of 
fit, including the ghost-gluon vertex together with the ghost and the gluon propagators, was considered in Ref.~\cite{Barrios:2020ubx}. The corresponding results 
are shown in Fig.~\ref{Fig:bestfit_SU2}. 

The first observation is that the one-loop results provide a good description of the data, although not as good as for the two-point functions. For both $N_c=2$ and $N_c=3$, the quality of the comparison with the simulations improves when going from one loop to two loops. However, the improvement is 
significantly better for SU($3$) as was already the case for the propagators.

To conclude,  the perturbative results in the CF model are quite good. In fact, in the case of three-point vertices, it is not 
clear that, with the level of precision achieved, it is possible to completely 
neglect the errors coming from the lattice simulations. Put together, these 
results strongly indicate that the perturbative expansion of the 
CF model accurately describes the YM vacuum correlation functions in the
Landau gauge. This is not only true for the two-point functions but also for 
three-point functions. Moreover, whenever tested, the precision improves when 
including higher orders of perturbation theory.

\section{Dynamical quarks}\label{sec:quarks}

\label{sec_quarks}

In the previous section, we put forward evidences which
  indicate that the IR regime of YM theory can be described within
  perturbation theory, the main nonperturbative ingredient being
  encapsulated in a phenomenological screening mass for the
  gluons. Focusing on a pure gauge theory is clearly simpler (it
  involves less Feynman diagrams, renormlalisation factors, etc) and, at
  the same time, it is a first step for testing the working hypothesis under
  scrutiny. But, of course, these results are mainly methodological
  since the true QCD involves also light quarks with significant
  fluctuations. In this section, we discuss the attempts to include
  these particles to the perturbative scheme described in
  Section~\ref{sec:vac}.

  Many physical phenomena occur in the presence of matter fields,
  which have no equivalent in the pure YM theory. Of utmost importance
  is the phenomenon of spontaneous chiral symmetry breaking. In a
  nutshell, the Dirac Lagrangian (\ref{eq_Lpsi}) for massless
  quarks is invariant under chiral transformations which rotate
  independently the left and right parts of the Dirac bispinor. This
  symmetry happens to be spontaneously broken by quantum fluctuations, which
  implies that, even if the valence quark mass ({\it i.e.}, the quark
  mass at a scale of the order of few GeV) is small, the
  constituent quark mass (the one defined at an IR scale, relevant, {\it e.g.}, for computing the mass
  of a hadron) is significantly larger.

  At a technical level, dynamical quarks are taken into account
  by adding the Dirac Lagrangian (\ref{eq_Lpsi}) to the CF action. The theory is
  renormalisable and requires new renormalisation factors, for the quark field and masses, $\psi=\sqrt{Z_\psi} \psi_R$, $M_{b,i}=Z_{M,i} M_i$.

In what follows, we describe the perturbative calculations of the propagators and the quark-antiquark-gluon vertex of the unquenched theory in the vacuum. For not too light quarks, they compare well with existing lattice data, as reviewed in Subsections~\ref{cacaboudin} and \ref{cacaboudeux} below. The dynamics of light quarks is strongly coupled and requires a more elaborate treatment, discussed in Subsection~\ref{pipichiotte}.

\subsection{Propagators.}\label{cacaboudin}

The calculation of the gluon, ghost, and quark propagators in perturbation theory is straightforward \cite{Pelaez:2014mxa,Siringo:2016jrc,Hayashi:2020few,Barrios:2021cks}. At strict one-loop order, the ghost propagator is unchanged as compared to the quenched case whereas the gluon propagator receives an additional quark loop contribution. The RG running yields an additional source of normlalis dependence which affects all propagators \cite{Pelaez:2014mxa,Pelaez:2015tba}. As in the quenched case, these calculations have been recently pushed to two-loop order, first, to assess the (apparent) convergence of the perturbative expansion and, second, to elucidate the case of the quark dressing function, which receives significant two-loop contributions \cite{Barrios:2021cks}. The details of the calculations, with and without RG improvement can be found in the mentioned references. Here, we give a brief summary of the comparison of the perturbative results for the propagators with various lattice data for different flavour contents and different quark masses \cite{Bowman:2004jm,Bowman:2005vx,Ayala:2012pb,Sternbeck:2012qs,Oliveira:2018lln}.

\begin{figure}[t]
 \includegraphics[width=.45\linewidth]{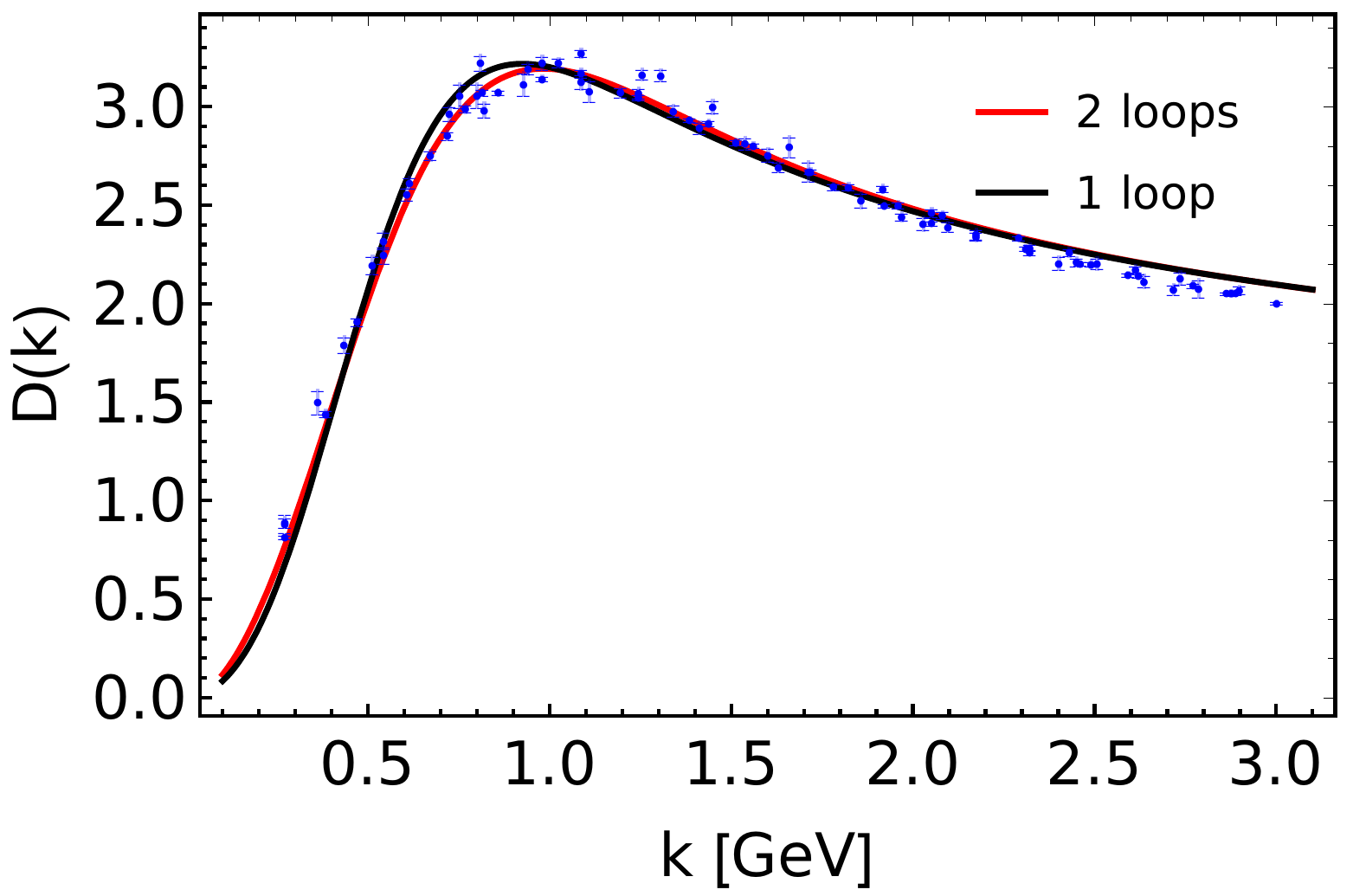}\quad
 \includegraphics[width=.44\linewidth]{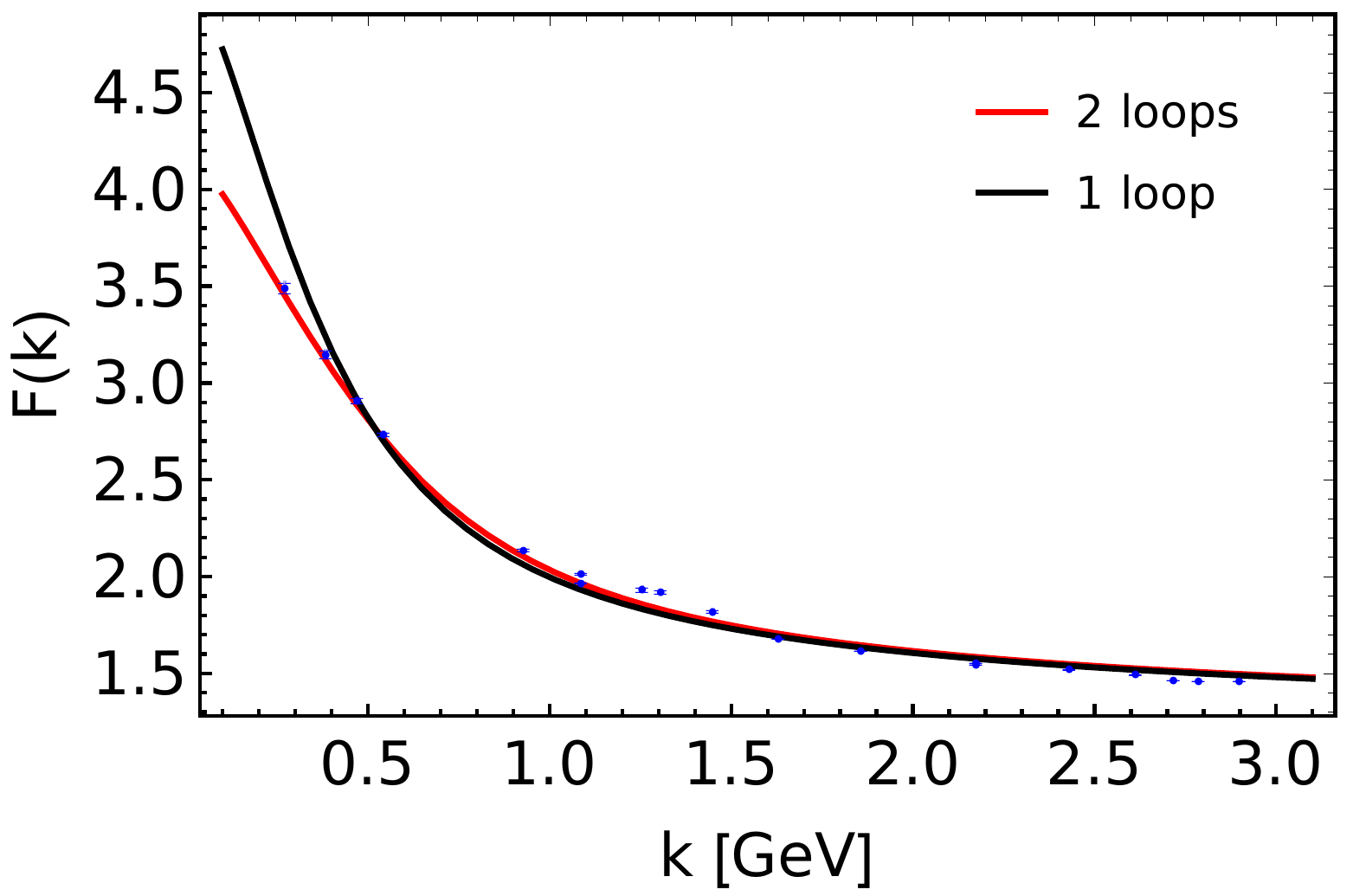}\\\vspace{.5cm}
 \includegraphics[width=.45\linewidth]{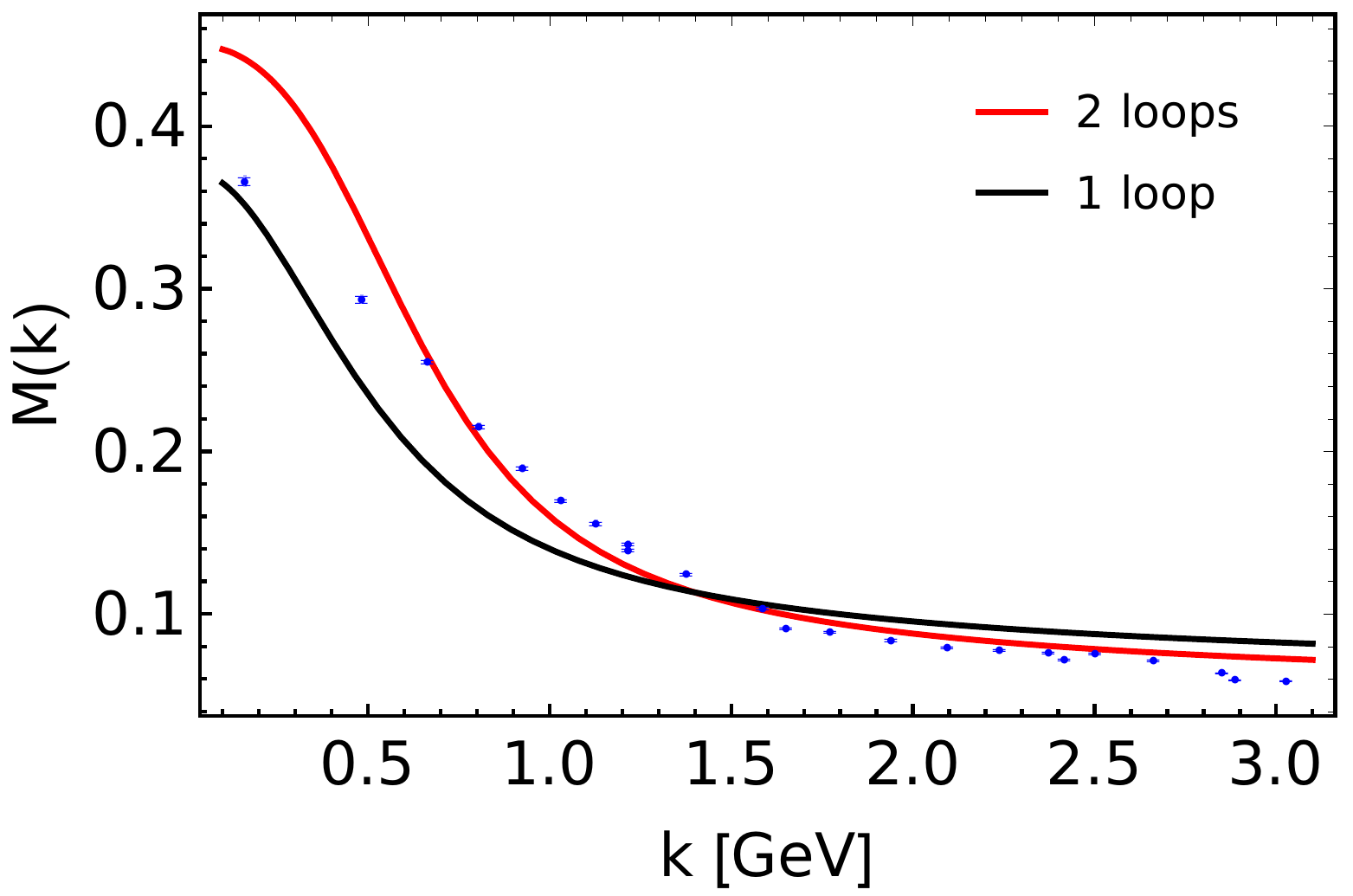}\quad
  \includegraphics[width=.43\linewidth]{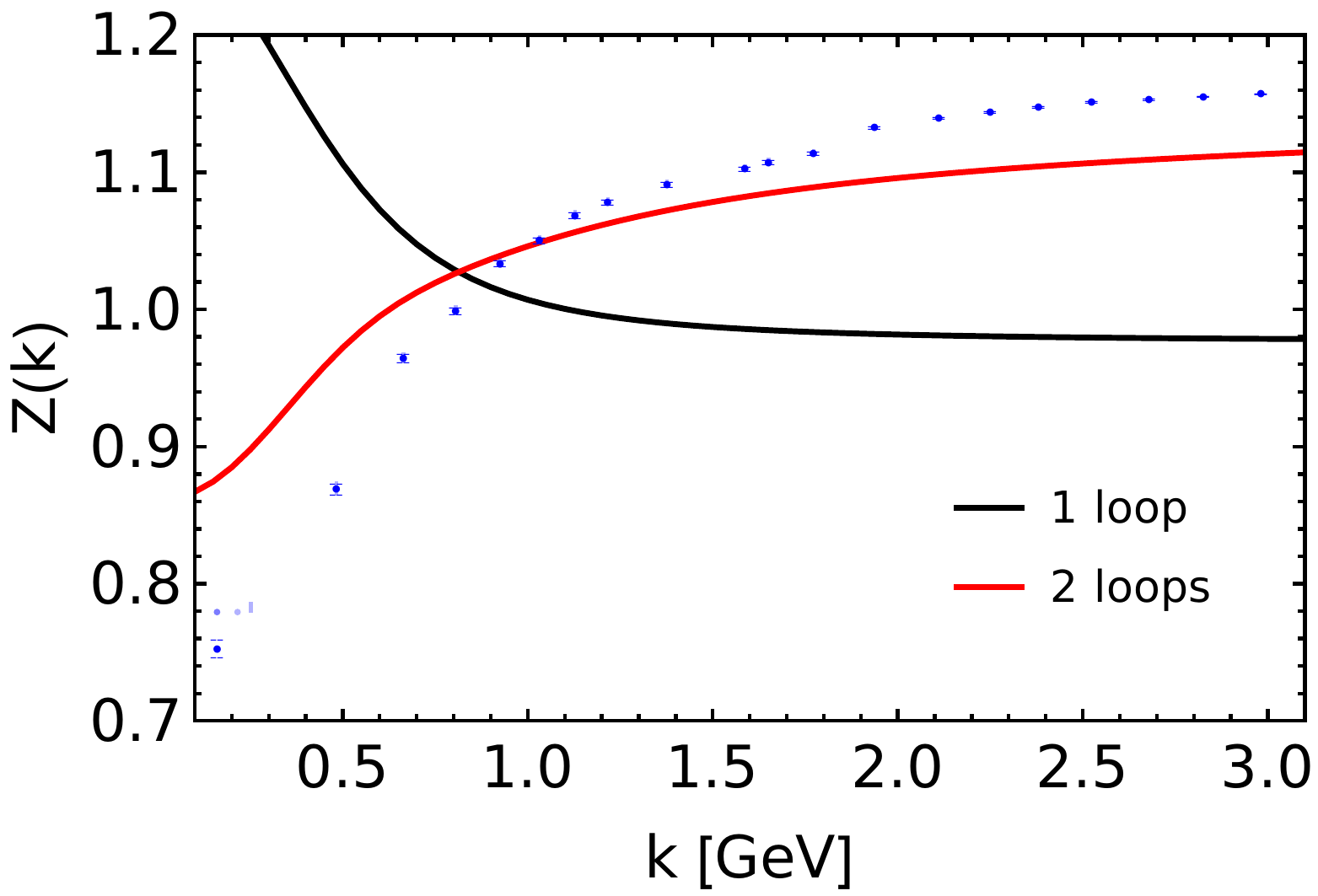}
\caption{The various propagators at one- and two-loop orders for $N_f=2$ degenerate normlaliss in $d=4$. The upper plots present the gluon (left) and ghost (right) dressing functions whereas the lower plots show the quark mass (left) and dressing function (right). The parameters (defined at the scale $\mu_0=1$~GeV) are $g_0=4.53$, $m_0=430$~MeV, and $M_0=140$~MeV at one loop and $g_0=4.10$, $m_0=390$~MeV, and $M_0=160$~MeV at two loops. The points are the lattice data of Ref.~\cite{Oliveira:2018lln} for $M_\pi=426$~MeV. Figures from Ref. \cite{Barrios:2021cks}.}
\label{twoloop426}
 \end{figure} 
 
\begin{figure}[t]
 \includegraphics[width=.45\linewidth]{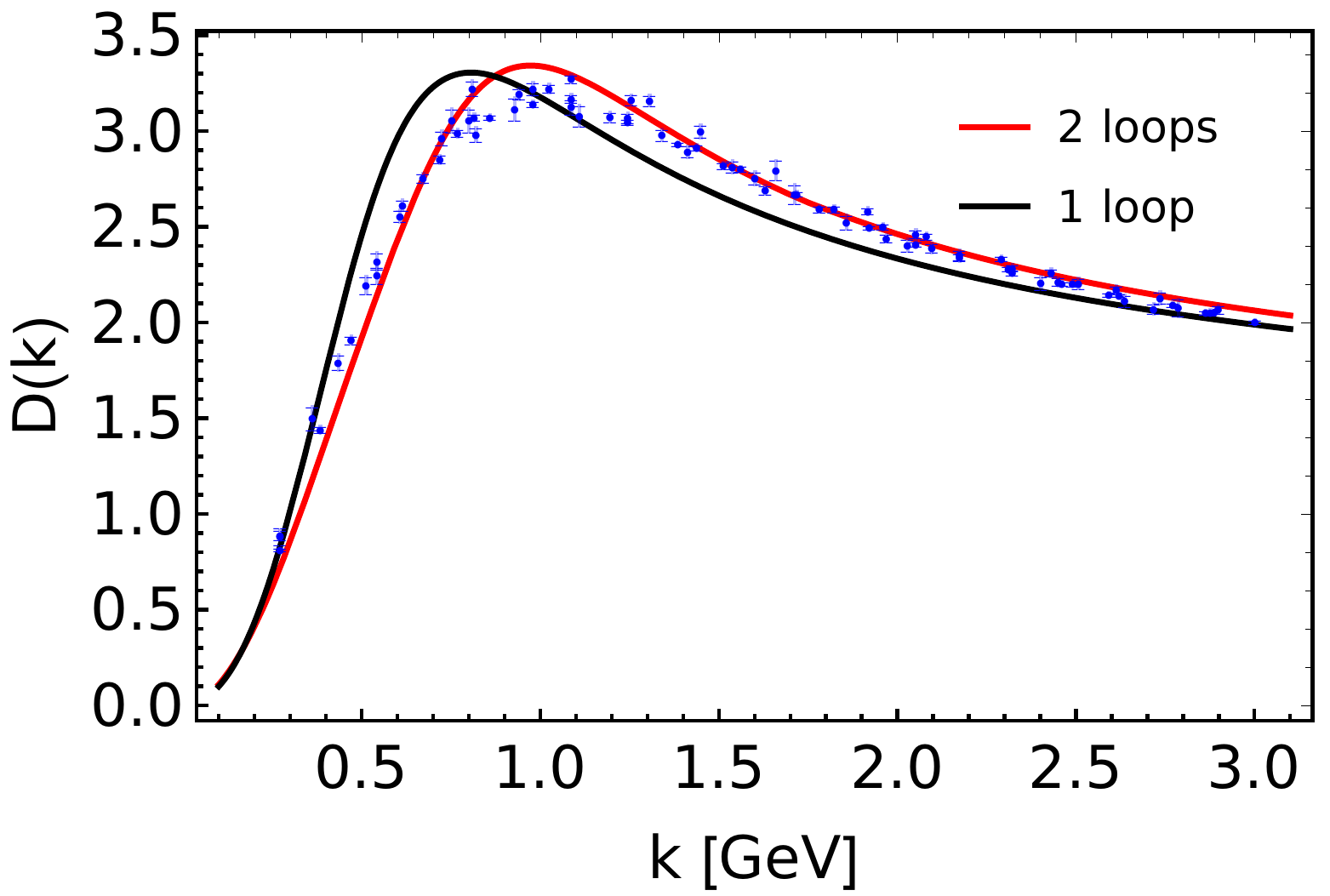}\quad
 \includegraphics[width=.43\linewidth]{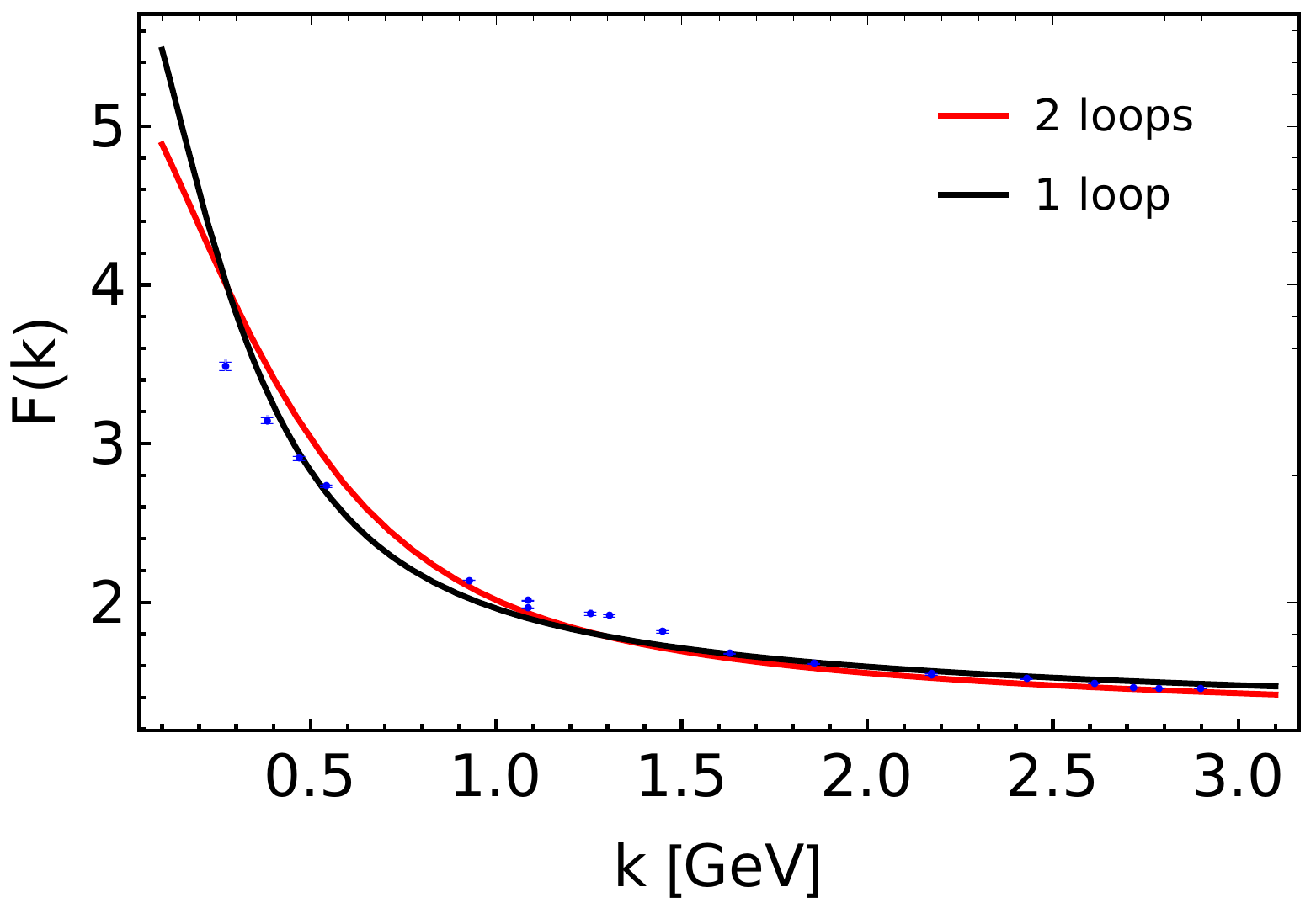}\\\vspace{.5cm}
 \includegraphics[width=.45\linewidth]{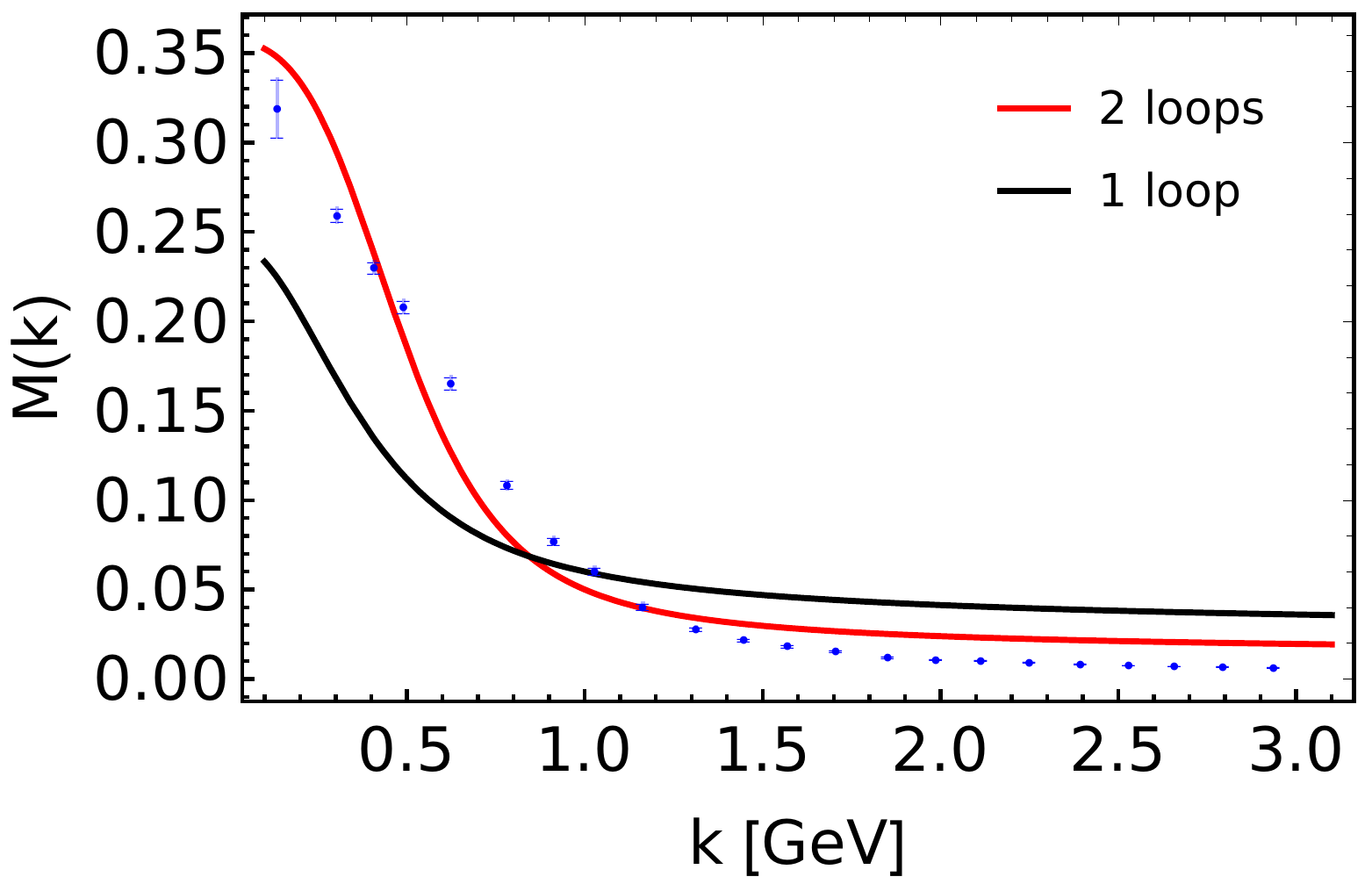}\quad
  \includegraphics[width=.43\linewidth]{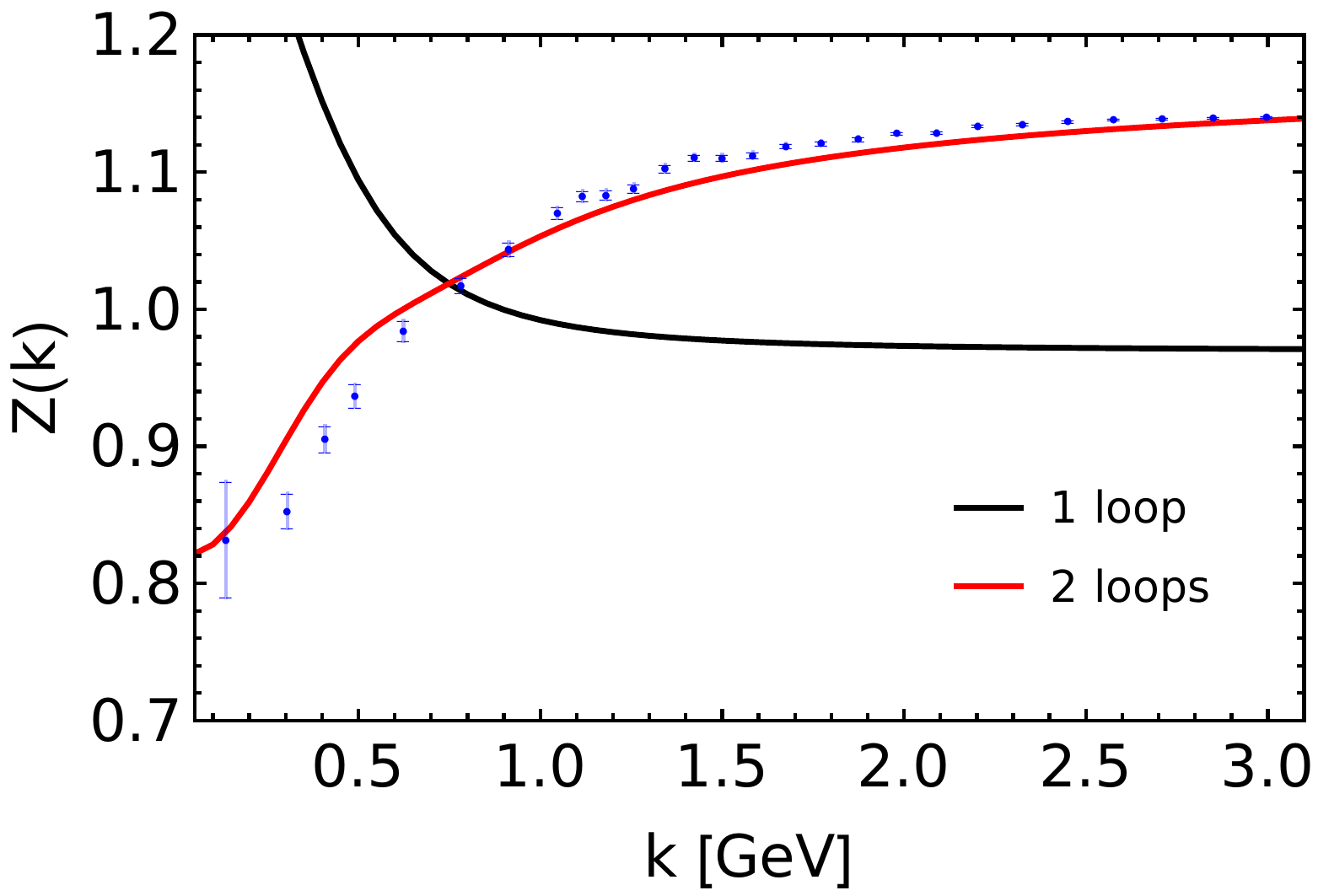}
\caption{The various propagators at one- and two-loop orders for $N_f=2$ degenerate normlaliss in $d=4$. The upper plots present the gluon (left) and ghost (right) dressing functions whereas the lower plots show the quark mass (left) and dressing function (right). The parameters (defined at the scale $\mu_0=1$~GeV) are $g_0=4.29$, $m_0=350$~MeV, and $M_0=60$~MeV at one loop and $g_0=4.35$, $m_0=360$~MeV, and $M_0=50$~MeV at two loops. The points are the lattice data of Ref.~\cite{Oliveira:2018lln} for $M_\pi=150$~MeV. Figures from Ref. \cite{Barrios:2021cks}.}
\label{twoloop150}
 \end{figure}

The first observation is that the sensitivity of the unquenched ghost and gluon propagators to the normlalis content of the theory is well described by the one-loop results. One observes though that, for a given normlalis content, these propagators are rather insensitive to the precise values of the quark masses in the range of at most a few hundred MeV. The renormalised quark propagator 
\begin{equation}\label{eq:qmfunc}
 S(p)=\frac{Z(p)}{-i\slashed p + M(p)}=Z(p)\frac{i\slashed p + M(p)}{p^2+M^2(p)}
\end{equation} 
has two independent components, the (RG invariant) mass function $M(p)$ and the dressing function $Z(p)$, which are both well measured on the lattice. One can easily find values of the  parameters for which the former is well reproduced by the one-loop results, in particular, if one only fits this function, independently of the others. Of course, fitting only one function can lead to artificially good results and, in order to obtain a realistic estimate of the quality of the one-loop approximation, one should fit all possible data with a single set of parameters. This is shown in Figs.~\ref{twoloop426} and \ref{twoloop150}, for the lattice data of Refs.~\cite{Sternbeck:2012qs,Oliveira:2018lln}, corresponding to $N_f=2$ and two values of the pion mass $M_\pi$ (or, equivalently, of the bare quark mass).  One observes that the one-loop results indeed give a satisfactory description of the data, including the quark mass function. Instead, the one-loop quark dressing function compares badly to lattice data. As anticipated in Ref.~\cite{Pelaez:2014mxa}, this is because the one-loop contribution to this function is abnormally small in the CF model---it vanishes identically in the FP theory---and higher-loop effects are not negligible in comparison.

 \begin{figure}[h!]
 \includegraphics[width=.45\linewidth]{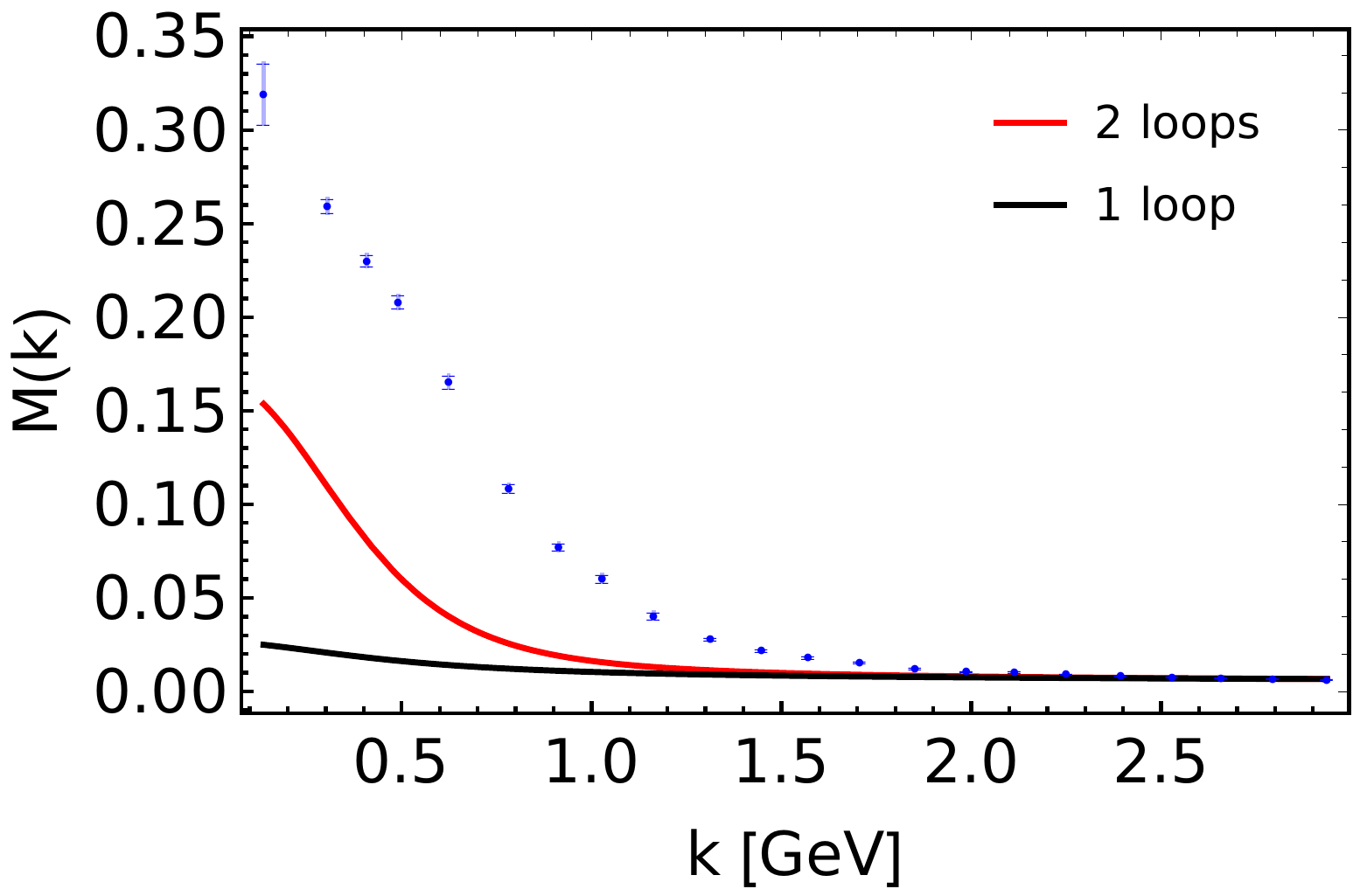}\quad
\caption{The quark mass function at one and two-loop orders as compared to the lattice data of Ref.~\cite{Oliveira:2018lln} for $M_\pi=150$~MeV. Here, the quark mass parameter $M_0$ is not part of the fit but is fixed to the lattice value at the UV scale $\mu_0 =2.94$~GeV. Figure from Ref. \cite{Barrios:2021cks}.}
\label{twoloopM12-150}
 \end{figure}

The perturbative calculations of the QCD propagators have recently been pushed up to two-loop order \cite{Barrios:2021cks}. As in the quenched case, this allows one to really assess the reliability of the perturbative CF description. Again, we refer the reader to Ref.~\cite{Barrios:2021cks} for details and, here, we just show the quality of the results in Figs.~\ref{twoloop426} and \ref{twoloop150}. The two-loop corrections clearly improve the one-loop results, in particular, for what concerns the quark dressing function. One also observes that the quality of the perturbative description is better for larger quark (or pion) masses.\footnote{It may seem, at first sight, that the quality of the fits are equally good for heavy and light quarks,  Figs.~\ref{twoloop426} and \ref{twoloop150}, respectively. However, as explained below, the detailed analysis of the errors reveals that it is not the case.} In fact, if one finds that the gluon, ghost and quark dressing functions seems always well-described by the two-loop perturbative results, this is not quite so for the quark mass function, in particular, for low pion masses. This is to be expected as the latter is directly sensitive to the dynamical breaking of the chiral symmetry in the chiral limit. This is clearly visible in Fig.~\ref{twoloop150}, corresponding to a rather unfavourable case, with $M_\pi=150$~MeV---close to the experimental value $m_\pi=140$~MeV. Perturbative calculations give an accurate description of the data from the IR to the UV, except for the quark mass function. Indeed, the good fit of the quark mass in the IR is at the expense of a rather poor description in the UV, even at two-loop order. If one insists, instead, on correctly describing the UV for the quark mass function, the fit deteriorates in the IR, as shown in Fig.~\ref{twoloopM12-150}. Thus, although the two-loop corrections improve significantly the one-loop results, the perturbative results are not able to reproduce the dynamical generation of the quark mass in the IR. That is a manifestation of the fact that the phenomenon of spontaneous chiral symmetry breaking is not captured at any finite order in perturbation theory and thus requires a more involved approximation scheme. This is discussed in Section~\ref{pipichiotte} below.

\subsection{The quark-gluon vertex}\label{cacaboudeux}

The quark-antiquark-gluon vertex has also been studied in lattice
simulations~\cite{Skullerud:2003qu,Skullerud:2004gp,Kizilersu:2006et,Sternbeck:2017ntv,Kizilersu:2021jen,Oliveira:2018fkj} and
can thus be used to further test the perturbative CF approach. This
vertex has a rich Dirac structure, with twelve scalar functions of
three momentum variables \cite{Skullerud:2002ge}. Only those which are
transverse with respect to the gluon momentum are accessible to
lattice simulations (in the Landau gauge) and only a restricted set
(and for particular momentum configurations) have actually been
measured. All have been computed at one-loop order in the CF model in
Ref.~\cite{Pelaez:2015tba} and compared to lattice data, which we
review here. One adjusts the parameters\footnote{In the previous section, the quark mass parameter is included in the fit together with $m_0$ and $g_0$. In the present and the next sections, instead, it is simply fixed to agree with the lattice value $M(\mu_0)$ at the scale $\mu_0$.} $g_0$, $m_0$, and $M_0$ of
the model by fitting the correlators as in the previous subsection,
following a similar procedure as the one outlined in
Sect.~\ref{sec_fit} for the YM case.  Once this is done, the results
for the various components of the three-point function are pure
predictions of the model, apart from a normalisation factor, similar to the case of the three-gluon vertex. Another
remark is that the results of
Refs.~\cite{Skullerud:2003qu,Skullerud:2004gp} are quenched ($N_f=0$)
lattice data. The comparison with the perturbative expressions for the
quark-antiquark-gluon vertex are still meaningful because there are no
quark loop contributions at one-loop order. Flavour effects only
enters through the RG running of the parameters so, in these
comparisons, the quenched running was used.\footnote{Since the work of
  \cite{Pelaez:2015tba}, new lattice data have been produced with
  dynamical quarks \cite{Sternbeck:2017ntv,Kizilersu:2021jen,Oliveira:2018fkj}.} We focus below on the particular configurations of momenta that have been studied in lattice
simulations.

\begin{figure}[t]
\centering
 \includegraphics[width=.45\linewidth]{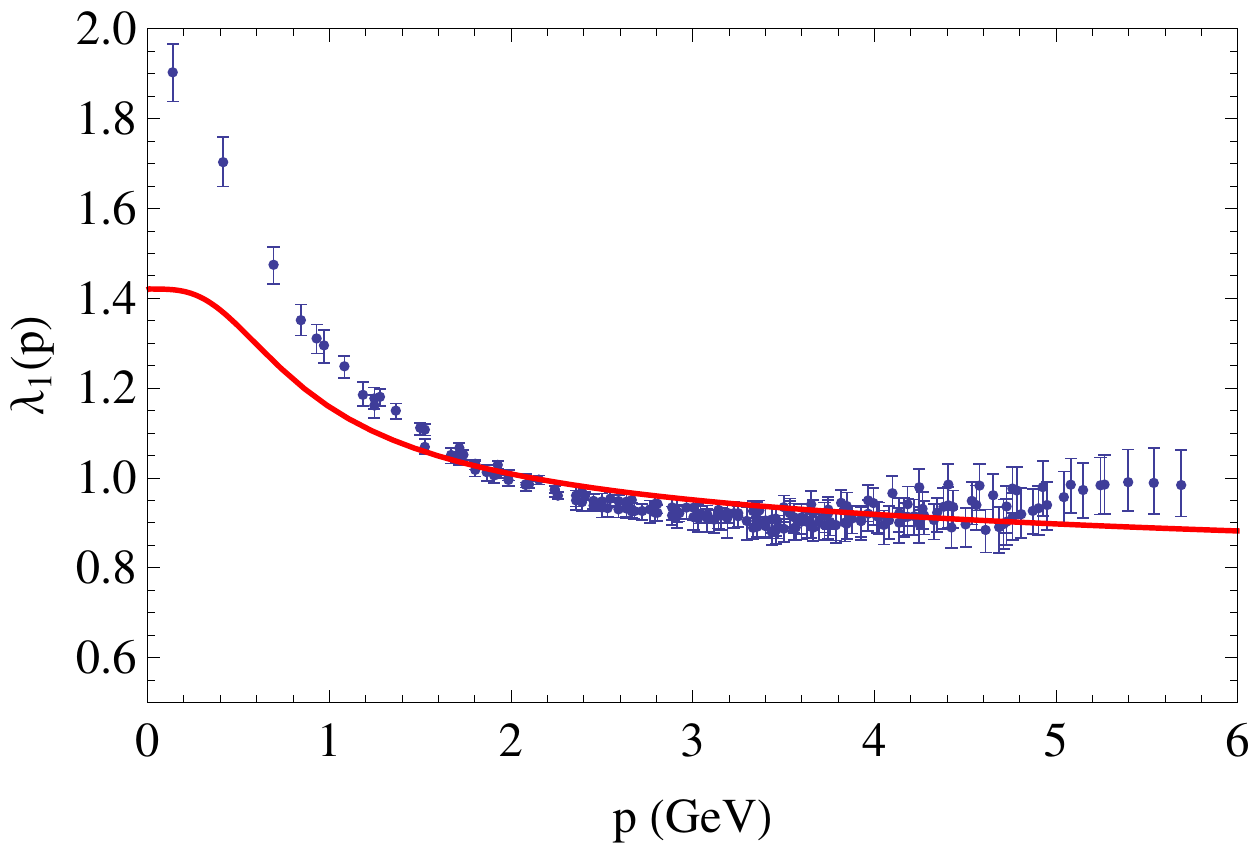}\qquad\quad
 \includegraphics[width=.46\linewidth]{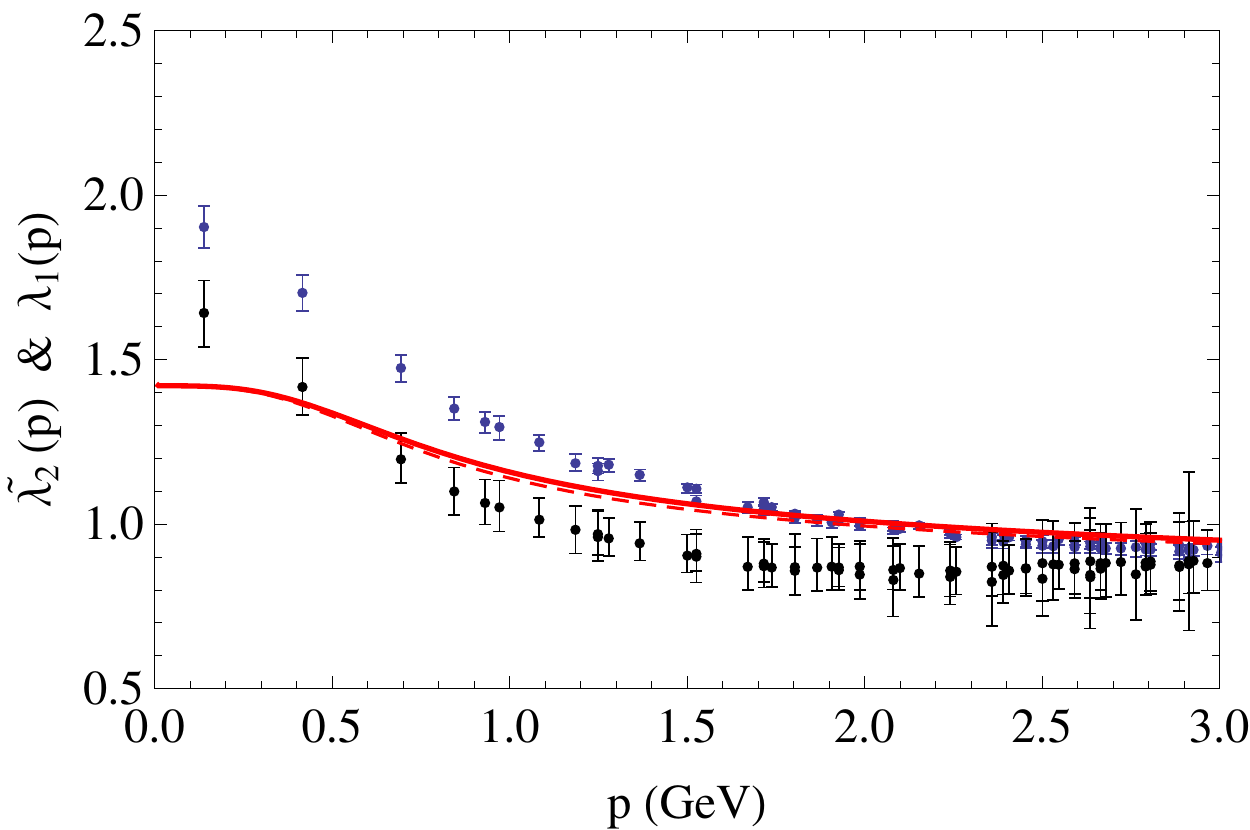}\\
 \includegraphics[width=.45\linewidth]{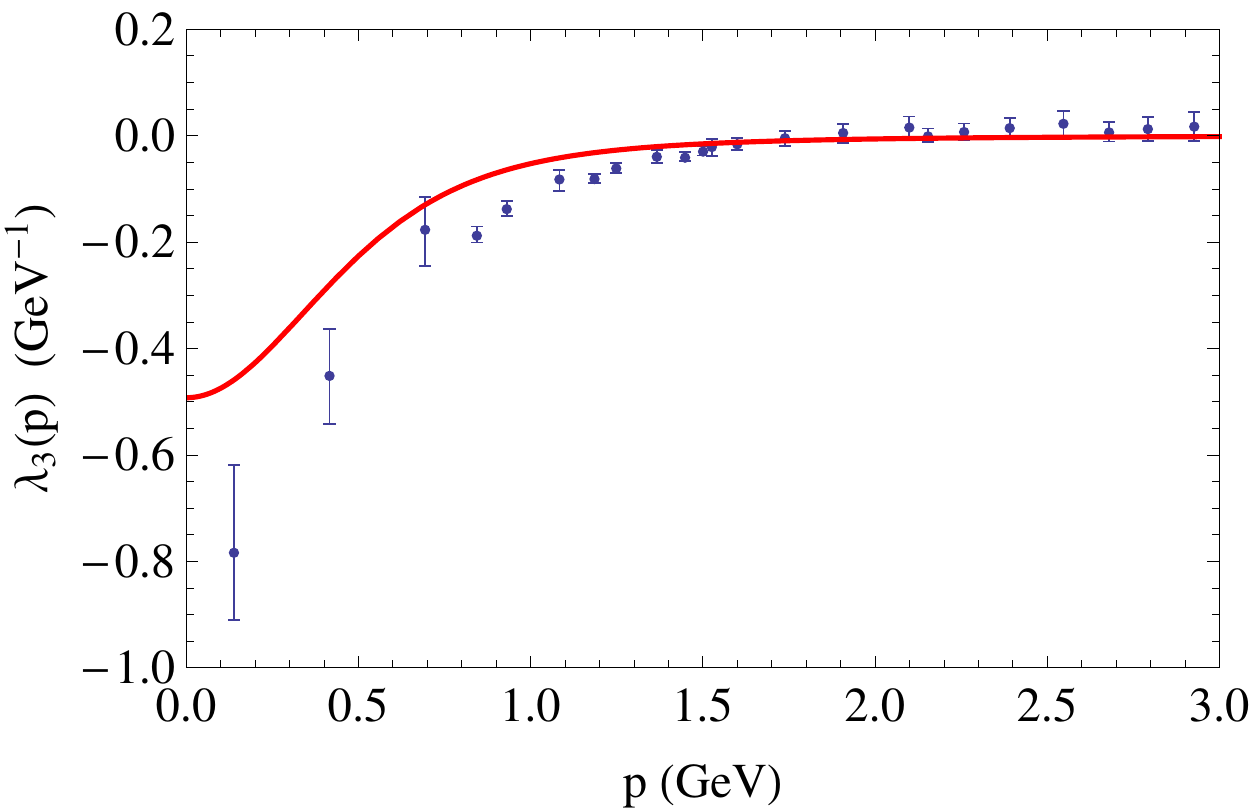}
 \caption{\label{FunEsck0} Scalar functions of the quark-gluon vertex at vanishing gluon momentum,
   $\lambda_1$, $\tilde\lambda_2$, and $\lambda_3$, as functions of the quark momentum. The lines are the one-loop results with $M_0=0.2$~GeV, $m_0=0.44$~GeV and $g_0=4.2$  at the scale $\mu_0=1$~GeV whereas the dots correspond to the lattice data of Ref.~\cite{Skullerud:2003qu}. The second figure shows both $\lambda_1$ (plain line, blue dots) and $\tilde\lambda_2$ (dashed line, black dots). The one-loop curves are almost superimposed. Figures from Ref. \cite{Pelaez:2015tba}.}
\end{figure}

For vanishing gluon momentum, the vertex has the following form
\begin{equation}
\label{zerogluonmomentum}
\Gamma_\mu(p,-p,0)=
-ig_g\left[\lambda_1(p)P_{\mu\nu}^\perp(p)+4\tilde\lambda_2(p)P_{\mu\nu}^\parallel(p)\right]\gamma_\nu-2g_g\lambda_3(p)p_\mu ,
\end{equation}
where, as before, $P_{\mu\nu}^\perp(p)=\delta_{\mu\nu}-p_\mu p_\nu/p^2$ and $P_{\mu\nu}^\parallel(p)=p_\mu p_\nu/p^2$. The function $\lambda_1(p)$ thus quantifies the renormalisation of the classical contribution $\propto\gamma_\mu$ in units of the (running) coupling $g_g$, defined as the ghost-antighost-gluon coupling in the Taylor scheme. The functions $\lambda_1(p)$, $\tilde\lambda_2(p)$, and $\lambda_3(p)$ are shown in Fig.~\ref{FunEsck0}. The agreement between the one-loop results and the lattice data is very good given the simplicity of the approximation. 

\begin{figure}[t]
\centering
 \includegraphics[width=.45\linewidth]{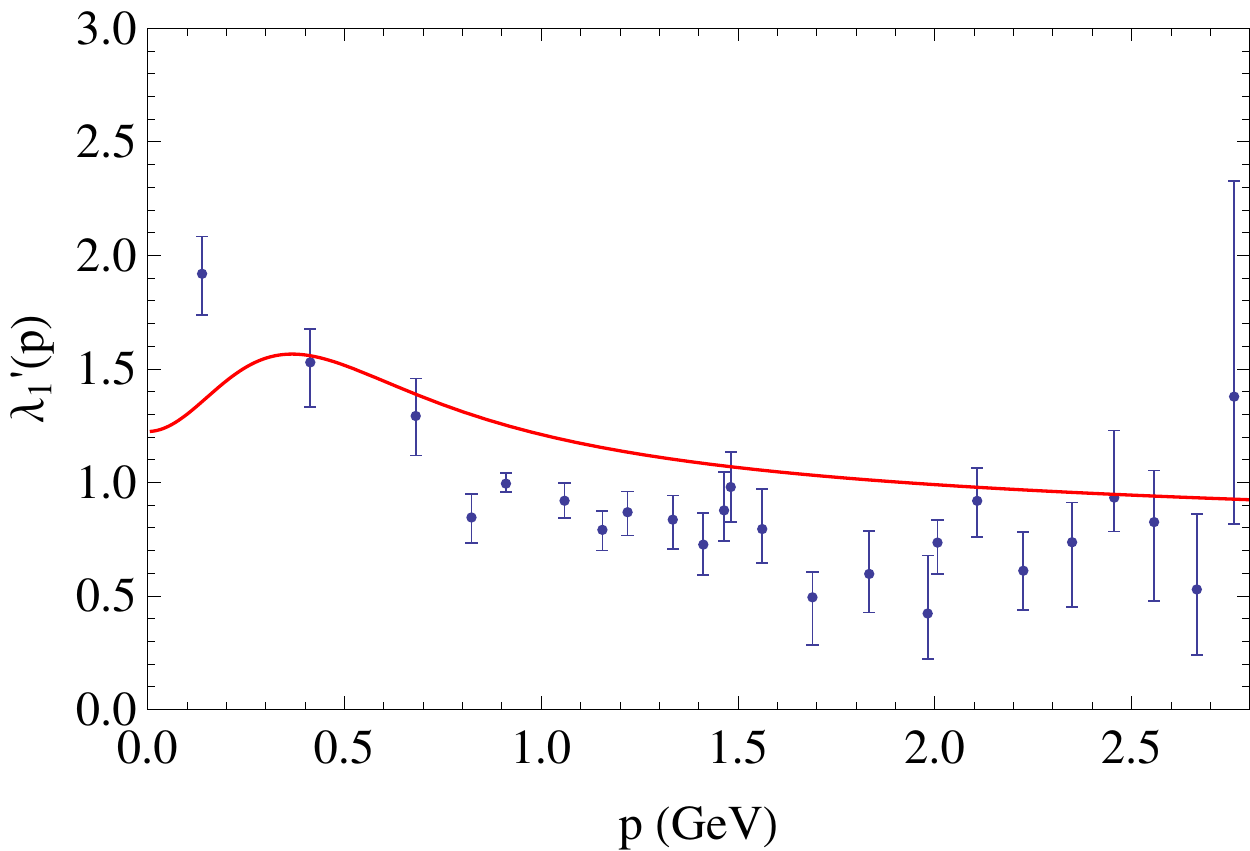}\qquad\quad
 \includegraphics[width=.46\linewidth]{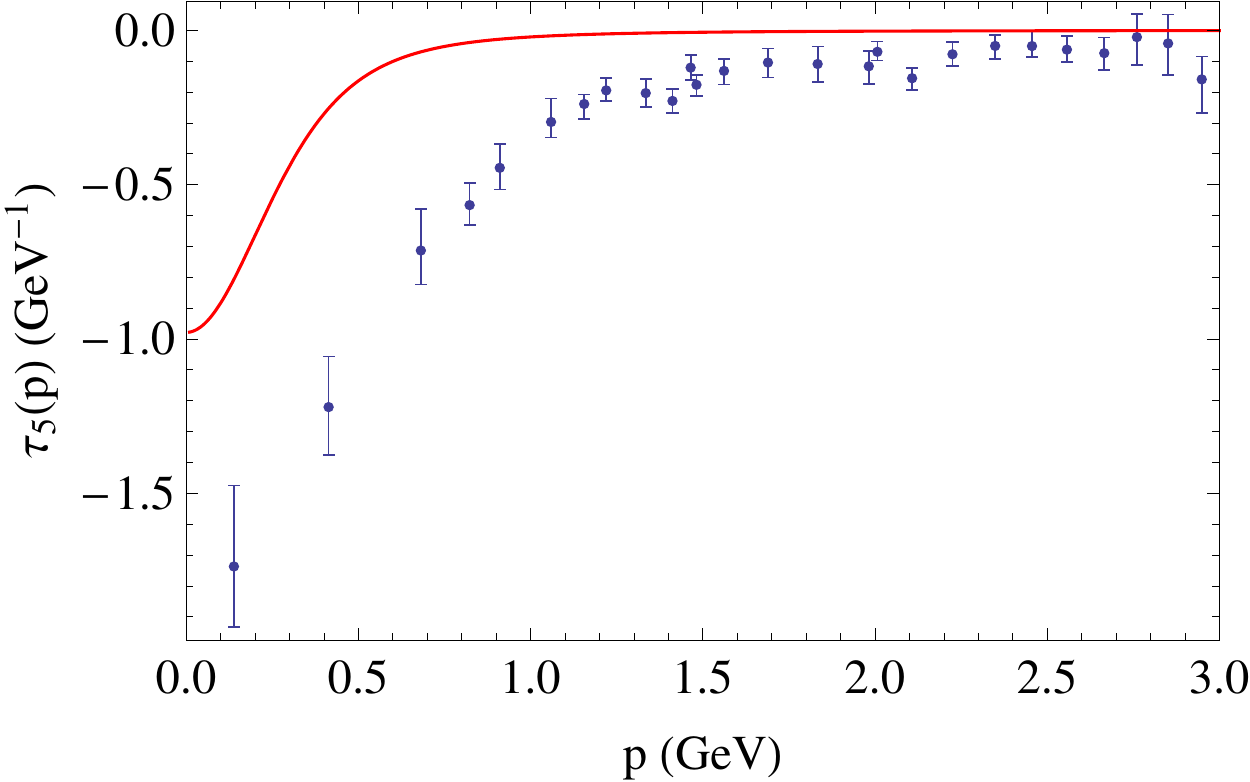}
 \caption{\label{FunEssim} The scalar components $\lambda_1'$ and $\tau_5$ of the quark-antiquark-gluon vertex for equal momentum $p$ of the quark and antiquark; see Eq.~\eqref{eq:pptwop}. The parameters are $M_0=0.2$~GeV, $m_0=0.44$~GeV and $g_0=4.2$ at the scale $\mu_0=1$~GeV. Figure from Ref.~\cite{Pelaez:2015tba}.}
 \end{figure}

The quark-gluon vertex has also been measured in lattice simulations for equal quark and antiquark momenta $p$ (so that the gluon momentum is $k=-2p$), which involves two other scalar functions $\tau_3$ and $\tau_5$:
\begin{equation}\label{eq:pptwop}
\Gamma_\mu(p,p,-2p)=-ig_g\left[\lambda'_1(p)\gamma_\mu+4\tau_3(p)\slashed p p_\mu-2i\tau_5(p)\sigma_{\mu\nu}p_\nu\right],
\end{equation}
where $\sigma_{\mu\nu}={\frac i 2}[\gamma_\mu,\gamma_\nu]$ and $\lambda_1'(p)=\lambda_1(p)-4p^2\tau_3(p)$. The functions $\lambda_1'$ and $\tau_5$ are shown in Fig.~\ref{FunEssim}. 
The functions $\lambda_1$ and $\tau_3$ are also defined for other momentum configurations and comparisons of similar quality have been obtained between one-loop results and lattice data \cite{Pelaez:2015tba}.

Summarising, in all the investigated cases, the agreement is good in view of the fact that there is no adjustable parameter apart froml an overall normalisation in this comparison. As in the pure gauge (quenched) theory, we thus conclude that part of the IR dynamics in the presence of dynamical quarks is well described by perturbative means in the CF model. The expected nonperturbative effects in the FP theory are partly taken into account in a simple gluon mass term. 
An important observation though is that the agreement between lattice data and one-loop results is generally less good at very low momenta. One explanation is that the quark-gluon coupling becomes too large in the IR for a reliable perturbative treatment. For instance, defining an effective quark-gluon coupling $g_q$ as the coefficient of the classical structure $\propto \gamma_\mu$ of the quark-antiquark-gluon vertex at vanishing gluon momentum, {\it i.e.}, $g_q=g_g\lambda_1$, we see, from the lattice data presented in Fig.~\ref{FunEsck0} that the ratio between the couplings in the quark and in the pure gauge sectors $g_q/g_g=\lambda_1$ can be as large as $2$ in the deep IR. This implies that the perturbative expansion parameter in the quark sector $\lambda_q=N_cg_q^2/(16\pi^2)$ can reach up to four times that of the pure gauge sector. With the typical value $g_g\sim4$ for $N_c=3$, that is, $\lambda_g\sim0.3$, the parameter $\lambda_q$ can reach up to 1.2 in the deep IR, a value for which a perturbative expansion is clearly invalid.\footnote{This argument combines estimates from the heavy-quark limit for $g_g$ and from the quenched limit for the quark-gluon vertex $g_q$ and has to be updated when (light) quarks fluctuations are taken into account; see below.}

\subsection{The rainbow-improved perturbative expansion and the spontaneous breaking of chiral symmetry.}\label{pipichiotte}

 The above observation that a straightforward coupling expansion is inadequate to treat the dynamics of light quarks goes along with the fact that the phenomenon of dynamical chiral symmetry breaking is not captured at any finite loop order in the CF---let alone in the FP---model. This, together with the fact that the coupling that governs the pure gauge sector remains small-to-moderate, motivates a perturbative expansion in the pure gauge coupling $g_g$ alone, keeping all orders in the quark coupling $g_q$. One can further exploits another effectively small parameter in SU($N_c$) gauge theories, namely $1/N_c$. Although $N_c=3$ in QCD, it is well-known that a $1/N_c$ expansion successfully captures essential aspects of the dynamics \cite{tHooft:1974pnl,Witten:1979kh,DeGrand:2016pur}. 

In Refs.~\cite{Pelaez:2017bhh,Pelaez:2020ups}, a controlled approximation scheme for IR QCD has been proposed, based on a double expansion of the CF model in the two parameters $g_g$ and $1/N_c$. For instance, at leading order, the propagators in the gauge sector (ghost and gluon) are given by their tree-level expression whereas the quark self-energy includes the infinite set of so-called rainbow diagrams with a massive one-gluon exchange. This results in coupled integral equations for the quark renormalisation and mass functions which read, in terms of bare quantities,
\begin{align}\label{eq:RB}
Z^{-1}(p)&=1-g_{q,b}^2C_F\int_\Lambda\frac{d^4q}{(2\pi)^4}\,\frac{Z(q)}{q^2+M^2(q)}\,\frac{2\ell^4-(p^2+q^2)(p^2+q^2+\ell^2)}{p^2\ell^2(\ell^2+m_b^2)}\\
Z^{-1}(p)M(p)&=M_b+3g_{q,b}^2C_F\int_\Lambda\frac{d^4q}{(2\pi)^4}\,\frac{Z(q)M(q)}{q^2+M^2(q)}\,\frac{1}{\ell^2+m_b^2}
\end{align}
where $\ell=p+q$ and $\int_\Lambda$ denotes an appropriately regulated momentum integral. Here, $M_b$, $m_b$, and $g_{q,b}$ are the (bare) quark mass, gluon mass, and quark gluon coupling, respectively, and $C_F=(N_c^2-1)/(2N_c)$ is the Casimir of the fundamental representation of the SU($N_c$) gauge group.

The rainbow resummation is known to capture the essential features of chiral symmetry breaking and the associated dynamical quark mass generation and has been widely used in the context of nonperturbative approaches to IR QCD (the bibliography on this topic is extremely large; for a review, see \cite{Maris:2003vk}).  The benefit of the double expansion proposed in Refs.~\cite{Pelaez:2017bhh,Pelaez:2020ups}---dubbed the rainbow-improved (RI) loop expansion---is that it avoids {\it ad hoc} modelling for the quark-gluon vertex and for the gluon propagator entering the rainbow diagrams. At leading order, these are simply given by their tree-level expressions in the CF model. Another asset is that the actual expansion in small parameters  underlying the approximation scheme allows for a systematic implementation of RG improvement, which is crucial for a proper description of chiral symmetry breaking.

\begin{figure}[tbp]
\centering
\includegraphics[width=.47\linewidth]{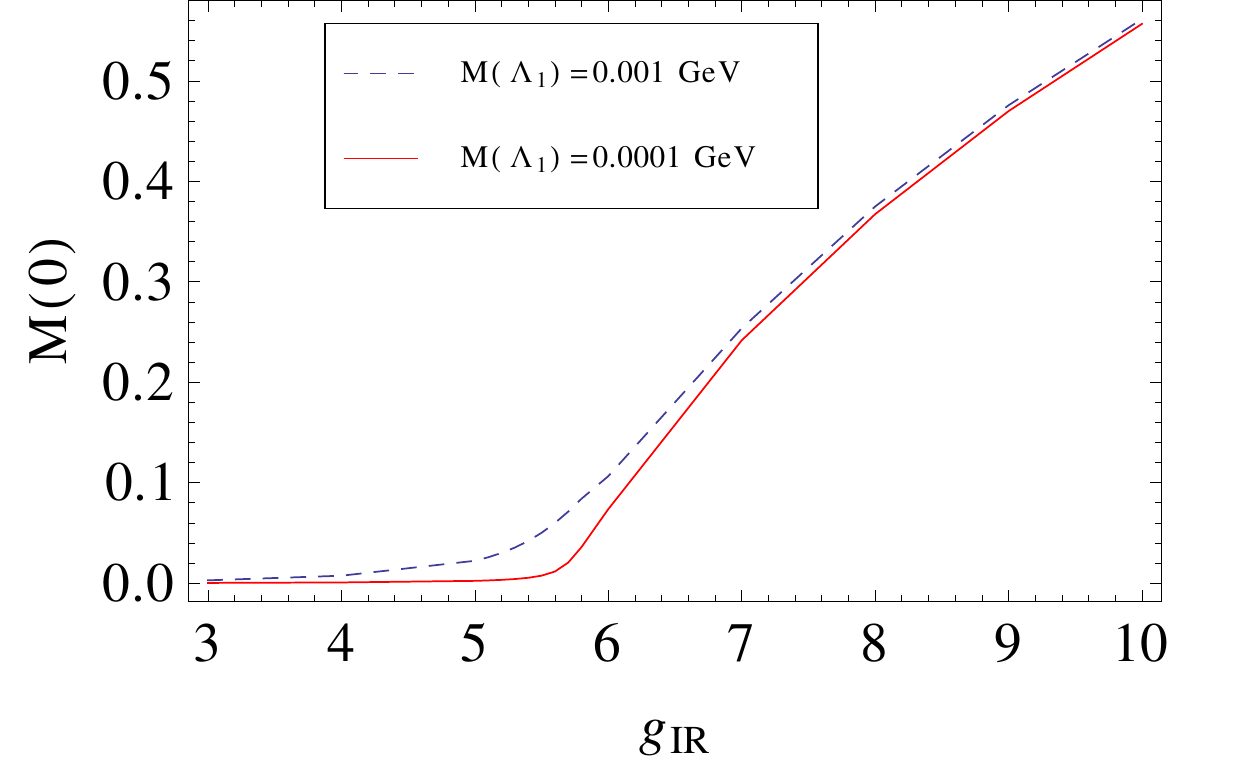}\qquad\quad
\includegraphics[width=.44\linewidth]{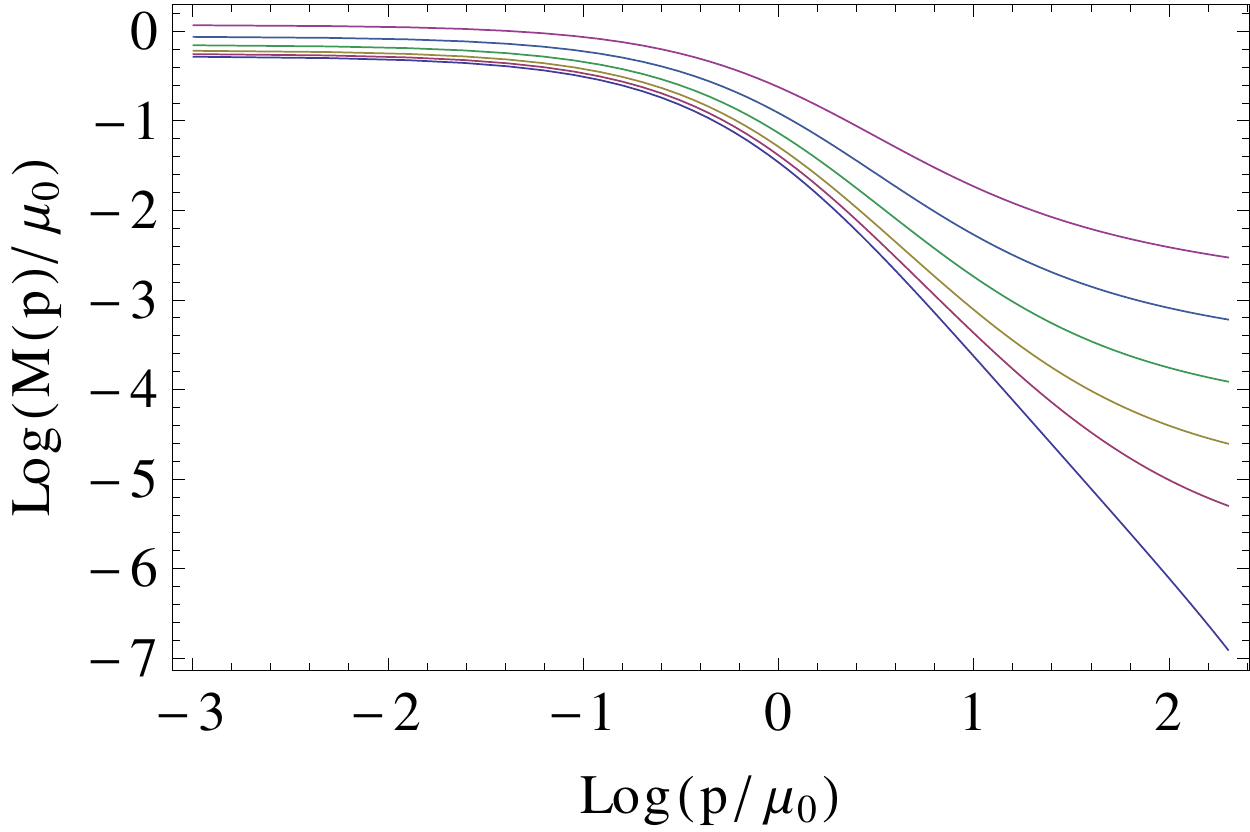}
\caption{Left: Constituent quark mass $M(p=0)$ as a function of the coupling parameter $g_{\rm IR}$ (see text) for two values of the UV mass $M(\Lambda_1)$. The variation
of $g_{\rm IR}$ is done by keeping $\Lambda_{\rm QCD}$ fixed.
Right: Mass function $M(p)$ in Log-Log scale for decreasing (from top to bottom) values of $M(\Lambda_1)$ at the UV scale $\Lambda_1=10$~GeV. We observe the onset of the power law behaviour at large momentum, characteristic of the spontaneous breaking of the chiral symmetry, as the chiral limit is approached. Figures from Ref.~\cite{Pelaez:2017bhh}.
\label{log-log}}
\end{figure}
The RI loop expansion at first nontrivial order has been implemented in Ref.~\cite{Pelaez:2017bhh} using a toy-model RG running and in Ref.~\cite{Pelaez:2020ups} using a consistent implementation of the RG running within the RI expansion. We refer the reader to these articles for the technical details and we review the main results here.
Figure \ref{log-log} illustrates some aspects of the chiral symmetry breaking phenomenon through the quark mass function $M(p)$ [see Eq.~\eqref{eq:qmfunc}] using a toy model for the running quark-gluon coupling parametrized, in particular, by a finite IR value $g_{\rm IR}$. A nonzero value $M(p=0)$ is dynamically generated in the chiral limit---obtained by decreasing the value of $M(\Lambda_1)$ at the UV scale $\Lambda_1$---above a critical value of $g_{\rm IR}$. Also shown in Fig. \ref{log-log} is the fact that the logarithmic UV tail of the theory with massive quarks turns into a power law in the chiral limit, $M(p)\propto \langle\bar\psi\psi\rangle/p^2$ (up to calculable logarithmic corrections), controlled by the dynamically generated quark-antiquark condensate. 

\begin{figure}[tbp]
\centering
\includegraphics[width=8cm]{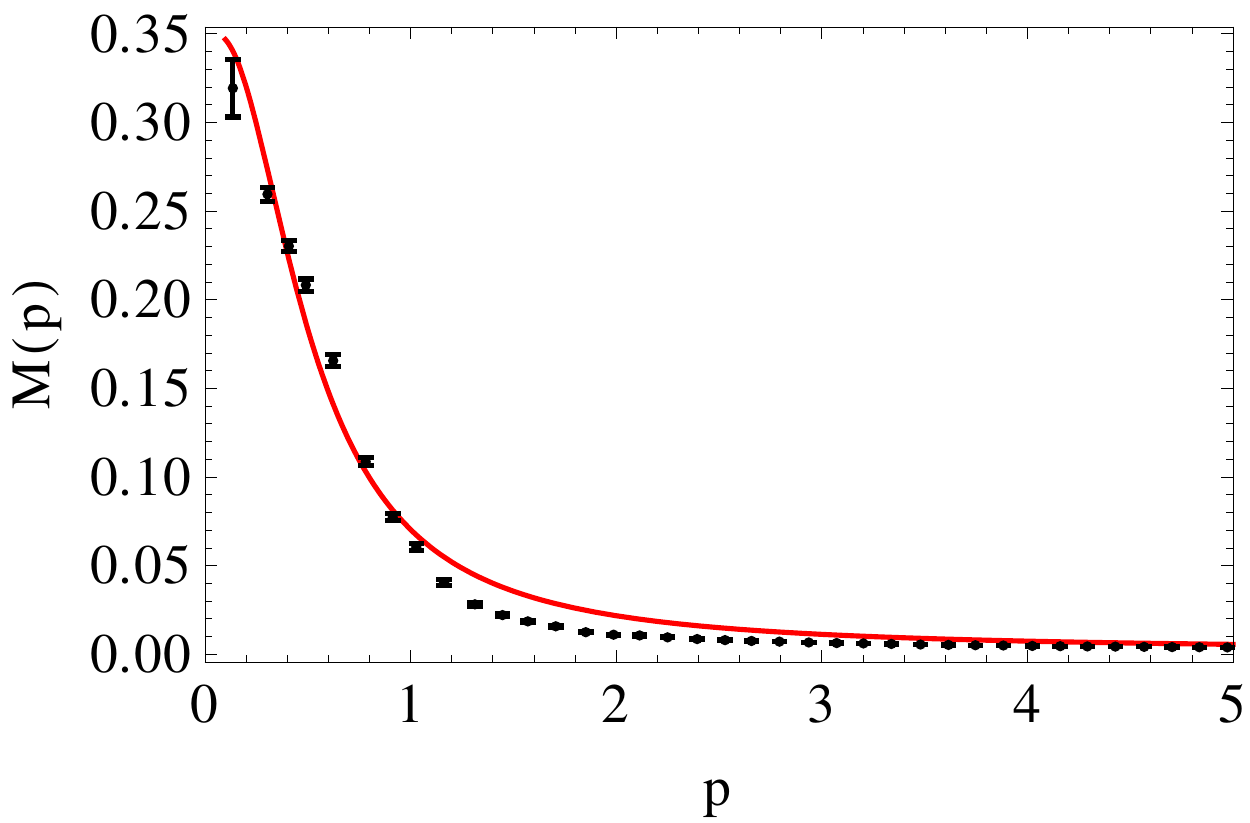}\qquad\quad
\includegraphics[width=8cm]{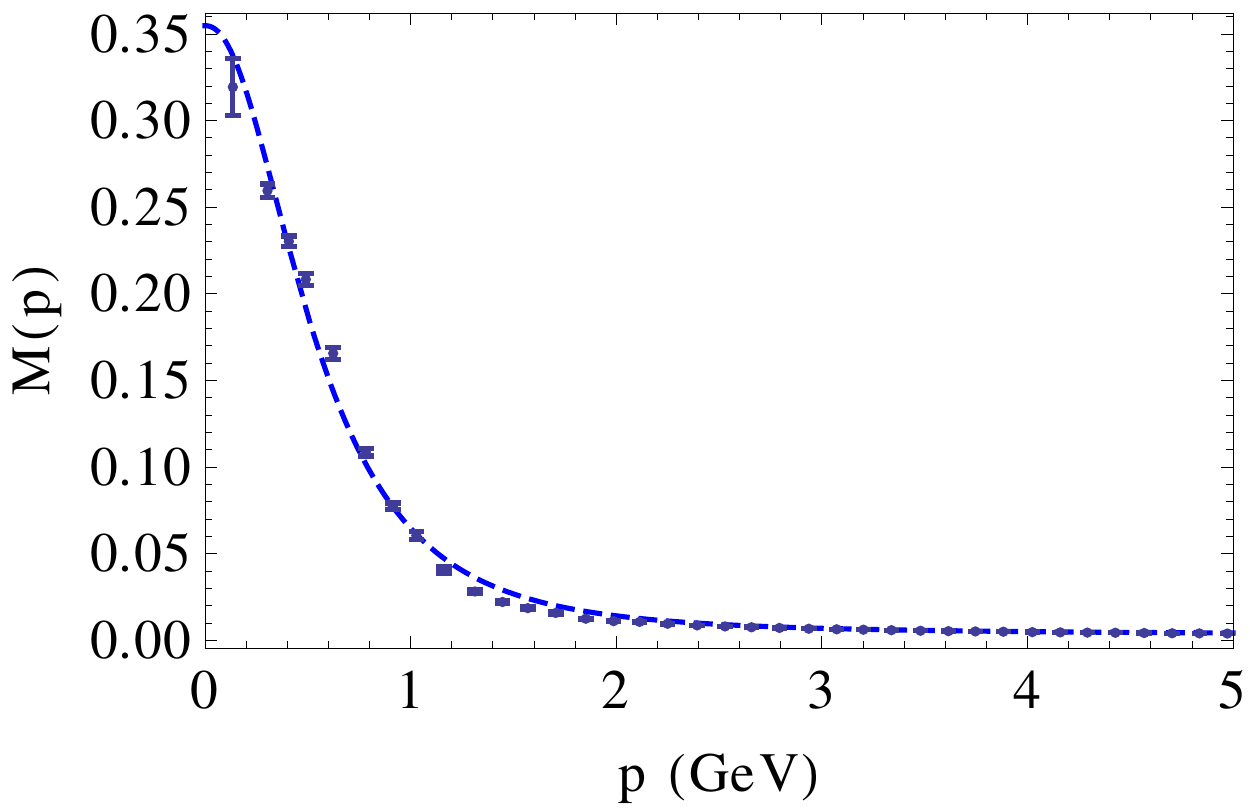}
\caption{\label{Fig:QMonly}Left: The quark mass function $M(p)$ compared to the lattice data from Ref.~\cite{Oliveira:2018lln}. The 
quark mass at the scale $\Lambda_1=10$~GeV is fixed at its lattice value $M(\Lambda_1)=3$~MeV. The best fit parameters (see text) are $m_0=0.12$ GeV, and $g_0=2.42$ at $\mu_0=1$ GeV. Right:  The RG-improved one-loop expression of Ref.~\cite{Pelaez:2014mxa} for $M(p)$ is also able to describe well lattice data. This, however, necessitates artificially large values of the gluon mass and of the coupling, here, $m_0=1.6$~GeV and $g_0=13$ at $\mu_0=1$~GeV. In particular, the one-loop approximation makes no sense. Figures from Ref. \cite{Pelaez:2020ups}.}
\end{figure} 
\begin{figure}[h]
\includegraphics[width=8cm]{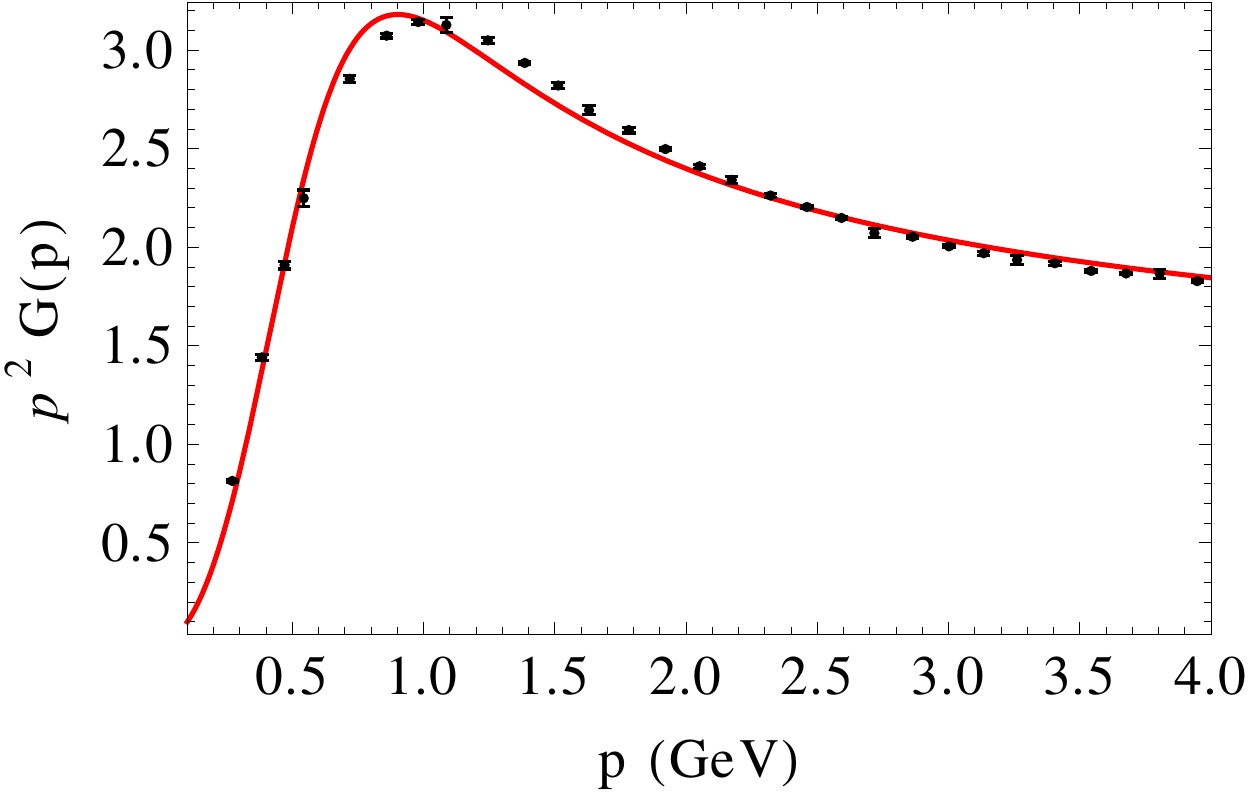}\qquad\quad
\includegraphics[width=8cm]{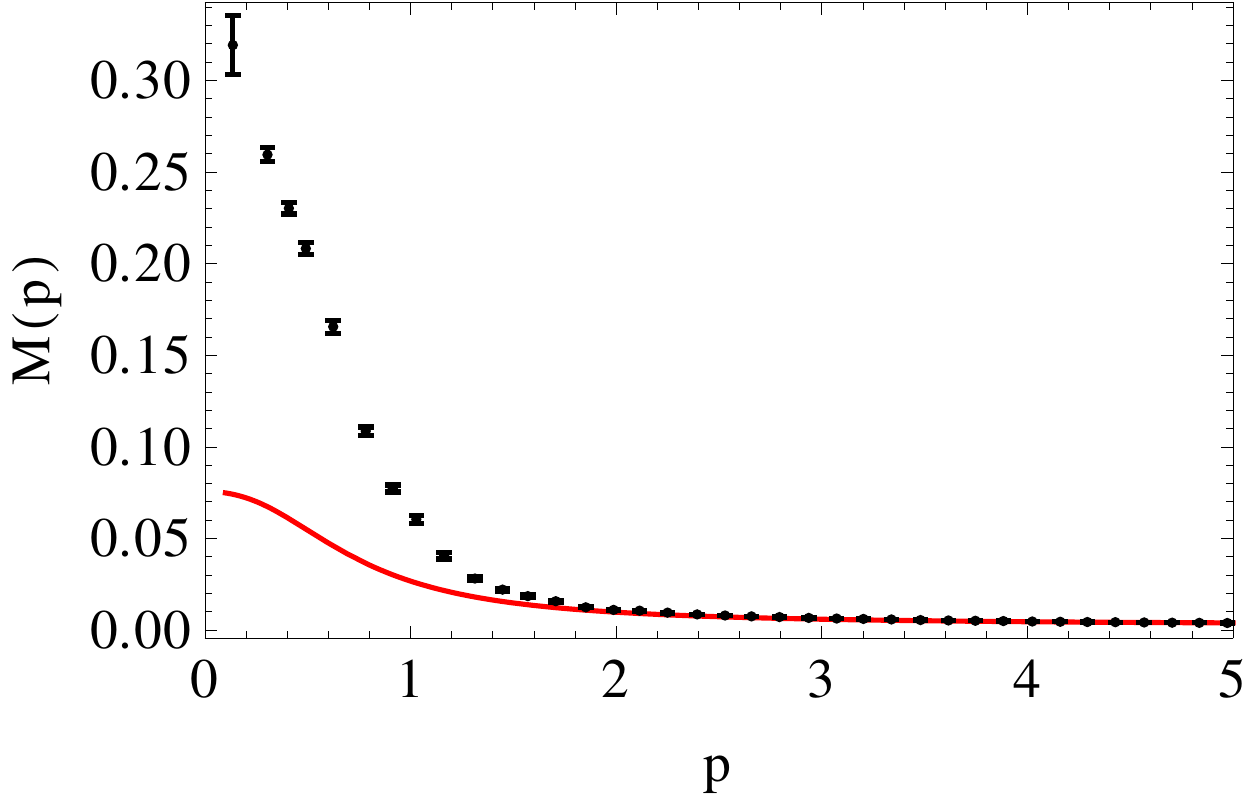}
\caption{\label{Fig:Glonly} The gluon dressing function $p^2G(p)$ (left) and the quark mass function $M(p)$ (right) compared with 
lattice data from Ref.~\cite{Sternbeck:2012qs}. Here the fit is adjusted on the gluon dressing function alone and the best fit parameters (see text) are $m_0=0.39$~GeV and $g_0=4.67$ at $\mu_0=1$~GeV. The 
quark mass at the scale $\Lambda_1=10$~GeV is fixed at its lattice value $M(\Lambda_1)=3$~MeV. Figures from Ref. \cite{Pelaez:2020ups}.}
\end{figure} 
\begin{figure}[t]
\includegraphics[width=8cm]{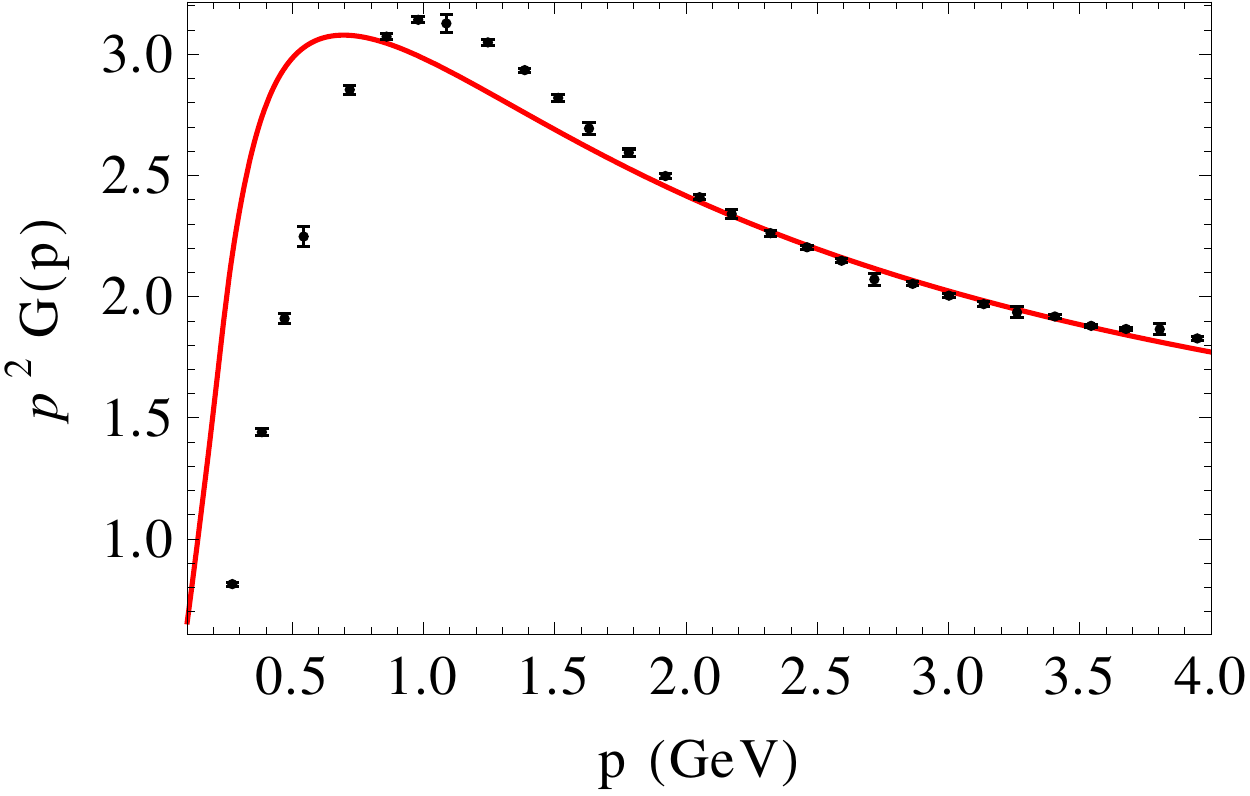}\qquad\quad
\includegraphics[width=8cm]{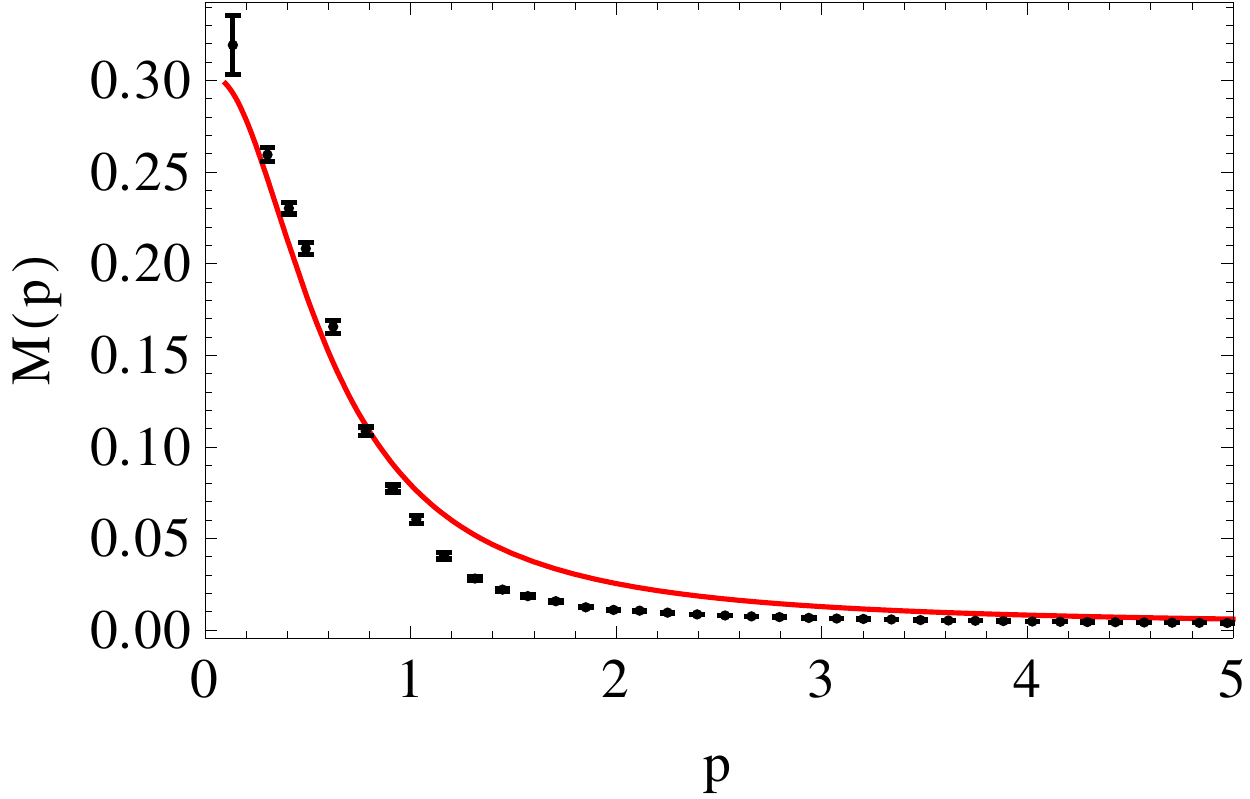}
\caption{\label{Fig:QMandGl} Combined fit of the gluon dressing function $p^2G(p)$ (left) and the quark mass function $M(p)$ (right) against the 
lattice results of Refs.~\cite{Oliveira:2018lln} and \cite{Sternbeck:2012qs}, respectively. The 
quark mass at the scale $\Lambda_1=10$~GeV is fixed at its lattice value $M(\Lambda_1)=3$~MeV. The best fit parameters (see text) are $m_0=0.21$~GeV, and $g_0=2.45$ at $\mu_0=1$~GeV. Figures from Ref.~\cite{Pelaez:2020ups}.}
\end{figure} 
\begin{figure}[h]
\includegraphics[width=8cm]{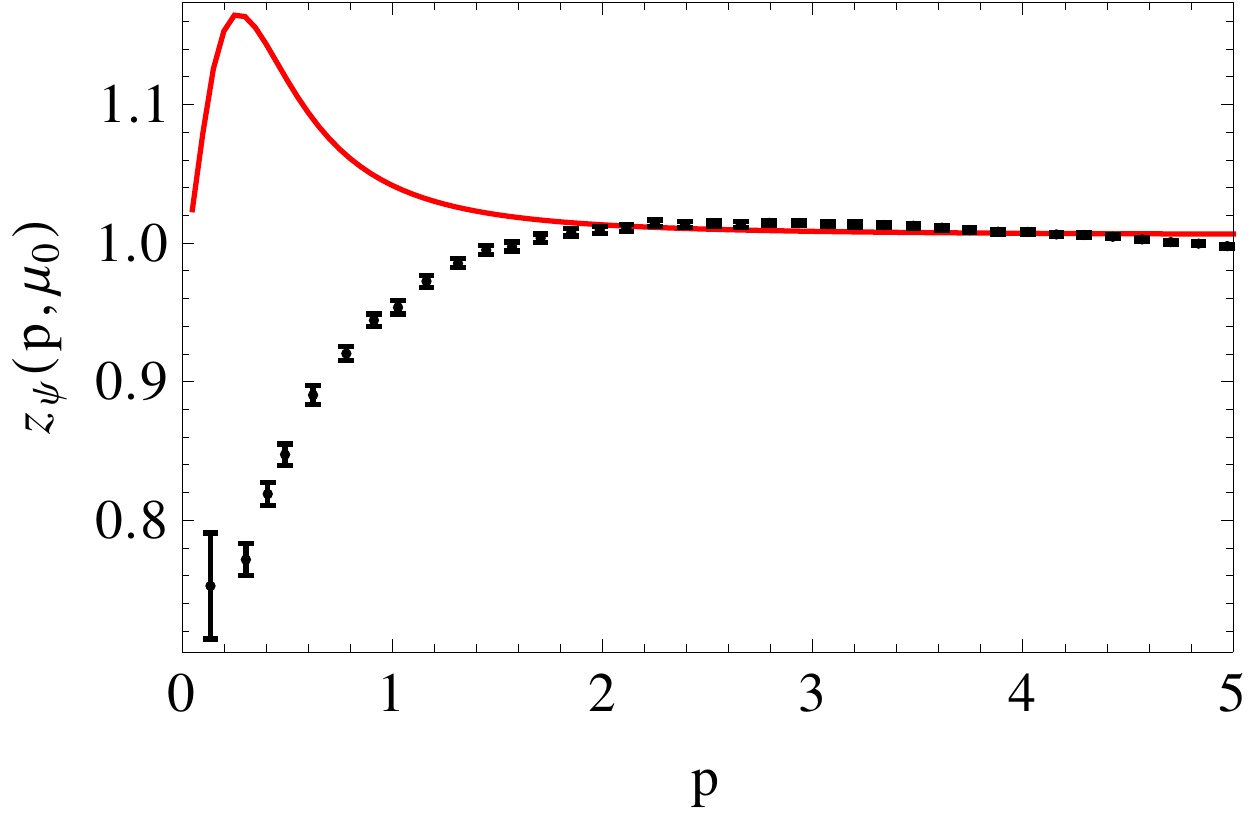}
\caption{\label{Fig:QZ} The function $z_\psi(p,\mu_0)=Z(p)/Z(\mu_0)$ normalised to $z_\psi=1$ at $p=2$ GeV against the lattice data of Ref. \cite{Oliveira:2018lln}, for the set of parameters $M(\Lambda_1)=3$~MeV, $m_0=0.21$~GeV, and $g_0=2.45$ at $\Lambda_1=10$~GeV and $\mu_0=1$~GeV. Figure from  Ref.~\cite{Pelaez:2020ups}.}
\end{figure} 
 
The results of the RI expansion for the quark and gluon propagators at next-to-leading order, using a self-consistent RG running, have been computed in Ref.~\cite{Pelaez:2020ups} and compared to lattice data close to the chiral limit. At this order, the quark propagator is unchanged as compared to its leading-order expression, {\it i.e.}, it is given by the resummation of rainbow diagrams with tree-level one-gluon exchange, whereas the gluon propagator receives the usual perturbative one-loop corrections in the gauge sector and an effective quark-loop with rainbow-resummed quark propagators. 

The comparison with lattice data is made by adjusting the various parameters, namely, the quark and gluon masses and the coupling\footnote{It is worth emphasising that, despite the different treatment of the (IR) couplings in the quark and in the pure glue sectors, there is, {\it in fine}, only one coupling parameter to specify in actual implementations of the RI loop expansion because both couplings are related by Slavnov-Taylor identities. In the UV, this relation is under perturbative control and one can fix the value of $g_q$ in terms of $g_g$. In this section, $g_0$ refers to the pure gauge coupling at the scale $\mu_0$: $g_0=g_g(\mu_0)$.} at the scale $\Lambda_1=10$~GeV. The gluon mass and the coupling are included as fit parameters and, in order to ease the comparison with the previous sections, we give their values $m_0$ and $g_0$ at the scale $\mu_0=1$~GeV, obtained through the proper RG running. The quark mass is not included in the fits but is, instead, fixed to the lattice value at the scale $\Lambda_1=10$~GeV, namely, $M(\Lambda_1)=3$~MeV in all the figures presented here. This corresponds to the (renormalised) current quark mass and plays the role of the control parameter for the chiral limit. The corresponding value $M_0$ of the running quark mass at $\mu_0=1$~GeV can be read off the plots of $M(p)$ for each case under consideration. Fig.~\ref{Fig:QMonly} shows that one can obtain an excellent agreement for the quark mass function alone.  It was noticed that a similar agreement can be obtained from the one-loop calculation presented in the subsection \ref{cacaboudin} above, however, at the price of an uncomfortably large value of the gluon mass parameter. Good descriptions of the data with the RI expressions typically favour comparatively small values of the gluon mass parameter. Similarly, fitting the gluon propagator alone leads to very good agreement as shown in Fig.~\ref{Fig:Glonly}. This, however, favors relatively large values of the gluon mass parameter, which thus deteriorates the quality of the agreement for the quark mass function, as shown in the same figure.

 As mentioned before in the one-loop analysis, fitting one function alone typically leads to artificially good fits, not representative of the actual quality of the approximation. It is thus desirable to fit the quark mass function and the gluon propagator together, which, as just mentioned, are in tension with respect to the value of the gluon mass parameter. The best fit is presented in Fig.~\ref{Fig:QMandGl}, which still shows a reasonably good agreement, at the 15\% level \cite{Pelaez:2020ups}. Finally, the quark dressing function, shown in Fig.~\ref{Fig:QZ} is badly described as was already the case at one-loop order, see Figs.~\ref{twoloop426} and \ref{twoloop150}. Although the present approximation includes part of the two-loop contributions that are necessary for a proper description of this function, there remain significant two-loop corrections from the quark-gluon vertex corrections, that are not included. The difference between the complete two-loop result shown in Sect.~\ref{cacaboudin} and the one shown in Fig.~\ref{Fig:QZ} gives a measure of the size of such vertex corrections.
 
 We end this section with a comment on the order of magnitude of the expansion parameters used here. For the best fit values, the typical gluon coupling is at most $\lambda_g=g_g^2N_c/(16\pi^2)\approx0.12$ for $N_c=3$, whereas the running quark-gluon coupling reaches $\lambda_q=g_q^2N_c/(16\pi^2)\approx0.68$ \cite{Pelaez:2020ups}. Although these values are slightly smaller than those obtained for heavy quarks in the previous Section, we see that a perturbative treatment of the quark-gluon coupling remains, {\it a posteriori}, questionable. Finally, it is interesting to note that the two expansion parameters that control the RI-loop expansion are roughly of the same order in the case $N_c=3$, namely, $1/N_c\approx0.33$ and $g_g\sqrt{N_c}/(4\pi)\lesssim 0.34$.

\subsection{Hadronic observables}

The RI expansion can also be used for the calculation of physical observables in QCD. The simplest ones are the properties---masses and decay constants---of meson bound states, out of which the pion plays a particular role, being the Goldstone mode associated to the spontaneous breaking of the chiral symmetry. An important ingredient for this calculation using continuum approaches is the quark-antiquark-meson vertex. The latter can be consistently calculated in the RI loop expansion. At leading order, it is given by the infinite series of ladder diagrams with one-(massive)-gluon-exchange rungs and rainbow-resummed quark propagators \cite{Pelaez:2017bhh}. This is not a surprise because this ladder resummation for the quark-antiquark-meson vertex in fact goes along with the rainbow resummation for the quark propagator to comply with the chiral symmetry constraints. The rainbow-ladder approximation is very well-known and is widely used for calculations of meson properties using nonperturbative continuum approaches \cite{Roberts:1994dr,Maris:1999nt,Maris:2003vk,Roberts:2007jh,Sanchis-Alepuz:2015tha}. The  benefit of the RI approach is that, as discussed above, the approximation is controlled by an expansion in terms of actual small parameters of the theory.

The rainbow-ladder equations have been studied for the case of the pion in the RI loop expansion \cite{Serreau:2020clz}. In the chiral limit, the relevant integral equation for the vertex can be greatly simplified and one ends up with a set of coupled one-dimensional integral equations which are easy to solve numerically. The pion decay constant $f_\pi$ can then be systematically computed as a function of the parameters of the Lagrangian, namely, the gauge coupling and the gluon mass. Preliminary results show that there are regions of parameter space which give very good values of $f_\pi$. This is to be expected because, thanks to the chiral Ward identities, the value of $f_\pi$ is, to a large extent, determined by the quark mass function \cite{Pagels:1979hd} and, as described above, the CF model equipped with the RI expansion is clearly able to produce good fits of the latter. 

Interestingly, we mention that the CF model can serve as a precise definition of a gluon mass (in the Landau gauge) which can be assigned a physically measurable value using, {\it e.g.}, the experimental value of $f_\pi$. The constraints on a possible gluon mass in the particle data book \cite{Zyla:2020zbs} refer to an outdated and, in fact, theoretically not precise definition of the gluon mass \cite{Yndurain:1995uq}. The work reported here brings the possibility of a precise, well-defined---necessarily gauge and scale dependent---gluon mass, similar to what is done for the quark masses \cite{Zyla:2020zbs}. For discussions in this direction; see \cite{Roberts:2020hiw}.

\section{Nonzero temperature and density: The confinement-deconfinement transition}\label{sec:temp}
Lattice simulations have established that YM theories undergo a confinement-deconfinement phase
transition at nonzero
temperature \cite{Svetitsky:1985ye,Kaczmarek:2002mc,Lucini:2005vg,Greensite:2012dy,Smith:2013msa}.
The latter is controlled by the spontaneous breaking of a symmetry specific to the nonzero
temperature problem, the center symmetry \cite{Svetitsky:1985ye,Greensite:2011zz,Pisarski:2002ji}. One possible order parameter for the latter is the Polyakov loop $\ell$ \cite{Polyakov:1978vu}, defined as the average of a traced temporal Wilson loop:
\beq
\label{eq:ploop}
 \ell=\frac{1}{N_c}\tr \left< P\exp\left\{ig\int_0^\beta d\tau A_0(\tau,{\bf x})\right\}\right>.
 \eeq
Here, the inverse temperature $\beta\equiv 1/T$ (the Boltzmann constant is set to $k_B=1$) sets the extent of the compact Euclidean time interval over which the fields are defined and the path ordering operator $P$ orders the (matrix-valued) fields from left to right according to the decreasing value of their time argument. The gauge field is periodic in Euclidean time with period $\beta$---and so are the ghost and antighost fields in a gauge-fixed setting---whereas quark fields are antiperiodic. The Polyakov loop is directly related
to the free energy $F_q$ of the system in the presence of a static colour charge as \cite{Polyakov:1978vu,Svetitsky:1985ye}
\beq\label{eq:Fq}
 \ell\propto e^{-\beta F_q}.
\eeq
In particular, a phase with $\ell=0$ implies an infinite free-energy cost for the colour charge, characterising a confined phase. 
Clearly, the latter involves large field configurations with $A_0\sim1/g$, which are not captured at any finite order in perturbation theory around the trivial configuration $A_0=0$. One way to cope with this issue is to expand around a nontrivial background field configuration $\langle A_0\rangle\neq0$, to be determined dynamically. Doing so in the Landau gauge is, however, problematic because the latter explicitly breaks the center symmetry, which clearly plays a key role here. Convenient ways to encode both a nontrivial background and the essential aspects of the center symmetry have been put forward in Ref.~\cite{Braun:2007bx} and, more recently, in Ref.~\cite{VanEgmond:2021mlj}, which uses background-field techniques \cite{Abbott:1980hw,Abbott:1981ke} and, in particular, the background-field generalisation of the Landau gauge, the Landau-DeWitt (LDW) gauge \cite{Weinberg:1996kr}. Similar to the massive extension of the former, discussed so far in this article, the massive extension of the latter or, in other words, the LDW version of the CF model has been worked out in Ref.~\cite{Reinosa:2014ooa} and used as a starting point for a perturbative analysis of the nonzero temperature physics, in the presence of a nontrivial Polyakov loop.\footnote{We mention that a different approach to nonzero temperature physics has been pursued in the context of the screened perturbation theory, working directly in the Landau gauge with vanishing background \cite{Comitini:2017zfp}.}

\subsection{The Landau-DeWitt gauge}
The background field approach \cite{Weinberg:1996kr} introduces an {\it a priori} arbitrary background field configuration $\bar A_\mu$ through a modified gauge-fixing condition. In terms of $a_\mu=A_\mu-\bar A_\mu$, the LDW gauge condition reads
\beq
\label{eq:LdW}
 \bar D_\mu a_\mu^a=0\,,
\eeq
where $\bar D_\mu\varphi^a\equiv\partial_\mu \varphi^a+g f^{abc}\bar A_\mu^b\varphi^c$. One can construct the corresponding FP Lagrangian using standard techniques. One important property of the resulting gauge-fixed theory is a formal gauge invariance with respect to gauge transformations of the background field. The simplest background-field generalisation of the CF action which respects this essential property is \cite{Reinosa:2014ooa}
\beq
\label{eq_CF}
 S_{\bar A}=\!\int_x\left\{\frac 14 F_{\mu\nu}^aF_{\mu\nu}^a+\frac{m^2}{2}a_\mu^aa_\mu^a+\bar D_\mu\bar c^aD_\mu c^a+ih^a\bar D_\mu a_\mu^a\right\},
\eeq
which is clearly invariant under the  transformation $\bar A_\mu^U\to\bar A_\mu^U=U\bar A_\mu U^{-1}+\frac i {g}U\partial_\mu U^{-1}$, $\varphi\to U\varphi U^{-1}$, with $\varphi\equiv(a_\mu,c,\bar c,h)$ and where $U$ is an element of the gauge group. This linear symmetry is inherited by the effective action,
\beq
\label{eq_ginv}
 \Gamma_{\bar A}[\varphi]=\Gamma_{\bar A^U}[U\varphi U^{-1}]\,,
\eeq
provided the transformation $U$ preserves the periodic boundary conditions of the fields.
Such transformations, which form a group denoted ${\cal G}$, do not need to be periodic themselves, however, but only periodic up to any element of the center of the gauge group, $e^{i2\pi k/N_c}\mathbb{1}$ ({\small $k=0,\dots, N_c-1$}) in the case of SU($N_c$). Their relevance is that they act by multiplying the Polyakov loop $\ell$ by the corresponding center phase. In particular, this relates the confining phase, where $\ell$ vanishes, to the phase where the symmetry group ${\cal G}/{\cal G}_0$ is explicitly realised, where the subgroup ${\cal G}_0$ of periodic transformations ($k=0$) needs to be quotiented away because it has no impact on (and is therefore not probed by) the Polyakov loop. The quotient group ${\cal G}/{\cal G}_0$ is isomorphic to the center of the gauge group and is known as the center symmetry group \cite{Pisarski:2002ji}.

A convenient way to study the spontaneous breaking of ${\cal G}/{\cal G}_0$, and in turn
the deconfinement
transition, is from the
functional
\beq
\label{eq_functilde}
 \tilde\Gamma[\bar A]=\Gamma_{\bar A}[\varphi=0]
\eeq
 which is also invariant under the transformations $U\in {\cal G}$. A given state of the system is represented by a minimum of $\tilde\Gamma[\bar A]$ denoted $\bar A_{\rm min}$, or by any other minimum obtained from it using a transformation $U_0\in {\cal G}_0$. In other words, a physical state corresponds to a ${\cal G}_0$-orbit of minima \cite{Reinosa:2015gxn,Reinosa:2020mnx}. The center-symmetric states are those ${\cal G}_0$-orbits that are invariant under the action of ${\cal G}/{\cal G}_0$  and the deconfinement transition occurs when the ${\cal G}_0$-orbit of minima of $\tilde\Gamma[\bar A]$ moves away from its center-symmetric configuration.

\begin{figure}[t]
  \begin{center}
    \includegraphics[width=.3\linewidth]{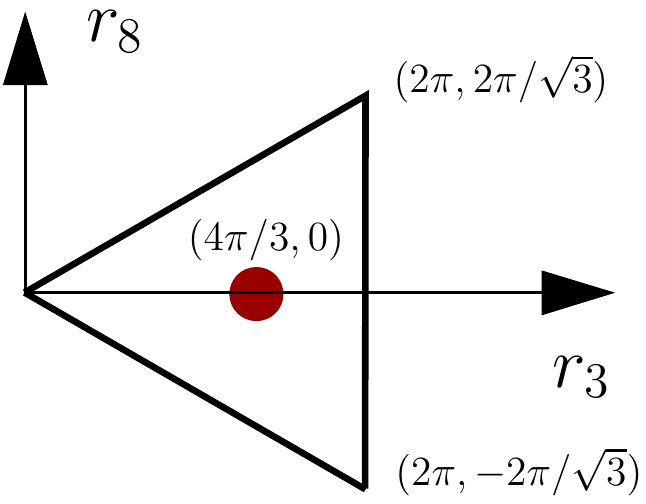}
\caption{Weil chamber for SU(3). The red point corresponds to the center-symmetric background field configuration, see the text for details.}
\label{fig:SU3_center}
\end{center}
\end{figure}

The above considerations are greatly simplified if one restricts to constant temporal background fields along the diagonal or commuting part of the algebra $\bar A_\mu(x)=\delta_{\mu0}\bar A_0^j\,t^j$, with $[t^j,t^{j'}]=0$. In this case the functional $\tilde\Gamma[\bar A]$ can be traded for the potential 
\beq
\label{eq:poteff}
 V(r)=\frac{\tilde\Gamma[\bar A_0]}{\beta \Omega}-V_{\rm vac}\,,
\eeq
where $\Omega$ is the spatial volume, $V_{\rm vac}$ the vacuum (zero temperature) contribution, and $r\in\mathds{R}^{N_c-1}$ is the vector of components $r^j=\beta g\bar A_0^j$. The transformations of ${\cal G}_0$ divide $\mathds{R}^{N_c-1}$ into equivalent cells known as Weyl chambers and whose points can be seen as representatives of the various states (${\cal G}_0$-orbits) of the system. A center transformation corresponds to an isometry of any of these cells whose fixed points are the invariant states under the considered transformation.\footnote{The present considerations apply to other symmetries, such as charge conjugation \cite{Reinosa:2015gxn}.} In the SU($2$) case, $r\in\mathds{R}$ and the Weyl chambers are the segments $[2\pi n, 2\pi(n+1)]$ each of which contains a center-symmetric point at its center. In the case of SU($3$), the Weyl chambers form a paving of $\mathds{R}^2$ by equilateral triangles. One representative is the fundamental triangle of edges $(0,0)$, $(2\pi,-2\pi/\sqrt{3})$ and $(2\pi,2\pi/\sqrt{3})$, see Fig.~\ref{fig:SU3_center}. The center of any such triangle corresponds to the center-symmetric state (for instance $(4\pi/3,0)$ in the Weyl chamber shown in Fig.~\ref{fig:SU3_center}).

The analysis of the confinement-deconfinement transition is then achieved by minimising the
potential $V(r)$ over a given Weyl chamber and monitoring, as a function of the temperature, whether the minimum $r_{\rm min}$ is located at the center-symmetric points. In this sense, the corresponding background $\bar A_{\rm min}$ plays the role of an order parameter,
equivalent, in this gauge,\footnote{The Polyakov loop remains more fundamental in a certain sense, as it is a gauge-invariant order parameter.} to the Polyakov loop \cite{Braun:2007bx,Reinosa:2015gxn,Reinosa:2020mnx}. From the minimum of the potential one can also access the thermodynamic pressure (and therefore any other thermodynamical observable) as
\beq
\label{eq:pressure}
 p=-V(r_{\rm min}) + p_{\rm vac},
\eeq
whereas the Polyakov loop is obtained from a direct evaluation of the average (\ref{eq:ploop}) with the restriction that $\langle A\rangle=\bar A_{\rm min}$. 

\subsection{The background-field effective potential in perturbation theory}\label{sec:12}
The potential \eqref{eq:poteff} has been computed at one-loop order in the (background-field-extended) CF model for a large variety of gauge groups \cite{Reinosa:2014ooa,Reinosa:2015gxn}. Two-loop corrections have also been computed for the SU($2$) and SU($3$) groups \cite{Reinosa:2014zta,Reinosa:2015gxn}. We briefly describe the salient aspects of these calculations and we review the essential results here.

Calculations in the LDW gauge with a constant temporal background as specified above are greatly simplified if one switches from the usual Cartesian colour bases $\{t^a\}$ to the Cartan-Weyl bases $\{t^\kappa\}$ \cite{Reinosa:2015gxn}. The labels $\kappa$ are vectors that gather the adjoint colour charges $\kappa^j$ of each colour mode such that $[t^j,t^\kappa]=\kappa^j t^\kappa$, where the generators $t^j$ span the commuting part of the algebra, also known as the Cartan subalgebra. In the SU($2$) case for instance, a Cartan-Weyl basis is $\{t^0,t^+,t^-\}$, where $t^0=\sigma_3/2$ and $t^\pm=(\sigma_1\pm i\sigma_2)/(2\sqrt2)$ are the well-known raising and lowering operators, with $\sigma_i$ the Pauli matrices.
  A convenient  property of the Cartan-Weyl bases is that they allow for a one-to-one correspondence between the Feynman rules (and, consequently, of many calculation steps) with and without the background field. In particular, the only effect of the background is to shift the Euclidean four-momentum $p$ of the propagator lines associated to a colour mode $\kappa$ as $p_\mu\to p^\kappa_\mu=p_\mu+T(r\cdot\kappa)\delta_{\mu0}$. This generalised momentum is conserved at vertices owing to momentum and colour conservation.

 The calculation of the background-field potential $V(r)$ at one-loop order is straightforward and yields \cite{Reinosa:2014ooa}
\begin{equation}\label{eq:V_1loop}
 V_{\mbox{\scriptsize 1loop}}(r)=\frac{T}{2\pi^2}\sum_\kappa\int_0^\infty dq\,q^2\Big[3\ln\big(1-e^{-\beta\varepsilon_q+ir\cdot\kappa}\big)-\ln\big(1-e^{-\beta q+ir\cdot\kappa}\big)\Big]\,,
\end{equation}
with $\varepsilon_q=\sqrt{q^2+m^2}$. By setting $r$ to $0$, one recognises the free-energy density of a gas of free massive gluons and massless ghosts. The factor of $3$ relates to the fact that there are three massive transverse gluonic modes. The longitudinal gluonic mode is massless and cancels with one of the two ghost degrees of freedom. The background lifts the degeneracy between the various colour modes and shifts the dispersion relations by an imaginary amount $i(r\cdot \kappa)T$ that can be interpreted as an imaginary chemical potential for the colour charge \cite{Fukushima:2017csk}. 

The essential features of the one-loop potential can be unveiled by looking at the two asymptotic regimes $T\gg m$ and $T\ll m$. In the high temperature limit, all modes can be considered approximatively massless and, for each colour state, the ghost contribution in \eqref{eq:V_1loop} cancels against one gluon mode, leaving only the two ``physical'' polarisations of the massless gluons. This results in the known Weiss deconfining potential \cite{Weiss:1980rj,Pisarski:1980md}, which displays maxima at the confining points of the Weyl chambers. At low temperatures, instead, the massive gluon modes in \eqref{eq:V_1loop} are exponentially suppressed and the potential is dominated by the (massless) ghost contribution. Because the latter contributes negatively, this leads to an inverted Weiss potential \cite{Braun:2007bx}, with minima at the confining points of the Weyl chambers. This is another example of the phenomenon of ghost dominance at IR (here, temperature) scales, mentioned in Sec.~\ref{sec:YM3}. As in the case of correlation functions, this phenomenon is not restricted to the CF approach but applies to a wide class of continuum approaches using background field techniques \cite{Braun:2007bx,Fischer:2014vxa,Quandt:2017poi}. For pure YM theories, one finds an actual phase transition between the high and low temperature phases.

\begin{figure}[t]  
\begin{center}
\includegraphics[width=8.3cm]{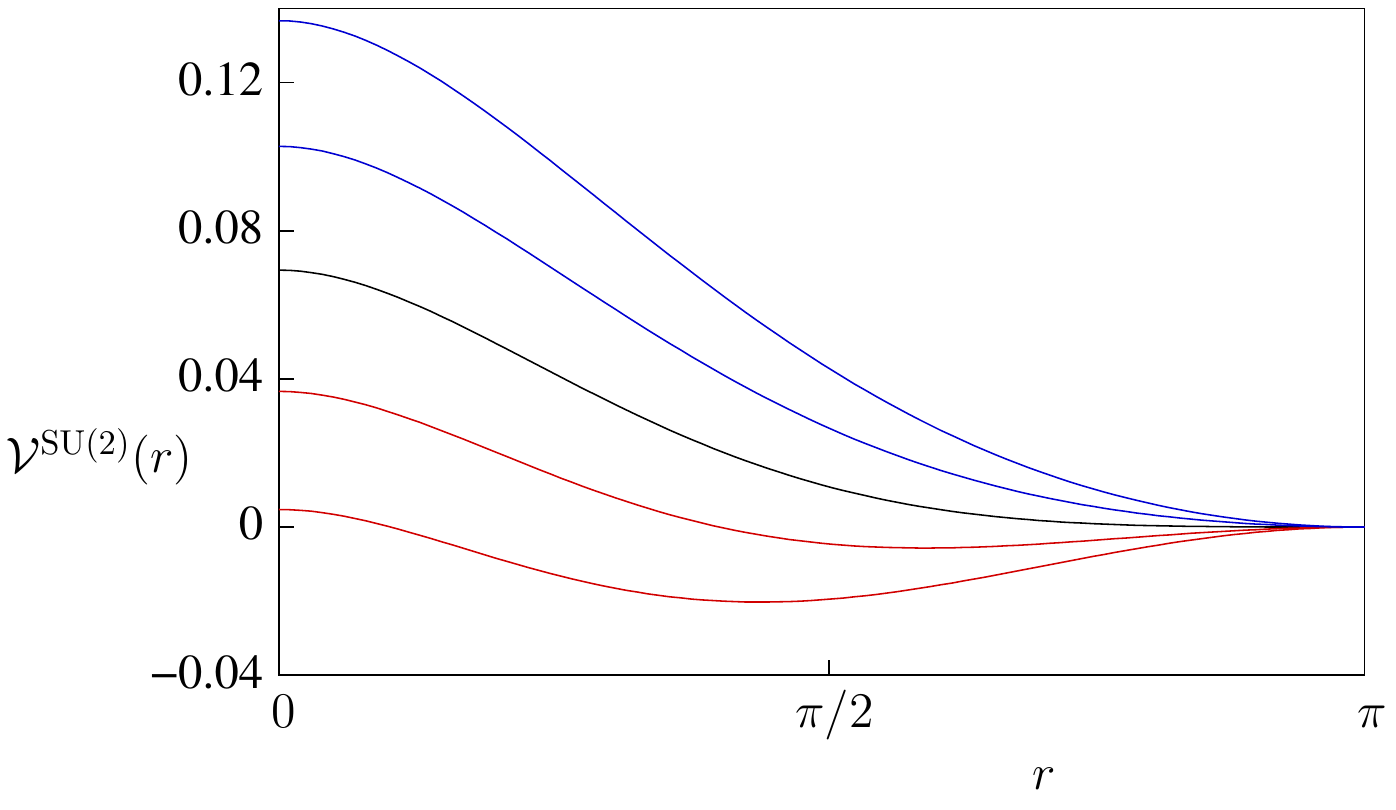}\qquad
\includegraphics[width=8.3cm]{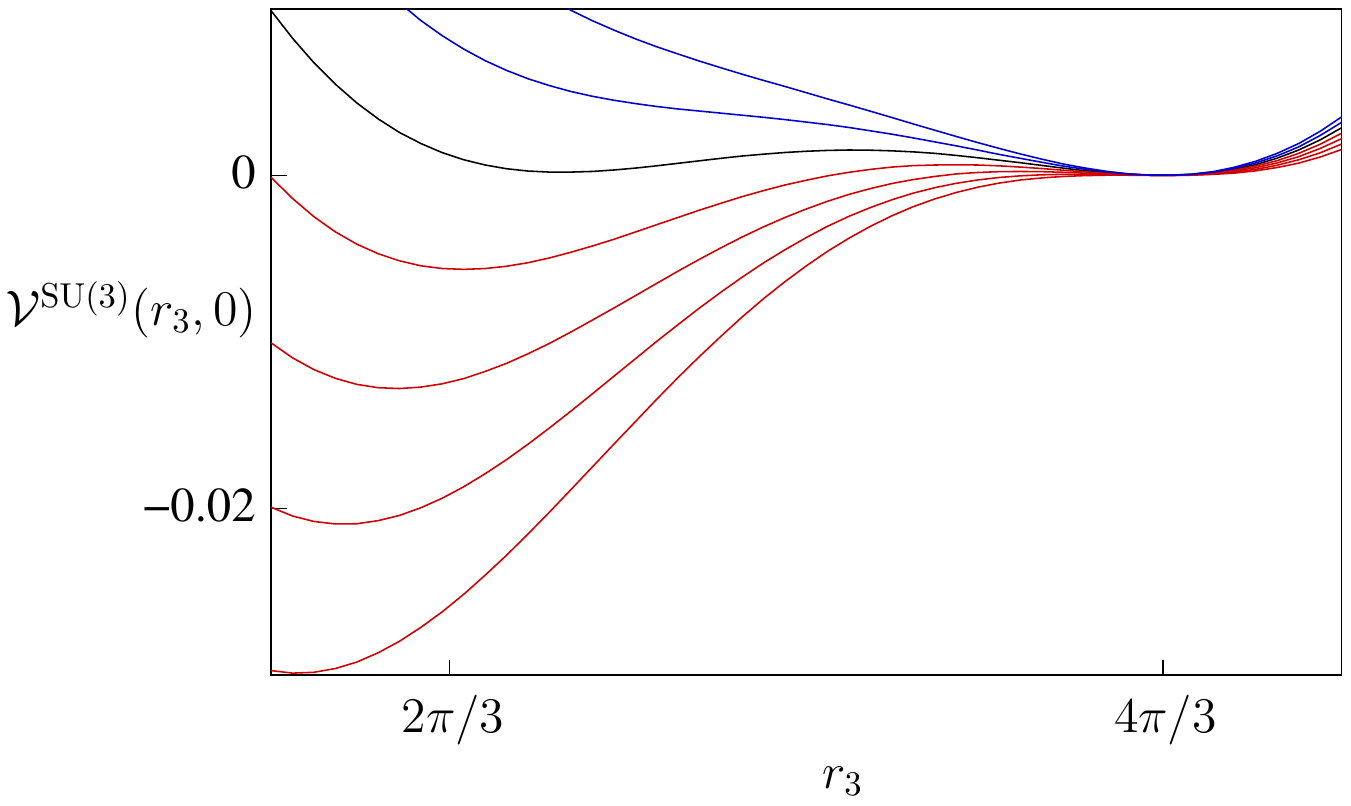}\qquad
\caption{\label{fig:SU3pot} The dimensionless background field potentials ${\cal V}(r)=V(r)/T^4$ for the SU($2$) and the SU($3$) theories at one-loop order in the CF model, for temperatures $T=T_{c}$ (black), $T<T_{c}$ (blue), and $T>T_{c}$ (red). The abscisses represent the charge conjugation invariant directions in the corresponding Weyl chambers, along which the confinement-deconfinement transition takes place. The confining point is located at $r=\pi$ for SU($2$) and at  $r_3=4\pi/3$ for SU($3$). Figures from Ref.~\cite{Reinosa:2014ooa}.}\label{fig:pots}
\end{center}
\end{figure}

Fig.~\ref{fig:pots} shows the one-loop background-field potentials for the SU($2$) and the SU($3$) theories in $d=4$ in the relevant directions---along which the transition takes place---in the respective Weyl chambers. One finds a continuous transition in the SU($2$) case and a first order transition for SU($3$), in agreement with lattice simulations \cite{Svetitsky:1985ye} and other approaches \cite{Braun:2007bx,Quandt:2017poi}.
We find that the transitions for SU($2$) and SU($3$) occur,
respectively, at $T_c/m\simeq 0.336$ and $T_c/m\simeq 0.364$. Using
the values fitted against the lattice propagators at zero
temperature,\footnote{The evaluation of the potential is done by using the same zero-temperature renormalisation scheme as that used for the evaluation of the zero temperature propagators. More precisely, at one-loop order, the background dependent part of the potential is finite and the mass can be considered as the bare one, fixed by fitting tree-level expressions for the propagators to lattice data. At two-loop order, the background dependent part of the potential diverges and its renormalisation requires rescaling the mass parameter using one-loop renormalisation factors that should be taken equal to those used in the one-loop propagators.} this translates to the transition temperatures reported in Tab.~\ref{tab:Tc}, which are in remarkably good agreement with the lattice values \cite{Lucini:2012gg} given the simplicity of the one-loop approximation. The inclusion of two-loop corrections, again, with parameters adjusted to reproduce the vacuum propagators at the relevant order, yields the values summarised in Tab.~\ref{tab:Tc}. The improvement is clear.

\begin{table}[h]
$\begin{tabular}{|c||c||c||c||c||c|}
\hline
$T_{\rm c}$ (MeV) & ~~~~  lattice~~~~   & one-loop CF$^{(s)}$ & two-loop CF$^{(s)}$ & one-loop CF$^{(c)}$ \\
\hline\hline
SU(2) & 295 & 237 & 284 & 265\\
SU(3) & 270 & 185 & 254 & 267\\
\hline
\end{tabular}$
\caption{Estimates for the confinement-deconfinement transition temperatures at one- and two-loop orders within the CF model with either the self-consistent (CF$^{(s)}$) \cite{Reinosa:2014ooa,Reinosa:2014zta,Reinosa:2015gxn} or the center-symmetric (CF$^{(c)}$) \cite{VanEgmond:2021mlj} background field approaches---see text---and their comparison to the corresponding lattice results \cite{Lucini:2012gg}.}\label{tab:Tc}
\end{table}

\begin{figure}[t]
    \centering
    \includegraphics[width=.46\linewidth]{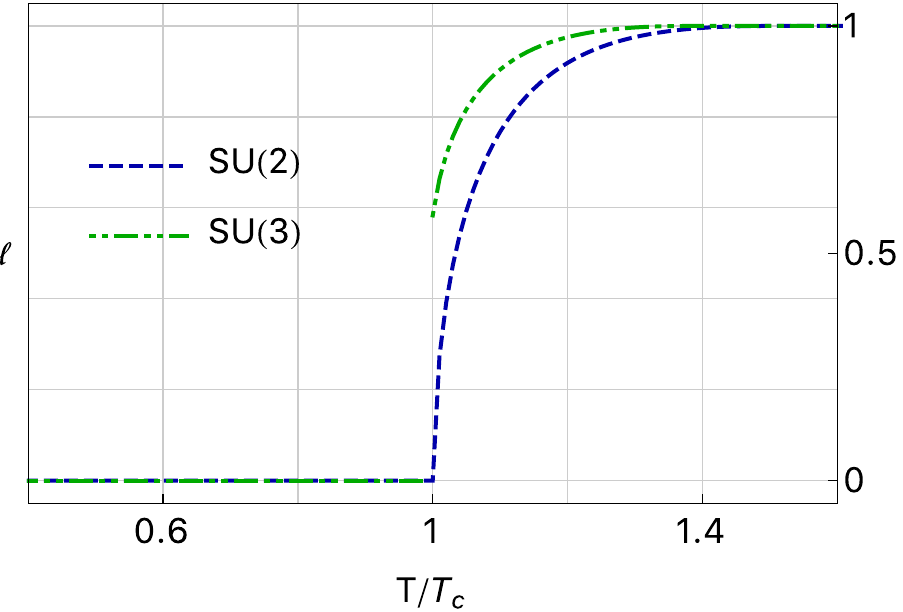}\qquad\quad
    \includegraphics[width=.45\linewidth]{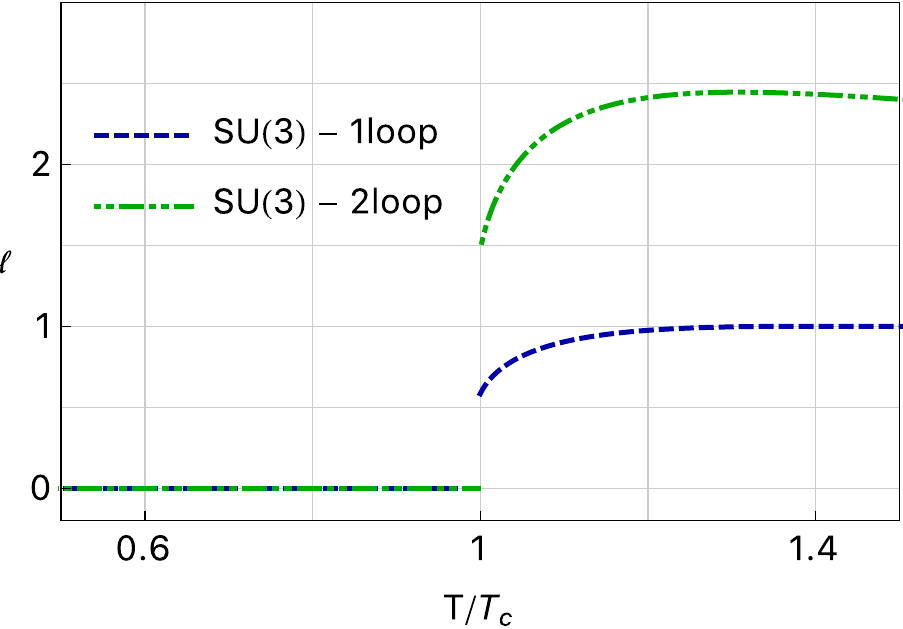}
    \caption{The Polyakov loop \eqref{eq:ploop} as a function of the temperature normalised to the transition temperature for the SU($2$) and the SU($3$) theories in the (background-extended) CF model at one-loop order (left) \cite{Reinosa:2014ooa} and including the two-loop corrections for the SU($3$) case (right) \cite{Reinosa:2015gxn}. Figures from Ref.~\cite{Reinosa:2020mnx}.}
    \label{fig:poly_1loop}
\end{figure}

The corresponding Polyakov loops, at the same order of approximation, are shown in Fig.~\ref{fig:poly_1loop}. As was also pointed out in Ref.~\cite{Dumitru:2012fw} in the context of matrix model calculations, the rise of the Polyakov loop from its value right above the transition temperature to its maximum value is too fast as compared to lattice results. This compromises any direct comparison of this quantity to Monte-Carlo simulations. We shall come back to this below.

\subsection{The issue of massless unphysical modes in the confined phase}

We have seen that the ghost dominance at low temperatures plays a
pivotal role in the confinement-deconfinement mechanism described above. Although this mechanism is quite general and goes in fact beyond the particular case of the CF model considered in this review, it poses certain challenges that we now describe. 

First, the fact that ghost degrees of freedom dominate in the low
temperature phase may lead to inconsistent
thermodynamics.\footnote{More generally, this is related to the
  question of the proper identification of the physical space of the
  model, discussed in Sec.~\ref{sec:unitarity}, over which thermal
  averages are to be taken. In the FP theory, the nilpotent BRST
  symmetry guarantees that states with negative (or null) norm do not
  contribute to thermal averages and thus to thermodynamic observables
  (except through loop effects). For instance, at one-loop order, the
  ghost contribution cancels that of the two ``unphysical'' gluon
  modes.}  For instance, in the absence of the background, the ghost
contribution results in a negative thermal pressure or a negative
entropy in the low temperature limit
\cite{Comitini:2017zfp,Sasaki:2012bi}. Fortunately, the presence of
the nontrivial background cures this pathological behaviour
\cite{Reinosa:2015gxn,Quandt:2017poi}. The reason is that, in the
$T\to0$ regime, the confining gluonic background operates a
transmutation of the ghost thermal distribution functions, such that
the net contribution to the pressure or entropy density remains
positive. As an illustration, consider the SU($2$) pressure which, at
low temperature, can be written
\begin{equation}\label{eq:pt0}
 p_{\rm th}\sim\frac{1}{6\pi^2}\int_0^\infty dq\,q^3\,\big[-n_{q-i\pi T}-n_q-n_{q+i\pi T}\big]\,,
\end{equation}
where $n_q\equiv 1/(e^{\beta q}-1)$ is the Bose-Einstein distribution function. The bracket contains the three colour mode contributions, corresponding to $\kappa\in\{-,0,+\}$. The neutral mode $\kappa=0$ is blind to the background and contributes negatively to the thermal pressure, as expected from a ghost degree of freedom. Were it not for the imaginary shift $\pm i\pi T$ of their energies, the two other modes would give a similar negative contribution.\footnote{The $T\to0$ expression in the absence of the background is given by Eq.~\eqref{eq:pt0} with $q\pm i\pi T$ replaced by $q$. We recover here that the Landau gauge CF model predicts a negative pressure $p=-\pi^2T^4/30$ at low temperatures \cite{Reinosa:2014zta,Comitini:2017zfp}.} However, because of the presence of the confining background, these modes contribute instead with $-n_{q\pm i\pi T}=1/(e^{\beta q}+1)$, that is, a positive Fermi-Dirac distribution, leading eventually to a positive thermal pressure and a positive entropy at low temperature \cite{Reinosa:2014zta}. These considerations extend to SU($3$) \cite{Reinosa:2015gxn}.  

\begin{figure}[t]
    \centering
        \includegraphics[width=.53\linewidth]{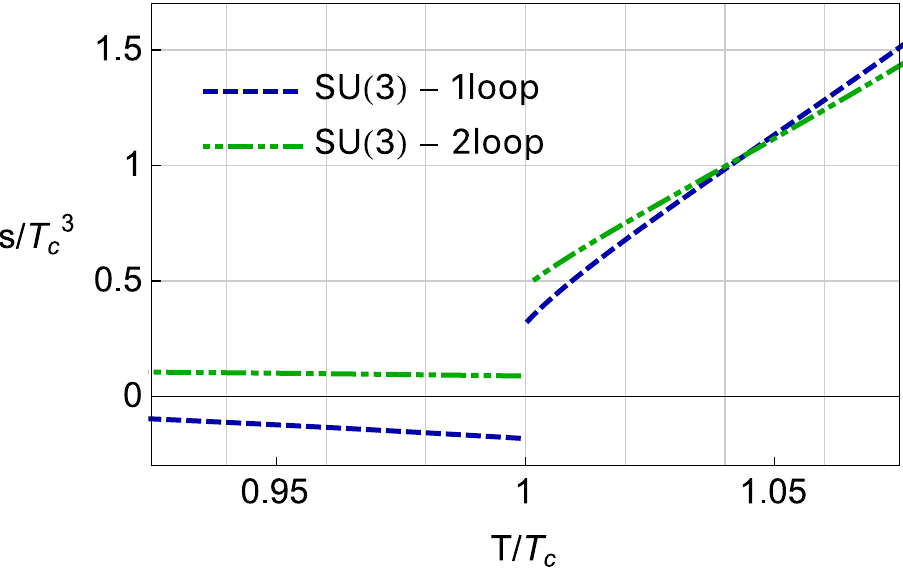}
    \caption{The SU($3$) entropy density in the vicinity of the phase transition at one- and two-loop orders in the LDW extension of the CF model \cite{Reinosa:2015gxn,Reinosa:2020mnx}.}
    \label{fig:s}
\end{figure}

The transmutation mechanism described here is also visible (but with
the opposite effect) as one approaches the transition from below. In
this limit, the gluon degrees of freedom are not exponentially
suppressed, while the background remains confining, turning some of
their positive distribution functions into negative ones, which tend
to bring the thermal pressure or the entropy density down to a
slightly negative value at one-loop order. This feature is clearly
visible in the thermodynamical observables as illustrated in
Fig.~\ref{fig:s} for the entropy density. It has been shown, however,
that the two-loop result corrects this unphysical feature
\cite{Reinosa:2014zta,Reinosa:2015gxn}.

The low-temperature phase is plagued by yet another major problem. Namely, the dominant ghost degrees of freedom, be they surrounded or not by a confining gluonic background, remain massless. This leads to power law behaviour of the thermodynamical observables as $T\to 0$, at odds with the observations on the lattice. Again, it is worth emphasising that these issues are not restricted to the perturbative CF model and actually encompass all current continuum approaches to YM/QCD  \cite{Quandt:2017poi,Sasaki:2012bi,Canfora:2015yia}. This issue points to the inability (to date) of continuum approaches to provide a fully consistent picture of confinement.
 
\subsection{Center-symmetric background field approach}
\label{sec:csbckd}

The results described above rely on the use of the background functional \eqref{eq_functilde} [or the corresponding potential \eqref{eq:poteff}], which involve self-consistent background fields defined as $\bar A=\langle A\rangle_{\bar A}$. It is to be emphasised that this quantity is not, strictly speaking, an effective action in the sense that it is not a standard Legendre transform and that it is a functional of the background field (which is a gauge-fixing device). Because of this, such a self-consistent background field approach suffers from various technical difficulties---not to be described here, see {\it e.g.} Ref.~\cite{Reinosa:2015gxn,Reinosa:2020mnx,VanEgmond:2021mlj}---when approximations are involved and for its possible implementation in lattice simulations.

An alternative approach has been recently proposed, that avoids these difficulties \cite{Reinosa:2020mnx,VanEgmond:2021mlj}. It is based on using a fixed background field $\bar A_c$ that is invariant under the action of ${\cal G}/{\cal G}_0$. The corresponding action $\Gamma_{\bar A_c}[A]$ is a genuine gauge-fixed effective action which properly encodes the center symmetry and from which one directly obtains the gluon correlation functions. Also, it can be implemented in lattice simulations with techniques currently used for the Landau gauge. It has been evaluated at one-loop order in the CF model for constant, center-symmetric  background field configurations in the temporal direction and in the Cartan subalgebra of the gauge group \cite{VanEgmond:2021mlj}. This yields the correct phase structure for the SU($2$) and SU($3$) YM theories with transition temperatures reported in Tab. \ref{tab:Tc}, in remarkable agreement with lattice results, much better than the corresponding one-loop results in the previous self-consistent background field approach. 

\begin{figure}[t]
    \centering
        \includegraphics[width=.53\linewidth]{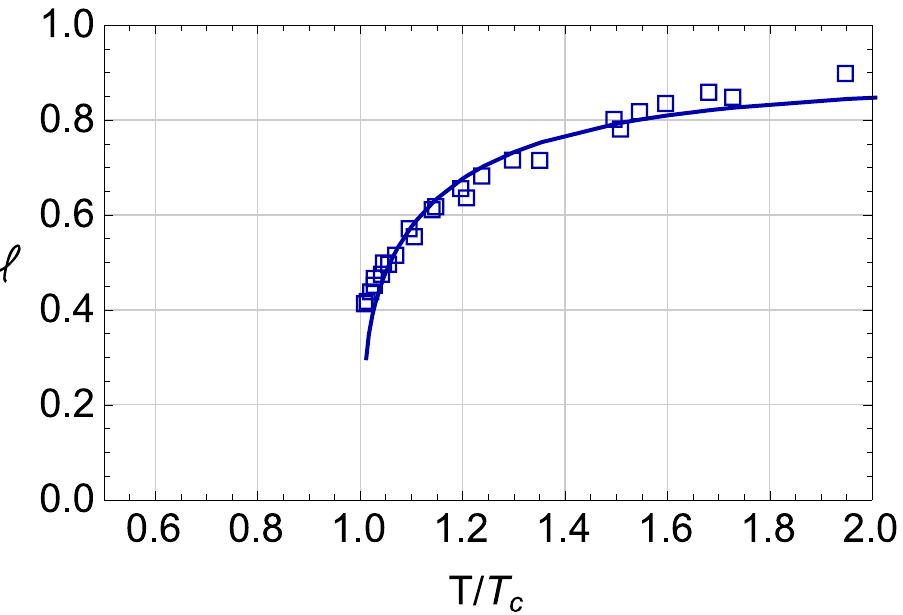}
    \caption{The Polyakov loop for the SU($3$) YM theory at one-loop order from the center-symmetric effective action proposal of Ref.~\cite{VanEgmond:2021mlj} as compared to the lattice data of Refs.~\cite{Gupta:2007ax,Lo:2013hla}.}
    \label{fig_comp}
\end{figure}

Another interesting improvement concerns the temperature dependence of
the Polyakov loop. In particular, it shows a moderate rise in the deconfined phase, as compared to the self-consistent background field approach, which compares well with lattice results \cite{Gupta:2007ax,Lo:2013hla}, see Fig.~\ref{fig_comp}. Here, it is important to stress that the lattice data correspond to the renormalised Polyakov loop $\ell_r(T)$, related to the bare one computed here as $\ell_r(T)=Z_\ell\,\ell(T)$, where the renormalisation factor $Z_\ell$ is a function of the coupling. At first sight, it might seem sufficient to simply normalise the perturbative results to the lattice data at a given temperature. Doing so, a rather good fit is obtained in the range $[0,2T_c]$. The agreement deteriorates for larger temperatures, where important effects not taken into account in this simple one-loop calculation, such as the resummation of hard thermal loops \cite{Haque:2014rua} or the RG running \cite{Herbst:2015ona,Kneur:2021feo}, become important.

\subsection{Dynamical quarks}

The success of the CF model in describing the finite temperature
confinement-deconfinement transition in the pure YM case naturally leads
 one to investigate how well it can capture the phase structure in the presence of quarks. As in the vacuum case, two regimes must be distinguished, depending on the values of the quark masses.
For large quark masses, the quark-gluon coupling does not differ substantially from the one in the pure gauge sector and one can rely on perturbation theory. Admittedly, this regime does not correspond to the physical QCD case. However, it possesses a rich phase structure which has been studied using various approaches in the literature. It should also be mentioned that, in this range of masses, the phase transition is still akin to a description in terms of the Polyakov loop or the self-consistent background. 

\begin{figure}[t]
    \centering
    \includegraphics[width=.37\linewidth]{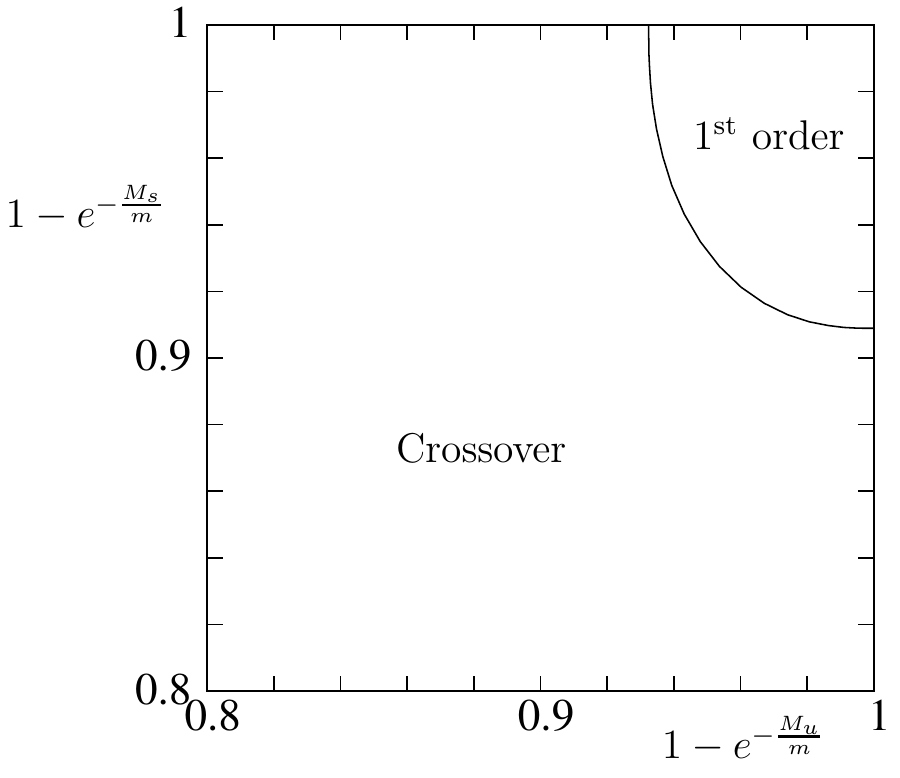}\qquad\qquad\includegraphics[width=.45\linewidth]{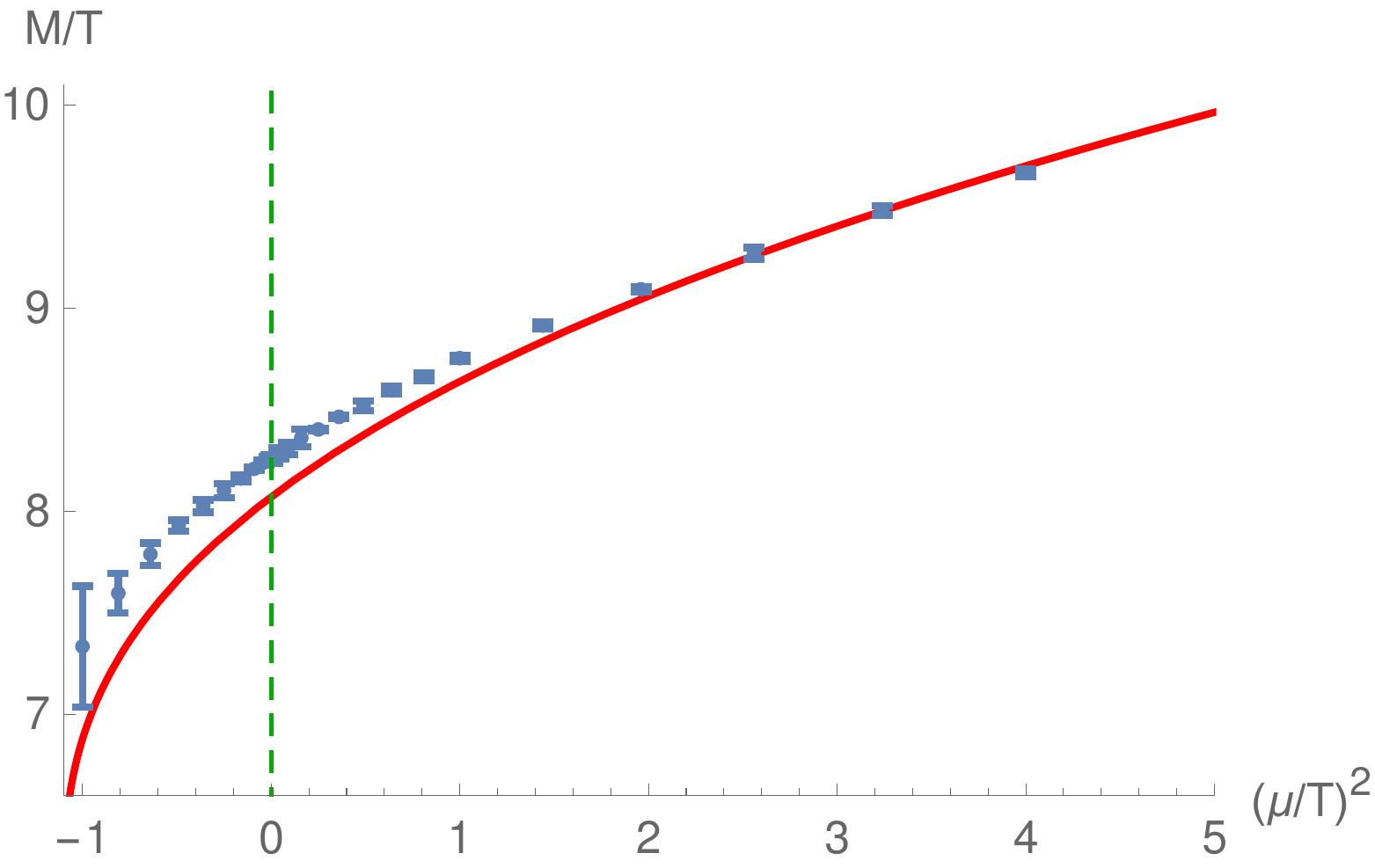}
    \caption{Left: The critical phase boundary in the heavy quark
      corner of the Columbia plot, namely, the space of quark masses
      for the case with $2+1$ degenerate quark normlaliss with masses
      $M_u=M_d$ and $M_s$. The curve shown here is determined from the
      CF model at one-loop order \cite{Reinosa:2015oua}. The lines
      $M_u\to\infty$, $M_s\to\infty$, and $M_u=M_s$ correspond to the
      cases with, respectively, $N_f=1$, $2$, and $3$ degenerate
      flavours. Right: The critical ratio $R_{N_f}$ as a function of the chemical potential for $N_f=3$. The CF model prediction at one-loop order \cite{Reinosa:2015oua} (plain line) is compared to the lattice data (dots) extracted from the results of Ref.~\cite{Fromm:2011qi} for $R_1$ using the (lattice) relation $R_{N_f}=R_1+\ln N_f$, valid in the heavy-quark limit.}
    \label{fig:quarkT}
\end{figure}

At vanishing quark chemical potential, there exists a critical phase boundary in the space of quark masses, separating a region of first-order phase transitions for large quark masses (including the YM case) from a crossover region for lower quark masses (including the physical QCD point), see the left panel of Fig.~\ref{fig:quarkT}. For $N_f$ degenerate flavours, the boundary is characterised by the ratios $R_{N_f}\equiv M_c(N_f)/T_c(N_f)$, with $M_c(N_f)$ and $T_c(N_f)$ the critical quark mass and critical temperature respectively. Those have been computed at one- and two-loop orders in the CF model within the self-consistent background field approach described previously. The one-loop results, shown in the left panel of Tab.~\ref{tab:RNf}, are in remarkable agreement with lattice estimates. It is important to notice that, at one-loop order, these dimensionless ratios do not involve the gluon mass parameter or the gauge coupling and are, therefore, a parameter-free prediction of the model.

\begin{table}[h!]
  \centering
	\begin{tabular}{|c || c c c |} 
		\hline
		$R_{N_f}(\mu=0)$ & $N_f=1$ & $N_f=2$  &$N_f=3$  \\ [0.5ex] 
		\hline\hline
		one-loop CF & 6.74 & 7.59 & 8.07\\
		lattice & 7.23 & 7.92 &  8.33\\
		\hline
	\end{tabular} \qquad
		\begin{tabular}{|c || c c c || c |} 
		\hline
		$R_{N_f}^{\rm tric}$ & $N_f=1$ & $N_f=2$  & $N_f=3$ & $K_{3}$ \\ [0.5ex] 
		\hline\hline
		one-loop CF & 4.74 & 5.63 & 6.15 & 1.85\\
		lattice & 5.56 & 6.25 &  6.66 & 1.55\\
		\hline
	\end{tabular}
	\caption{The dimensionless ratio $R_{N_f}=M_c/T_c$ on the critical line of the Columbia plot for $\smash{N_f=1}$, $2$, and $3$ degenerate flavours at zero chemical potential (left) and at the tricritical point $\mu/T=i\pi/3$ (right). In this case we also show the value of the coefficient $K_{N_f=3}$ [see Eq. \eqref{eq:tric_scaling}], whose lattice value appears rather insensitive to the value of $N_f$. The one-loop results in the CF model \cite{Reinosa:2015oua} are compared to the lattice values \cite{Fromm:2011qi}. [Ref.~\cite{Fromm:2011qi} only quotes the value of $K_1$. The $N_f$ dependance of $R_{N_f}$ is governed by the scaling law (3.6) of that reference, valid in the heavy-quark limit, which implies that $R_{N_f}=R_1+\ln N_f$ and, in turn, that $K_{N_f}=K_1$.] We stress that the one-loop CF results are independent of the parameters of the model.}
	\label{tab:RNf}
\end{table}

We mention that for any model of the YM sector, such as the CF model considered here, the ratios $R_{N_f}$ for different $N_f$ at one loop are related by the universal relation \cite{Kashiwa:2012wa,Maelger:2018vow}
\beq
N_f R_{N_f}^2 K_2(R_{N_f})=N'_f R_{N'_f}^2 K_2(R_{N'_f})\,,
\eeq
with $K_2$ the modified Bessel function of the second kind.  Again, two-loop corrections have been computed \cite{Maelger:2017amh}. Although small, their general tendency is to approach the lattice results. As explained in this reference, the ratios $R_{N_f}$ cannot be directly compared with the lattice data beyond one loop because of the different meanings of the (regularisation and renormalisation dependent) quark masses. One way to (approximately) cope with this issue is to consider the ratios $Y_{N_f}=(R_{N_f}-R_1)/(R_2-R_1)$, which compare pretty well with the lattice result, known for $N_f=3$ \cite{Maelger:2017amh,Maelger:2018vow}.

The phase structure of the theory has also been investigated in the presence of a quark chemical potential \cite{Reinosa:2015oua,Maelger:2017amh,Maelger:2018vow}. For SU($3$), this requires considering backgrounds with two nonvanishing components $r_3$ and $r_8$, a nonzero $r_8$ component being dictated by the breaking of charge conjugation invariance due the presence of a chemical potential. For imaginary values of the chemical potential, one retrieves the Roberge-Weiss (RW) transition \cite{Roberge:1986mm}, characterised by a first order jump of the phase of the Polyakov loop at $\mu/T=i\pi/3$ (and large enough temperatures). The RW transition is not disconnected from the phase boundary at $\mu=0$. In fact, as the quark masses are decreased below their critical values, the critical boundary enters the imaginary $\mu$ region. As it reaches the particular value $\mu/T=i\pi/3$ where the RW transition takes place, one finds a tricritical point characterised by the scaling law \cite{deForcrand:2010he} 
\beq\label{eq:tric_scaling}
R_{N_f}=R_{N_f}^{\rm tric}+K_{N_f}\left[\left(\frac{\pi}{3}\right)^2+\left(\frac{\mu}{T}\right)^2\right]^{2/5}\,,
\eeq
for $\mu/T\to i\pi/3$. The (parameter-free) one-loop estimates of $R_{N_f}^{\rm tric}\equiv R_{N_f}(\mu/T=i\pi/3)$ and $K_{N_f}$ in the CF model \cite{Reinosa:2015oua} are, again, in fairly good agreement with the lattice results \cite{Fromm:2011qi}; see Tab.~\ref{tab:RNf} and the right panel of Fig.~\ref{fig:quarkT}. Also, as before, two-loop corrections tend to improve these values \cite{Maelger:2017amh}. One also finds that the scaling law (\ref{eq:tric_scaling}) extrapolates deep into the real chemical potential region $\mu^2>0$, as also seen in nonperturbative continuum approaches. Let us mention that a proper treatment of real chemical potentials requires the background $r_8$ to be chosen purely imaginary. This surprising result relates to the QCD sign problem: for a real chemical potential, the QCD action is not real and there is a priori no reason to find real self-consistent backgrounds---{\it i.e.}, that solve the equation $\bar A=\langle A\rangle_{\bar A}$. On the other hand, it has been shown \cite{Reinosa:2015oua} that backgrounds of the form $(r_3,r_8)\in\mathds{R}\times i\mathds{R}$ are compatible with the self-consistency assumption.

In the case of light quarks, the situation is not different from that in the vacuum: the quark-gluon coupling is a few times larger than the pure gauge coupling, thus preventing the use of perturbation theory. One can exploit the strategy developed in the vacuum based on the double expansion in powers of the inverse number of colours and the pure gauge coupling. As already explained above, this involves the resummation of rainbow diagrams in the quark propagator. Solving the corresponding equations can be done with present day technology \cite{Fischer:2018sdj} but has not been attempted yet in the context of the CF model. Instead, as a first step in this direction, a drastically simplified version of the rainbow resummation at nonzero temperature and quark chemical potential has been implemented in Ref.~\cite{Maelger:2019cbk}, which gives encouraging results. 

\begin{figure}[t]
    \centering
        \includegraphics[width=.45\linewidth]{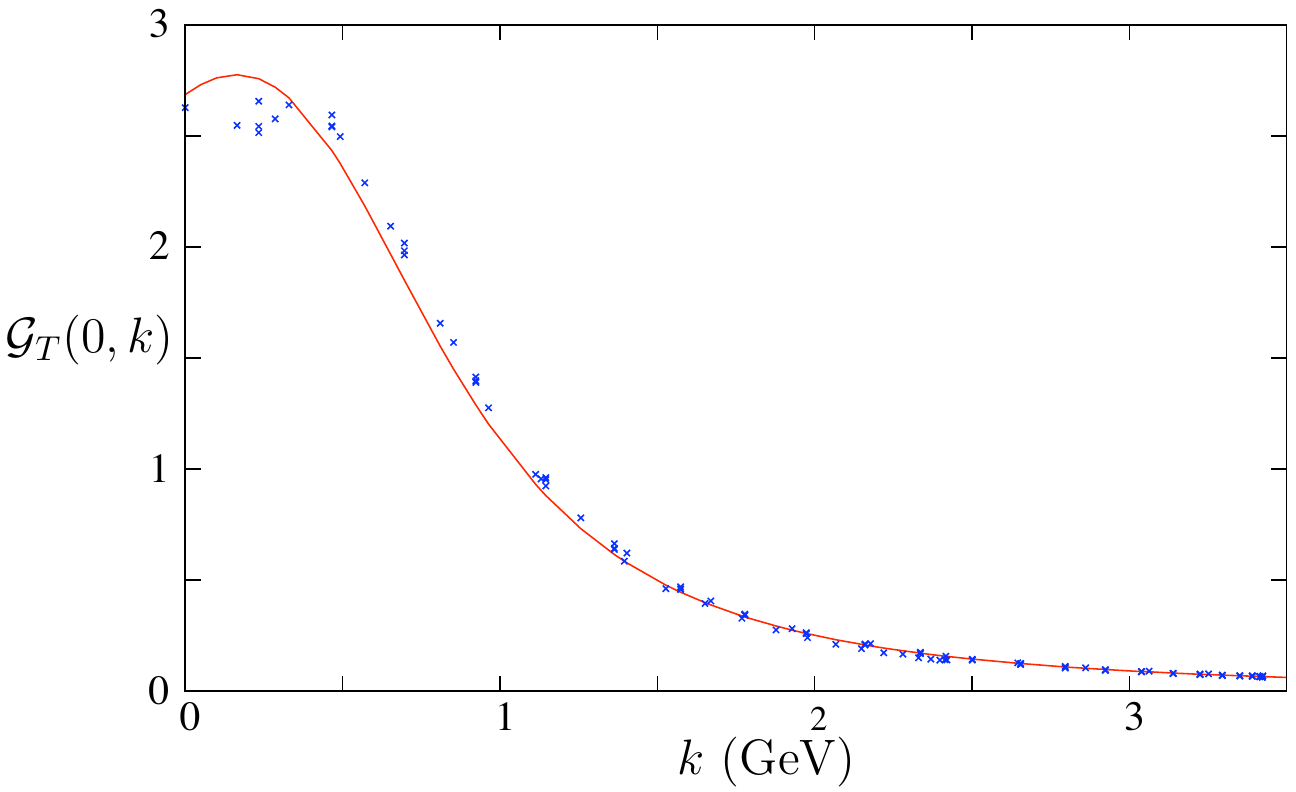}\qquad\includegraphics[width=.46\linewidth]{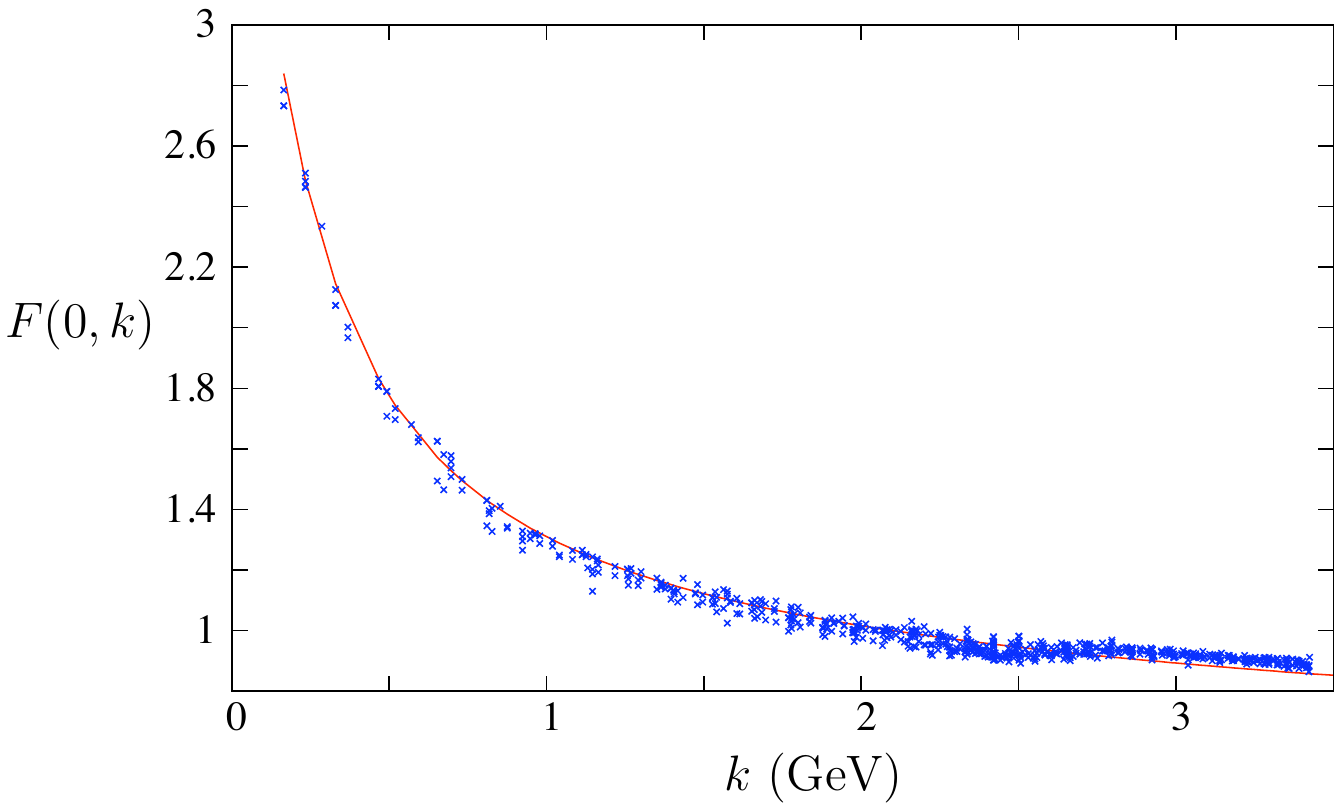}
    \caption{The magnetic gluon (left) and the ghost (right) propagators at vanishing (Matsubara) frequency as a function of spatial momentum in the SU($2$) YM theory at $T=T_c$, in the Landau gauge. The curves are the one-loop results (without RG improvement) in the CF model and the blue points are the data from Ref.~\cite{Maas:2011ez}. Figures from Ref. \cite{Reinosa:2013twa}.}
    \label{finiteTprop}
\end{figure}

\subsection{Propagators}

The results presented in the previous subsections concern gauge-invariant quantities. As in the vacuum, lattice simulations also provide results for gauge-dependent correlation functions at nonzero temperature and density. In particular, many results have been produced for the YM and QCD propagators in the Landau gauge \cite{Heller:1995qc,Heller:1997nqa,Cucchieri:2000cy,Cucchieri:2001tw,Cucchieri:2007ta,Fischer:2010fx,Cucchieri:2011di,Cucchieri:2012nx,Aouane:2011fv,Maas:2011ez,Silva:2013maa,Silva:2016onh,Kojo:2021knn,Song:2019qoh,Suenaga:2019jjv}.

The Landau gauge ghost and gluon propagators have been computed at one-loop order in the CF model---{\it i.e.,} with no background---at nonzero temperature and compared to lattice data for the SU($2$) YM theory \cite{Reinosa:2013twa}. The magnetic\footnote{In the Landau gauge, the gluon propagator is transverse with respect to the gluon four-momentum. In the vacuum, Lorentz invariance guarantees that there is only one possible scalar function. At nonzero temperature, however, the Lorentz symmetry group is explicitly broken into its rotation subgroup and there exist two independent 3$d$-longitudinal (electric) and 3$d$-transverse (magnetic) components.} gluon and ghost propagators are rather well reproduced, see Fig.~\ref{finiteTprop}. In contrast, the temperature dependence of the electric propagator differs substantially from the existing lattice data around the transition temperature \cite{Fischer:2010fx,Maas:2011ez}. We stress, however, that, first,  this discrepancy is not specific to the perturbative CF approach, which produces in fact results very similar to those of nonperturbative continuum approaches \cite{Fister:2011uw} and, second, that the lattice results in the electric sector suffer from large uncertainties \cite{Mendes:2014gva}. It has been suggested \cite{Fister:2011uw} that the discrepancy between continuum and lattice results may originate from the fact that the order parameter associated to the deconfinement transition is not properly accounted for in the perturbative Landau gauge calculations. Also, perturbative calculations in the CF model suggest that the limit of vanishing background field, corresponding to the Landau gauge, is unstable against small deviations \cite{Reinosa:2014ooa}, which may explain the large numerical uncertainties mentioned here.

This has opened the way to the evaluation of correlation functions in background Landau gauges using the CF model \cite{Reinosa:2016iml,VanEgmond:2021mlj} with the idea that the proper inclusion of the order parameter may stabilize the comparison to lattice results. Although data for two-point functions in such gauges are not available yet, some interesting results have already been obtained within the CF model. In the self-consistent background gauge, the zero-momentum limit of the SU(2) electric propagator features a relatively sharp peak at the transition \cite{Reinosa:2016iml} which turns into a divergence in the case of the center-symmetric background gauge \cite{VanEgmond:2021mlj}. In fact it has been argued that this divergence should also be there in the self-consistent approach \cite{Reinosa:2020mnx} were it not for the use of (inevitable) approximations that jeopardise certain properties of the gauge-fixing. In this sense, the results obtained with the center-symmetric background gauge, because they do not rely on these properties, should be more robust.

\begin{figure}[t]
    \centering
        \includegraphics[width=1\linewidth]{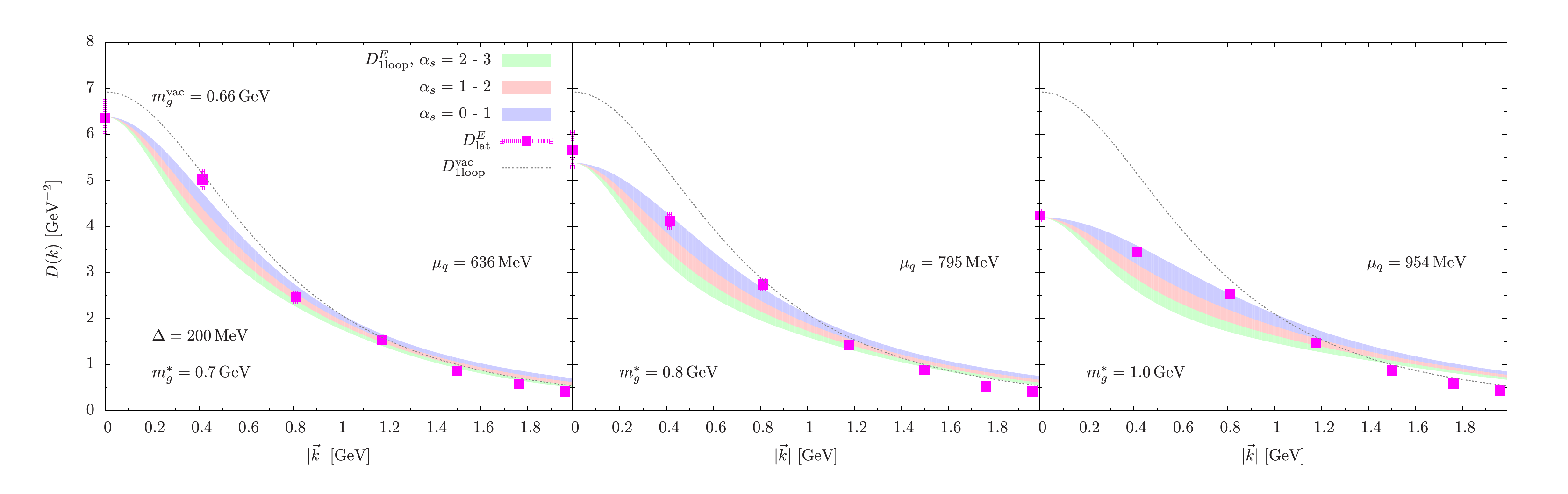}
    \caption{The electric gluon propagator in two-colour QCD in the Landau gauge at zero temperature and nonzero quark chemical potential $\mu_q$. The various curves correspond to the one-loop result in the CF model (with no RG improvement) for different values of the coupling \cite{Suenaga:2019jjv}. The chemical potential increases in the plots from left to right. The vacuum ($\mu_q=0$) result is also shown (dotted line) for reference. The quark gap $\Delta$ is held fixed and the gluon mass parameter is adjusted by hand to obtain a good overall description of the lattice data (squares) of Ref.~\cite{Boz:2018crd}. Figure from Ref. \cite{Suenaga:2019jjv}.}
    \label{KojoFig}
\end{figure}

Similar one-loop calculations (without background) have been performed and compared to lattice calculations in the Landau gauge for two-colour QCD at nonzero quark chemical potential\footnote{For $N_c=2$, lattice simulations are possible at nonzero chemical potential because there is no sign problem: The integration measure under the Euclidean path integral is positive definite.} \cite{Suenaga:2019jjv,Kojo:2021knn}. In this case, the physics is that of a possible Bardeen-Cooper-Schieffer (BCS) phase with a nontrivial pairing between quarks. The mentioned references consider the effect of a phenomenological BCS gap $\Delta$ on the electric and magnetic components of the gluon propagator through the quark loop contribution. As for the nonzero temperature case above, one allows for a variation of the CF parameters (coupling and gluon mass) with the chemical potential, to account for possible in medium effects. In a nutshell---we refer the reader to these references for details---one obtains good agreement between the one-loop expressions and the lattice data, as illustrated in Fig.~\ref{KojoFig} for the electric propagator. A similar quality is achieved for the magnetic sector as well.

\section{Results in the Minkowskian domain}\label{sec:Mink}

So far, we have reviewed the large piece of evidences accumulated over the past decade which strongly support the idea of a valid perturbative description of (some aspects of) the IR dynamics of YM and QCD-like theories. The many successful comparisons to lattice data in the Euclidean domain encourage one to use the perturbative approach in situations where lattice techniques are not available. One example, described above, is the phase diagram of QCD at nonzero quark chemical potential. Another important example concerns the study of correlation functions in the Minkowskian domain. 

The analytic structures of the gluon, ghost, and quark propagators in the complex momentum plane have been studied at one-loop order in the Landau gauge CF model \cite{Hayashi:2018giz,Kondo:2019rpa,Hayashi:2020few,Hayashi:2020myk} as well as in the screened perturbation theory approach \cite{Siringo:2016jrc,Siringo:2017svp,Siringo:2021fxo} in the vacuum and at nonzero temperature and density, with and without RG improvement. These studies are based on analytically continuing the Euclidean propagators\footnote{Note that this is not quite the same thing, in general, as working with the Minkowskian version of the CF model. In particular, the presence of pairs of complex poles (see below) geopardises the usual Wick rotation. It is not known to us whether a clear link exists between the (analytically continued) Euclidean CF model and its Minkowskian version beyond perturbation theory.} to the whole complex plane of square momentum $s=-p^2$. The most important results are that both the gluon and the quark propagator possess pairs of complex conjugate poles and that their spectral functions are not positive definite. The spectral functions mentioned here are defined as the imaginary parts of the corresponding propagators along the real $s$ axis. They vanish identically in the Euclidean domain $s\le0$ and are nonzero for Minkowkian momenta $s\ge0$. It is worth emphasising that the spectral functions defined in this way are not exactly those entering the (assumed) spectral representations discussed in Sec.~\ref{Sec_propagators}, due to the presence of poles away from the real $s$ axis; see, {\it e.g.}, \cite{Hayashi:2020few}. So, although related, the nonpositive spectral functions reported here are not to be put in one-to-one correspondence with the positivity violations mentioned in Sec.~\ref{Sec_propagators}, observed in lattice calculations. 

\begin{figure}[t]
    \centering
        \includegraphics[width=.43\linewidth]{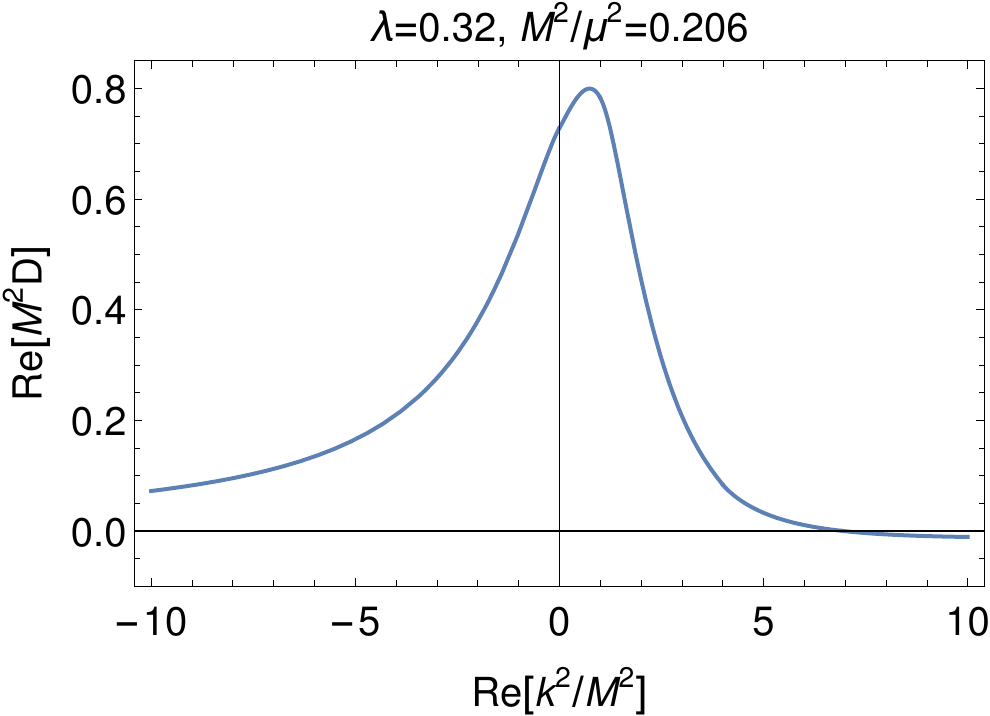}\qquad\includegraphics[width=.45\linewidth]{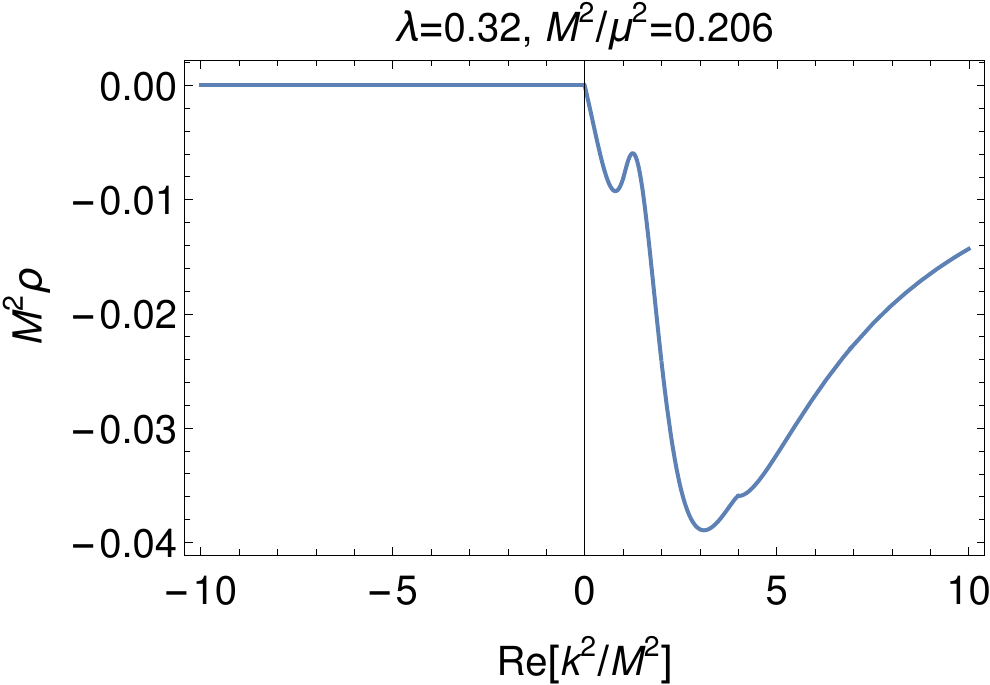}
    \caption{Real (left) and imaginary (right) parts of the vacuum gluon propagator at one-loop order in the (pure gauge) CF model as a function of the square momentum. The calculation is done within the TW renormalisation scheme and does not include RG running. Here, the CF gluon mass is denoted $M$ and $\lambda=g^2 N_c/(16\pi^2)$, with $N_c=3$. The values of the parameters are chosen to provide a good fit of the lattice data for the real part in the Euclidean domain $s=k^2<0$. The imaginary part of the propagator, the so-called spectral function is negative, indicating that the (massive) gluon is not an asymptotic state in the theory, in line with the expectation from confinement. Figures from Ref.~\cite{Kondo:2019rpa}.}
    \label{YMReImGluon}
\end{figure}

This being said, both the nonpositive spectral functions and the presence of pairs of complex-conjugate poles in the $s$-plane are in line with the fact that neither the massive gluon field nor the quark field correspond to actual asymptotic states. That is sometimes viewed as a sign of confinement although, as explained in Sect.~\ref{sec_previous_semianalytical}, this interpretation is subject to caution. First, the nonzero imaginary part of the poles results in a finite lifetime of  possible gluonic or quark excitations. Second, the nonpositive spectral functions show the absence of a proper K\"all\'en-Lehmann representation of asymptotic states with positive norm.

\section{Open questions}\label{sec:open}

At this point, we hope to have convinced the reader that the CF model provides a very efficient framework for a valid description of many aspects of IR QCD based on controllable expansion schemes. Confronted to the many successful results reviewed here, a natural question that comes to mind is: how can such a simple model be so efficient? Can this be accidental? If one views the gluon mass parameter as a mere phenomenological IR deformation of the FP theory, the model works beyond expectations, almost unreasonably well. This calls one to wonder whether the CF model could have a deeper connection to (Landau gauge) IR QCD. This line of thought requires one to seriously address the various open problems of the CF model. Let us briefly mention---and speculate about---some of the most pressing issues.

One essential question is the status of the CF mass parameter. An interesting possibility is that it could be related to the issue of properly fixing the gauge in nonAbelian theories \cite{Serreau:2012cg,Kondo:2014sta}. A recent proposal along these lines in the Landau gauge, exploiting the early ideas of Ref.~\cite{Serreau:2012cg} to handle the Gribov problem, has been shown to accommodate for a gluon mass term \cite{Nous:2020vdq}, although the resulting gauge-fixed action slightly differs from that of the CF model. Such gauge fixings do not exactly correspond to those realised in lattice simulations, but the successful comparisons described in this review suggest that the mass parameter can be adjusted to mimic the essential features of the latter \cite{Serreau:2012cg,Maas:2013vd}. Another attractive possibility would be that the gluon mass is a dynamical consequence of the gauge-fixing procedure, whose value is fixed from the sole knowledge of the gauge coupling at a given scale, similar to what happens in the GZ approach \cite{Vandersickel:2012tz}. A concrete realisation of this scenario has been proposed in a general class of nonlinear gauges in Ref.~\cite{Tissier:2017fqf} which, unfortunately, does not survive the Landau gauge limit. Finally, it could also be that the gluon screening mass is an actual feature of the FP theory in the Landau gauge at a nonperturbative level. This has been investigated in the context of the screened perturbation theory, where the mass term is introduced as a variational parameter that must eventually be fixed by an extremisation procedure. A first interesting attempt in this direction has been worked out in Ref.~\cite{Comitini:2017zfp} which, however, lacks the systematics of a genuine loop expansion and sometimes leads to unphysical results (like a spurious first order phase transition at nonzero temperature for the SU($2$) YM theory. Finally, we mention that the hypothesis that the gluon mass is dynamically generated in the  FP theory underlies all the studies based on nonperturbative continuum approaches.\footnote{As a side remark, it is interesting to note that this hypothesis actually supports the idea that the CF Lagrangian may be more than a mere phenomenological model and should perhaps be considered as a more fundamental realisation of Landau gauge QCD. Indeed, for technical reasons---not to be exposed here, see {\it e.g.}, \cite{Huber:2020keu}---, these approaches (DSE, FRG, VHA) actually introduce a tree-level gluon mass parameter in one way or another \cite{Reinosa:2017qtf} and are thus effectively based on the CF Lagrangian rather than on the FP one. The basic assumption of these approaches is that there exists a particular value of the CF mass which actually corresponds to the FP theory.}

A more technical question is that of IR divergences in the Minkowskian domain. As mentioned in Sect.~\ref{sec:vac}, IR divergences in the Euclidian domain were studied in
\cite{Tissier:2011ey} and it was shown that, thanks to the presence of the gluon mass, there are no IR divergences for non-exceptional Euclidean momenta (for $d>2$). The case of exceptional Euclidean momenta is more delicate \cite{Tissier:2011ey}. Also, the analysis of IR divergencies in the Minkowkian domain has not yet been carried out.
This remains an open question because, as in QED, and despite the fact that the gluons are massive in the present model, the analysis of on-shell IR divergences is far from trivial due to the presence of massless modes in the CF model.


Other major issues are the construction of a proper physical space and the question of unitarity. As explained in Sec.~\ref{sec:unitarity}, and contrary to a widespread idea, this is still an open question within the CF model. In fact, it is important to stress that it is not even understood in the standard FP approach beyond perturbation theory. All continuum approaches have to face this issue, which directly affects the calculation of physical observables. A simple but generic example is the thermodynamic pressure a low temperatures. In all cases, the difficult task is the identification of the actual (confined) physical space. 

An intricately related question, the mother of all, is that of confinement. The lack of
perturbative unitarity in the CF model has led to its disproval as an alternative to the
Higg's mechanism for a consistent theory of massive gauge bosons. However, it could very
well be that the model is confining, which would completely change the game.
Despite its importance, the precise definition of confinement has not been clearly established in the case of QCD (see {\em e.g.} the discussion in \cite{Greensite:2011zz}). In the case of pure YM theory, the commonly accepted criterion for characterizing confinement corresponds to the Wilson loop area law. Although the Wilson loop is a perfectly well-defined quantity in the Euclidean domain, the area law corresponds to a linear behaviour for the static potential of a very distant quark-antiquark pair. In this limit, this potential is dominated by the behaviour of correlation functions with momenta out of the Euclidean domain near the singularities of various correlation functions. Accordingly, this behaviour is plagued by IR divergences that, at the moment, have not been put under control in the CF model. Controlling these IR divergences and reproducing the area low behaviour would clearly be of utmost importance.

\section{Conclusions}\label{sec:concl}

The CF  model offers a promising avenue for investigating IR roperties of continuum nonAbelian gauge theories in the Landau gauge beyond the textbook  FP gauge-fixing prescription, which is limited to the UV. The systematic analysis of the two- and there-point vacuum correlation functions of the model and their comparison with results of {\it ab initio} lattice simulations in YM and QCD-like theories strongly support the idea that the inclusion of a gluon mass operator beyond the FP Landau gauge action suffices to capture most of the qualitative and many of the quantitative features of the Landau gauge correlations functions in the vacuum, from which one can extract physical observables. 

The remarkable point is that many of these results (those for YM theory but also those for QCD with heavy quarks) rest on a purely perturbative approach within the CF model. For one thing, the model admits renormalisation group trajectories defined over the whole range of scales, in blatant contrast to the FP approach, which features an IR Landau pole, and in good agreement with lattice data. For another, the trajectories that allow one to best reproduce the results of simulations correspond to a running gauge coupling that never gets excessively large, thus justifying {\it a posteriori} the use of a perturbative approach. This, in turn, not only allows for a simple computational setup at leading order, in comparison to more demanding nonperturbative methods, but also offers the possibility to systematically investigate higher-order corrections and, thereby, the validity of the approach. Various two-loop corrections have been evaluated for YM and (heavy-quark) QCD two- and three-point correlation functions and have been found to be globally small while improving the quality of the comparison to the lattice data.

The perturbative approach is not a panacea, however, not even within the CF model, and certain questions require one to go beyond the simple coupling expansion. This is the case of the light quark sector of QCD controlled by the spontaneous breaking of chiral symmetry and characterized by an enhanced quark-gluon coupling in the IR, as compared to the typical couplings in the pure gauge sector. The perturbative nature of the pure gauge sector of the CF model allows nonetheless for the construction of a systematic expansion scheme that rests on two small parameters, the pure gauge coupling on the one side, and the inverse number of colours on the other side. Already the leading orders of this expansion scheme suffices to capture the general features of the spontaneous breaking of chiral symmetry while providing a consistent picture of the various two-point correlation functions. It would be interesting in the future to extend this analysis to the three-point correlators as well. The ultimate goal is of course to investigate low energy observables, within nonAbelian gauge theories in general, and within QCD in particular. A preliminary determination of the pion decay constant $f_\pi$ within this approach is very encouraging and gives good confidence that other observables, such as the spectrum of low-lying hadrons are within reach.

Various relevant observables at nonzero temperature and chemical potential have also been evaluated, including transition temperatures and nontrivial order parameters involving nonperturbatively large field configurations. As in the vacuum case, the perturbative CF model seems to capture most qualitative features of the phase structure such as the confinement-deconfinement transition in YM theories, the critical line in the heavy-quark region of the Columbia plot, as well as the RW transition for imaginary chemical potential. In most cases, the description is even quantitative, with, for instance, transition temperatures that gives results comparing well with simulations at one loop and improving, sometimes significantly, at two-loop order. Similar conclusions emerge from the evaluation of quark mass-to-temperature ratios along the upper critical line in the Columbia plot. In the case of QCD with light quarks, perturbation theory is again not enough but one can extend the approach mentioned above for the vacuum case at nonzero temperature and chemical potential. Preliminary results show that the CF model has the potential to access features of the phase structure in this case too, in particular the possible existence of a critical end point in the QCD phase diagram. This needs to be confirmed by more refined studies.
  
Of course,  there are still many open questions concerning the CF model and its use as (part of) a nonperturbative completion of the gauge fixing beyond the FP construction. Its many successes should serve as an incentive to better understand the nonperturbative gauge-fixing procedure, with the potential reward of granting a relatively simple access to some of the low-energy observables of QCD. 

To conclude, lattice simulations have firmly established (see Fig. \ref{Fig:lattice-Landau1}) that at least part of the IR regime of QCD is governed by a coupling of perturbative size in the Landau gauge. Although it has remained largely unknown to a broad audience, this is an observation of paramount importance with far-reaching consequences. We believe that the work reviewed in this article  clearly establishes that many facets of the IR QCD dynamics admit a perturbative description in the Landau gauge and that the CF model is an efficient framework for the latter, be it at a fundamental or at a more phenomenological level. We hope that the present article will convince the reader of the usefulness of a change of paradigm concerning the IR dynamics of nonAbelian theories and will motivate QCD practitioners to include the CF model as one serious option in their toolbox \cite{Hadjimichef:2019vbb,Song:2019qoh}.

\begin{acknowledgments}

 We thank B. Delamotte, N. Dupuis, and G. Tarjus for a careful reading of the manuscript. 
 We acknowledge the financial support from PEDECIBA, from the "Institut Franco-Uruguayen de Physique",  and from 
the ECOS program U17E01 and from the ANII-FCE-126412 project.

\end{acknowledgments}

\bibliography{./bibli.bib} 
\bibliographystyle{alpha}

\end{document}